\documentclass[twocolumn,preprintnumbers,amsmath,amssymb]{revtex4}

\usepackage{graphicx}
\usepackage{dcolumn}
\usepackage{bm}
\usepackage{pdfpages}

\begin{document}


\title{An easy and efficient approach \\
for testing identifiability of parameters}

\author{Clemens Kreutz}
 \email{ckreutz@fdm.uni-freiburg.de}
\affiliation{ \vspace{0.5cm}
Freiburg Center for Systems Biology (ZBSA), \\
University of Freiburg, 79104 Freiburg, Germany
}


\begin{abstract}
The feasibility of uniquely estimating parameters of dynamical systems from observations is a widely discussed aspect of mathematical modelling. 
Several approaches have been published for analyzing identifiability. 
However, they are typically computationally demanding, difficult to perform and/or not applicable in many application settings.

Here, an intuitive approach is presented which enables quickly testing of parameter identifiability.
Numerical optimization with a penalty in radial direction enforcing displacement of the parameters is used 
to check whether estimated parameters are unique, or whether the parameters can be altered
without loss of agreement with the data indicating non-identifiability.
This \emph{Identifiability-Test by Radial Penalization (ITRP)} can be employed for every model where optimization-based fitting like \emph{least-squares} or \emph{maximum likelihood} is feasible and is therefore applicable for all typical deterministic models. 

The approach is illustrated and tested using 11 ordinary differential equation (ODE) models.
The presented approach can be implemented without great efforts in any modelling framework.
It is available within the free Matlab-based modelling toolbox \emph{Data2Dynamics} \citep{Raue2015}. 
Source code is available at https://github.com/Data2Dynamics.
\end{abstract}

\maketitle
\section{Introduction}
An essential step of mathematical modelling is estimation of parameters.
Although the methodology is not restricted to ordinary differential equation (ODE) models, 
the focus in this paper is on this class of models because they are frequently used to describe 
the dynamics of molecular compounds, e.g.~involved in signalling pathways or gene regulation networks.
In this setting, parameters represent abundances of cellular compounds or their interaction strengths,
but can also comprise scaling- or variance parameters for the measurements.
Defining dynamic models by translating molecular interaction maps based on biochemical rate laws 
can lead to large and over-parameterized models where 
the data does not provide enough information for uniquely estimating parameters. 
This issue has been termed \emph{non-identifiability} and occurs in all settings, where the level of detail of the model does not fit to the amount of experimental data. 

For small models non-identifiabilities can be detected by analytical approaches, e.g.~based on power series expansions \citep{Pohjanpalo78,Walter82},
calculation of transfer matrices \citep{Bellman70}, differential algebra \citep{Bellu2007, ljung94, Saccomani03}, similarity transformations \citep{Vajda81}, Lie-group theory \citep{Merkt15},
or by treating parameters as constant dynamic states and applying concepts from observability or controllability analyses \citep{Travis81}.
For more realistic settings, numerical methods have been published
which are based on the rank of the Jacobian \citep{Catchpole97,Karlsson2012} or Fisher-Information \citep{Hidalgo2001a,Viallefont98}.
Nonparametric transformations have been used in \citep{Hengl2007} to find non-identifiable parameter relationships based on multi-start optimization results.
In addition, the profile-likelihood has been suggested to investigate identifiability for given experimental data \citep{Raue2009, Raue2011}.
Since this approach is tailored to nonlinear systems as they frequently occur in systems biology and it provides statistically valid confidence intervals,
 this method might be the currently most frequently applied approach in this field.
However, calculation of likelihood profiles for all parameters is time-consuming, especially for large systems and there are recent efforts for developing computationally more efficient methods \citep{Raman17}.
Despite the multitude of approaches, the ongoing discussion and research in this field still indicates lack of efficient and broadly applicable approaches.

In this manuscript, penalized optimization is employed for testing of structural identifiability by an additional model fitting step.
The penalty enters like an additional data point which is used to pull in the parameter direction where the data provides least information.
This approach enables a fast and reliable procedure for identifiability analysis and thereby resolves a major bottleneck of mathematical modelling.
The applicability is demonstrated for two illustration- and nine application models.

\section{Methods}
In the following sections, the mathematical notation is introduced.
Since several definitions and terms of identifiability/non-identifiability exists, different terminologies are also briefly summarized.

\subsection{Model setting}
Although the presented method is applicable for any model where parameters are estimated by maximum-likelihood or any other optimization-based
fitting method, the discussion is restricted to ODE models in the systems biology context in the following.
In this setting, chemical reaction laws are frequently utilized to define rate equations $f$ describing the dynamics 
\begin{equation} 
\dot x(t) = f(x,u,\theta,t)
\end{equation}
of concentrations $x \in  \mathbb{R}^{n_x}$. 
Different stimulations or perturbations are represented in the model by inputs $u \in  \mathbb{R}^{n_u}$.
The inital values $x(0)$ are either known or defined as additional parameters, i.e.~$x(0)\subset \theta$.
The dynamic states are linked to measurements 
\begin{equation}
y_i = g_i(x,\theta) + \varepsilon_i \:\:, \:\:\varepsilon_i \sim N(0,\sigma_i^2)
\end{equation}
via observation functions $g_i \in  \mathbb{R}$, $i=1,\dots,N_{\text{data}}$ which might comprise scalings and/or transformations like a log-transformation.
In this formulation, $\varepsilon_i$ represents additive Gaussian noise although the approach is not restricted to this type of noise 
and other distributions can be considered by defining the likelihood $L(\theta) \equiv \rho(y|\theta)$ with the respective 
density function $\rho$ for the measurement errors.
In an easy setup, the magnitude of the experimental errors is constant, i.e. $\sigma_i \equiv \sigma$.
In general, an error model 
\begin{equation}
	\sigma_i = E_i(x,\theta)
\end{equation} 
e.g.~$E_i(x) = \sigma_\text{abs} + \sigma_\text{rel} \times x$, might depend on $x$ and could also contain parameters like absolute or relative noise levels.
Therefore, in a general formulation the parameter vector $\theta \in \mathbb{R}^{n_\theta}$ contains all unknown constants determining the dynamics, 
the predicted observations and the noise levels.

Note that a notation has been chosen where index $i$ enumerates individual measurements 
and each data point $y_i$ has an individual observation function $g_i$ for a specific experimental design \citep{Kreutz2009}
given by the time point, observed state(s) as well as possible assignments to inputs, offset- or scaling parameters and error model $E_i$.
This notation emphasize that measurements are often performed for different but sparse combinations of observables/measurement techniques, 
stimulation/perturbations and time points and is therefore commonly used in statistics, 
e.g.~for multivariate models like linear models, mixed effects models or survival models.

\subsection{Parameter estimation and objective function}
Maximum likelihood estimation 
\begin{equation}
\hat \theta_\text{MLE} = \arg \max_\theta \log L(\theta)
\end{equation}
has several beneficial statistical properties like asymptotic normality, consistency and efficiency \citep{cox94}.
For known Gaussian errors $\varepsilon_i \sim N(0,\sigma_i^2)$, least-squares estimation
\begin{equation}
	\hat \theta_{LS} = \arg \min_\theta \chi^2(\theta|y) 
\end{equation}
with the least-squares objective function
\begin{equation} \label{eq:chi2}
	\chi^2(\theta|y) = \sum_i \frac{ \left( y_i - g_i(\theta) \right)^2}{\sigma_i^2(\theta)}
\end{equation}
is a special case of maximum likelihood estimation since the estimate coincides with $\hat \theta_\text{MLE}$ 
because $\chi^2(\theta|y) = - 2 \log L(\theta|y) + const$.
If prior knowledge about some parameters is available, this can be accounted for by using a penalized log-likelihood
\begin{equation}
	\log L_\text{pen}(\theta) = \log L(\theta|y) + \sum_j \log \pi_j(\theta)
\end{equation}
which in the case of Gaussian priors $\pi_j \equiv \theta_j \sim N(\bar \theta_j,\bar\sigma_j^2)$ also yields a sum of quadratic terms
and can be treated like additional data points.
In the following, $V_\text{data}(\theta)$ is used as a general place-holder for the objective function which is without loss of generality assumed to be minimized
\begin{equation} 
	\hat \theta = \arg \min_\theta V_\text{data}(\theta)
\end{equation}   
for parameter estimation, i.e.~$V_\text{data}(\theta)$ might coincide with $\chi^2(\theta)$, $-2 \log L(\theta)$ or $-2 \log L_\text{pen}(\theta)$.
Independently of the chosen objective function and its signum, we term $V_\text{data}(\theta)$ as likelihood in the following.

\subsection{Parameter identifiability}
A variety of terms and definitions for non-identifiability are available in literature.
In general, non-identifiability refers to lack of information for uniquely specifying the parameters.
In this subsection, two complementary points of view are summarized.

\subsubsection{Mathematical point of view}
A series of papers consider a setting with predefined observables $g(t,x)$ but without specifying observation noise nor number and location of time points.
A widely used mathematical definition of non-identifiability \citep{Chis11} in this setting is:
A parameter $\theta_i$, is \emph{structurally locally identifiable} if for almost any $\theta_i$, there exists a neighbourhood $P$ such that if
\begin{equation} \label{eq:idMath}
\left . \begin{array}{c} \theta \in P \\  g(\theta_i^{(1)})  = g(\theta_i^{(2)}) \end{array} \right \} \Rightarrow \theta_i^{(1)} = \theta_i^{(2)} \:\:.
\end{equation}
If this property holds not only within a neighbourhood but for the whole parameter space, the parameter is termed \emph{structurally globally identifiable}.

This formulation considers analysis of identifiability of model parameters for a given set observation functions.
It is independent on the number and accuracy of data points and therefore fits well to applications settings 
where (almost) continuous and noise-free observations are feasible.
In this setting, identifiability refers to a unique mapping from the observed dynamics to parameters \citep{ljung99}.
In agreement with this, \citep{Bellman70} defined structural identifiability already in 1970 as a unique minimum of
\begin{equation}
	V_\text{Bellman}(\theta) = \int \left( y(t) - g(t,\theta)  \right)^2 dt \;  \:.\label{eq:bellman}
\end{equation}

In many application disciplines like cell biology, however, the number of data points is limited and the measurements exhibit a non-neglectable amount of noise.
Then, parameter identifiability depends on availability of measurements, e.g.~on the number of data points 
and on the exact combinations of measurement times, input and observation functions.
Moreover, non-identifiabilities occurring in parameters of the observation functions or error models have to be considered.
Therefore, the mathematical definition ($\ref{eq:idMath}$) or ($\ref{eq:bellman}$) does only partly capture the effects occurring in inverse problems
and can be reasonably extended by a formulation considering individual data points.

\subsubsection{Statistical point of view}
In a statistical formulation, parameter identifiability can be defined as a unique minimum of the log-likelihood \citep{Little10}, 
e.g.~in the least-squares setting as a unique 
minimum of 
\begin{equation}
	V_\text{data}(\theta) = \sum_i \frac 1 {\sigma_i^2} \left( y_i - g_i(t_i,\theta)  \right)^2 \label{eq:V}
\end{equation}
where summation is performed over all data points, i.e.~over all combinations of input, observation function and measurement time.
Moreover, as discussed above, summation can comprise additional terms originating from priors or estimation of noise levels.

For this setup, the profile likelihood 
\begin{equation} \label{eq:PL}
	\text{PL}_k (p) = \min_{\{\theta_{j \ne k} | \theta_k = p\}} V_\text{data}(\theta)
\end{equation}
has been suggested for assessing identifiability \citep{Raue2009} of a parameter $\theta_k$.
In ($\ref{eq:PL}$), all parameters are optimized except $\theta_k$ for different values $p$ for $\theta_k$. 
A parameter with a flat profile likelihood indicates non-unique parameter estimates because changing $\theta_k$ can be entirely compensated by refitting the other parameters.
Such flat directions in the parameter space indicate redundant parametrizations.
Therefore, the existence of flat likelihood profiles has been used to define structural non-identifiability \citep{Raue2009}.
This property depends on the set of given observations $g_i$, but 
is independent on the magnitude of observation noise $\sigma_i$ because scaling all $\sigma$'s only scales the objective function but does not impact existence of entirely flat manifolds.

Since the profile likelihood is applicable for any model which allows optimization-based estimation, 
it has become a standard approach in systems biology for assessing identifiability\footnote{Most citations for in \emph{Web of Science} when searching for ``identifiability analysis systems biology''}.
Using a proper threshold, the profile likelihood also enables the calculation of confidence intervals \citep{Kreutz2011}.
In some cases, it might occur that a unique minimum exists, but the profile likelihood does not exceed the confidence threshold in lower and/or upper direction.
Then, the confidence interval has infinite size.
This effect only occurs in the case of measurement noise and it vanishes in the limit $\sigma \rightarrow 0$.
Since averaging over $n$ replicates decreases the standard deviation $ \sigma \propto 1/\sqrt{n}$
this limit is asymptotically obtained by increasing the number of measurement replicates.
Because this effects only occurs due to practical limitations in generating a sufficient number of replicates, it has been used to define \emph{practical non-identifiability} \citep{Raue2009}.
Both cases are difficult to distinguish in applications since the likelihood might be locally flat but could still exceed a significance threshold
as shown later in section \ref{sec:applications}.
Since discrimination between both scenarios is only a terminological issue and not the focus of the article, we do not further discuss this aspect.
We use structural non-identifiability as synonym for locally flat likelihood
and focus on detection of this property in the following.

\section{Approach}
In the following, the major focus is the statistical point of view, i.e.~identifiability is investigated for an inverse setting with given experimental data.
In section \ref{sec:math}, the approach is adapted to investigate identifiability in the mathematical, i.e.~continuous and noise-free context.


\subsection{Testing identifiability}
Existence of redundant parametrizations, i.e.~presence of flat directions of the likelihood for a given data set, 
is investigated by penalized optimization.
After standard model fitting, i.e.~after parameters are estimated according to ($\ref{eq:V}$), 
we suggest usage of a penalized objective function 
\begin{equation} \label{eq:Vtot}
V_\text{tot}^R(\theta) = V_\text{data}(\theta) + V_\text{pen}^R(\theta)
\end{equation}
with
\begin{equation}  \label{eq:pen}
	V_\text{pen}^R(\theta) = \lambda \left( || \theta - \hat \theta ||_2 - R \right)^2  
\end{equation}in
to pull the parameter vector $\theta$ away from the estimated parameters $\hat \theta$.
The penalty term $V_\text{pen}^R(\theta)$ is quadratic and has its minimum at a circular manifold with radius $R$ centered around $\hat \theta$.
Parameters minimizing $V_\text{tot}^R(\theta)$ are denoted by $\theta^*$, i.e.~
\begin{equation}
	\theta^* = \arg \min_\theta	V_\text{tot}^R(\theta) \:\:.
\end{equation} 

The penalization strength $\lambda$ is chosen by default as $\lambda = 1/R^2$. 
Thereby, it holds $0 \le V_\text{pen}^R(\theta) \le 1, \forall \theta $ with $||\theta-\hat \theta||_2 \le R$ and the magnitude of the increase
\begin{equation} \label{eq:penalty}
	\Delta V^R = \min_\theta V_\text{tot}^R(\theta)  - V_\text{data}(\hat \theta)
\end{equation} 
of the objective function by penalization is after fitting in the interval $[0,1]$ and therefore easy to interpret.

$\Delta V^R$ is the major characteristic used to define the \emph{Identifiability-Test by Radial Penalization (ITRP)}.
In the case of structural non-identifiability, the parameters can be altered and thereby minimize the penalty without reducing agreement with the data.
Then, the penalty vanishes without increasing the data-related part $V_\text{data}(\theta)$ of the objective function.
Therefore $\Delta V^R=0$ indicates structural non-identifiability.
In contrast, $\Delta V^R>0$ indicates that the model is structurally identifiable since the parameter cannot be moved by an euclidean distance $R$ without loss of agreement with the data.
The \emph{Identifiability-Test by Radial Penalization (ITRP)} suggested in this manuscript consists of an additional fit based on ($\ref{eq:Vtot}, \ref{eq:pen}$) and evaluation whether there is an increase of the objective function ($\ref{eq:penalty}$).

\begin{figure*}[!tpb]
\includegraphics[width=\linewidth]{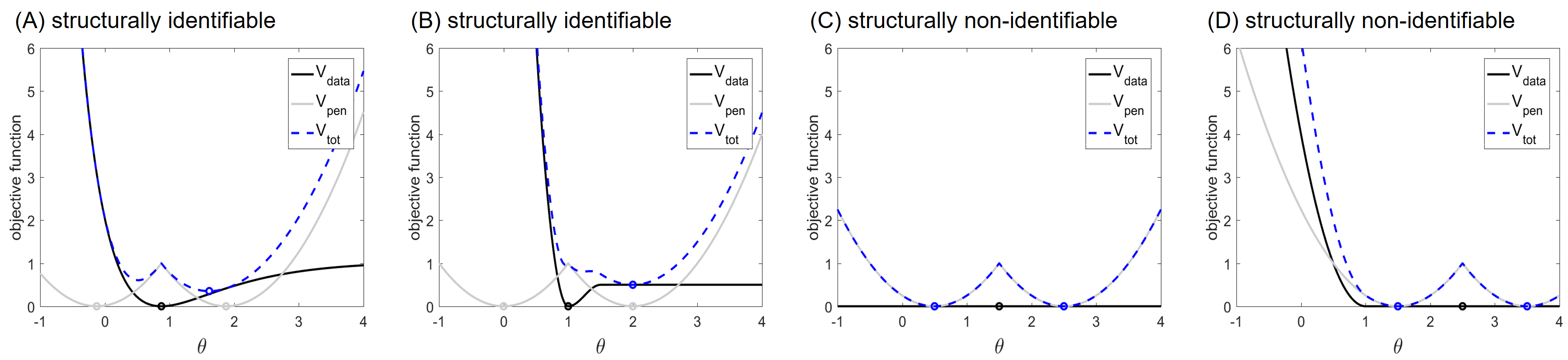}
\caption{Illustration of different scenarios for penalization radius $R=1$. 
In cases (A) and (B), there is a unique minimum (black dot), i.e.~the parameter is structurally identifiable for the given data set
and the penalty $V_\text{tot}$ increases in both cases. 
Scenario (B) also illustrates that agreement with the penalty should be assessed in terms of increase of $V$ 
and not based on distance in the parameter space.
In (C) and (D), the data-dependent part of the objective function is flat, for case (D) only towards large numbers.
In these two scenarios, the penalty can be satisfied without loss and $V_\text{tot}(\theta^*) = V_\text{data}(\hat \theta)$.  \label{fig:scenarios}
Panel (D) provides a hint for potential dependency on the penalization radius $R$ because the minimum of $V_\text{tot}$ in lower direction would vanish if $R$ is too large. In general, local non-identifiability is only detected if the radius is not too large.}
\end{figure*}

Because quadratic terms can be most efficiently optimized, the $L_2$-norm $|| . ||_2$ has been used in ($\ref{eq:pen}$) 
for calculating the distance between $\theta$ and $\hat \theta$ as well as for penalizing the distance to the target radius $R$ .
An alternative would be the $L_1$-norm which could be used to enforce that the parameters $\arg \min V_\text{pen}(\theta)$
are exactly at the sphere with radius $R$. 
This however depends on $\lambda$ and therefore requires a proper choice of $\lambda$.
Moreover, optimization could be hampered due to non-continuous derivatives which would slow down the approach.

Fig.~\ref{fig:scenarios} shows possible scenarios for the trade-off between penalty $V_\text{pen}$ and data agreement $V_\text{data}$.
Panel (B) corresponds to an identifiable setting with a unique minimum but a flat plateau which results in $\theta^*=\hat \theta$.
This scenario shows that $\Delta V^R$ is better suited for assessing identifiability 
than evaluating whether $||\theta^*-\hat \theta||$ is equal to $R$ .

\begin{figure*}[!tpb]
\includegraphics[width=\linewidth]{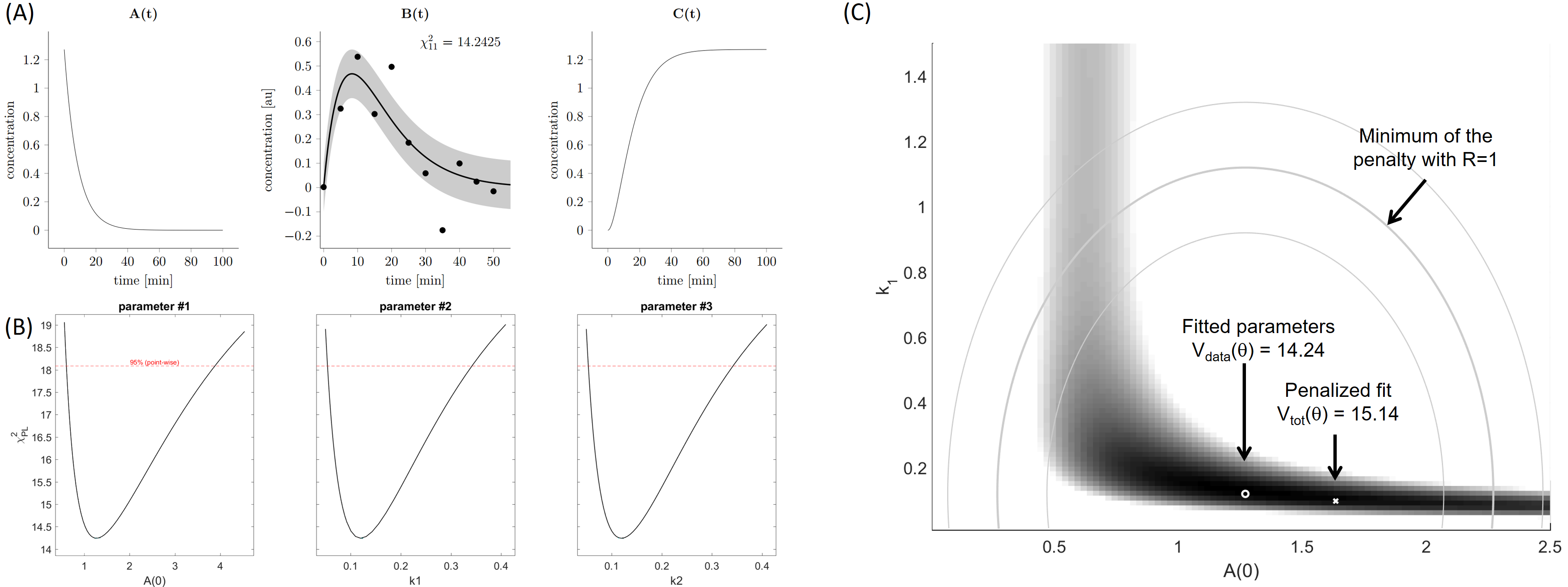}
\caption{Panel (A) shows the dynamics of the identifiable illustration model as well as the data. Gray shading indicates the size of the measurement errors.
The likelihood profiles shown in panel (B) have a unique minimum indicating parameter identifiability.
The same outcome is obtained by the penalization-based identifiability test as shown in panel (C).
Shading indicates the dependency of agreement between model and data from two parameters. The third parameter was optimized for all combinations.
Penalized fitting moves the estimated parameters towards the penalty. The resulting increase of the objective function $V_\text{tot}>V$ indicates identifiabilities. \label{fig:illu1}}
\end{figure*}

\subsection{Implementation}
Fitting a model by numerical optimization requires integration of the ODEs and an implementation of the objective function $V_\text{data}(\theta)$ as defined in ($\ref{eq:V}$).
For efficient numerical optimization, this function should also calculate the derivatives $\frac{\text dV_\text{data}}{d\theta}$. 
This is available in typical modelling toolboxes.
In the following, we provide equations for implementing the \emph{ITRP} introduced in the previous section.

For implementing the \emph{ITRP}, the standard objective function used to fit a model 
has to be augmented via $V_\text{tot} = V_\text{data}(\theta) + V_\text{pen}$ 
by adding the penalty term $V_\text{pen}$
and by respectively adapting the derivative 
\begin{equation}
\frac d {d\theta} V_\text{tot} = \frac d {d\theta}V_\text{data}(\theta) + \frac d {d\theta} V_\text{pen}
\end{equation}
with 
\begin{equation}
	\frac{d}{d\theta} V_\text{pen} = 2\lambda \left( \theta - \hat \theta \right)  \:\:.
\end{equation}
The Hessian is given by 
\begin{equation}
	\frac{d^2}{d\theta^2} V_\text{tot} = \frac {d^2} {d\theta^2}  V_\text{data}(\theta) + \frac {d^2} {d\theta^2} V_\text{pen}
\end{equation}
with 
\begin{equation}
	\frac{d^2}{d\theta^2} V_\text{pen} =  2 \lambda \:\:.	
\end{equation}

Some least-squares optimization routines like {\tt lsqnonlin} \citep{Coleman96,MatlabOTB} use data residuals
\begin{equation}
	\text{res}_i = \frac{y_i - g_i}{\sigma_i}
\end{equation}
and a Jacobian $J_{ij} = \frac d {d\theta_j} \text{res}_i$ with
\begin{equation}
	\frac d {d\theta_j} \text{res}_i = -\frac 1 {\sigma_i} \frac d {d\theta_j}g_i
\end{equation}
for optimization instead of a scalar objective function. 
These algorithms internally calculate the sum of squared residuals within the optimization routine and approximate the Hessian matrix by $J^\top J$.
For applying the identifiability test, the residual vector has to be augmented with the square-root of ($\ref{eq:pen}$), i.e.~by
\begin{equation}	
	\text{res}_\text{pen} = \sqrt\lambda  \left( \sqrt{\sum_i \left( \theta_i -  \hat\theta_i \right)^2}  - R\right)
\end{equation}
and the derivatives with
\begin{equation}
	\frac{d}{d\theta_j}  \text{res}_\text{pen} = \sqrt\lambda \frac{ \theta_j - \hat\theta_j}{ \sqrt{ \sum_i  \left( \theta_i -\hat \theta_i \right)^2} }  
\end{equation}

Since the presented approach is numeric, a threshold $\delta$ is required to decide whether $\Delta V^R$ is larger than zero.
A proper choice of $\delta$ depends on the accuracy of optimization e.g.~on the termination thresholds.
In our examples, 
we chose $\delta=1e$-$3$ which worked for all application examples.
For properly choosing $\delta$, we suggest to use a termination criterion for optimization which is based on minimal changes 
of the objective function $V$, e.g.~{\tt TolFun=1e-6} in Matlab notation \citep{MatlabOTB} 
instead of threshold based on parameter changes ({\tt TolX} in Matlab notation).  

In principle, a single penalized fit is sufficient to detect non-identifiability.
However, to increase the robustness of the outcome with respect to non-converging fits we chose a multi-start strategy 
with five fits with different initial guesses throughout the manuscript.
In the Supplementary Information we show the dependency on the number of fits and show that for all models two initial guesses, one using $\hat \theta$ as starting point and one random choice, are sufficient to perform the \emph{ITRP}.

\subsection{Parameter subsets}
In some applications, non-identifiability might only matter for a specific subset $\Theta_\text{sub}$ of the parameters, e.g.~the dynamic parameters.
Then, the exact values of other parameters, e.g.~scaling parameters, might be of minor concern.
In such a situation, only the parameters of interest $i \subset \Theta_\text{sub}$
should be used to define the penalty and ($\ref{eq:penalty}$) becomes
\begin{equation} \label{eq:Vsubset}
	V^R_\text{pen} = \lambda  \left( \sqrt{\sum_{i \subset \Theta_\text{sub}} \left( \theta_i -  \hat\theta_i \right)^2}  - R\right)^2
\end{equation}

\subsection{Iterative analysis} \label{sec:recursive}
The parameter component 
\begin{equation} \label{eq:least}
	i^* = \arg \max_i | \theta_i^* - \hat \theta_i|
\end{equation} 
which the largest change due to penalized optimization can be termed as the \emph{least identifiable parameter}.
In the case of non-identifiability, this parameter index indicate a non-identifiable parameter,
although the result might not be unique in the case of several non-identifiabilities.
Fixing such a parameter 
enables investigation of the remaining non-identifiabilities.
By repeatedly applying this procedure, the number of non-identifiabilities can be found, i.e.~the number of parameters which have to be fixed 
(or estimated elsewhere) for obtaining an entirely identifiable model.
This procedure is illustrated in the results section \ref{sec:applications}.

\subsection{Investigating mathematical identifiability \label{sec:math}} 
Data-based non-identifiability ($\ref{eq:idMath}$) can be seen as a necessary but not sufficient prerequisite for mathematical non-identifiability 
which is based on continuous, noise-free observations.
The \emph{ITRP} introduced above can be adapted to also investigate mathematical identifiability.
For this purpose, the limiting case $\sigma \rightarrow const., N_\text{data}\rightarrow \infty$ has to be considered.
This can be seen by comparing $\chi^2$ in ($\ref{eq:chi2}$) with the integral difference ($\ref{eq:bellman}$).
The constant value used to replace measurement uncertainties $\sigma$ is relevant for the \emph{ITRP} 
from the numerical point of view for distinguishing increasing from non-increasing objective functions $\Delta V^R$, 
i.e.~has to be chosen properly in relation to the magnitude of the threshold $\delta$.

We used the accuracy of numerical integration which is roughly specified by absolute and relative tolerances $atol$ and $rtol$ of the numerical ODE intergration algorithm. 
In the Supplementary Information, mathematical identifiability is investigated for a pathway model with
\begin{equation} \label{eq:sigma}
	\sigma = N_\text{sim}\left(\text{\emph{atol}} + \text{\emph{rtol}} \times x \right)  
\end{equation}
where \emph{atol} refers to the absolute integration tolerance and \emph{rtol} to the relative.
Since each time point where the dynamics is evaluated contributes to the objective function and 
the outcome should not dependent on the number of chosen data points $N_\text{sim}$ used to evaluate the dynamics.

\subsection{Scope and restrictions} 
The procedure is only applicable if the objective function $V_\text{data}(\theta)$ used for parameter estimation is deterministic.
This means, that the procedure does not reliably work for models with a stochastic dynamics.
Moreover, the \emph{ITRP} as presented above requires estimated parameters $\hat \theta$ as starting point.
Non-optimality of this parameters could be indicated by negative $\Delta V_\text{tot}$.

The presented approach rely on a reliably working optimization procedure.
If optimization does not reliable work, flat directions are not found by penalized optimization.
For the standard procedure using a single $R$, this problem cannot be distinguished from an identifiable setting because in
both cases the objective function increase due to penalization.
In contrast, calculating the dependency of $V_\text{tot}$ on $R$ on an interval might
indicate an optimization problem by a non-smooth outcome (see \ref{sec:cont}).
Another way to ensure that optimization works reliable enough, is to artifically introduce a non-identifiability, 
e.g.~by replacing a parameter $\theta_i$ by a product $\theta_i \times \theta_i'$ of two parameters 
and then check as a positive-control whether such a non-identifiability is found.

Prespecified bounds for the parameters can be considered by restricting optimization to the feasible region.
In fact, most of the application examples have parameter bounds which prevent failure of ODE integration.
The definition of the feasible parameter space can be considered as part of the model structure.

\section{Results}
\begin{table}[!t]
\begin{tabular}{@{}lrrcl@{}}\hline
Name	 &   $n_\theta$ 	& $N_{\text{data}}$ &  Identifiable? &  Publication\\\hline
ABC	 & 	  3	& 11$^*$ & Yes & \cite{Kreutz2011}\\
ABC\_rel	 & 	  4	& 11$^*$ & No & \\
Bachmann	 & 	  113	& 542 & No & \cite{Bachmann2011}\\
Becker	 & 	  16	&  85 & Yes & \cite{Becker2010}\\
Boehm	 & 	  9	& 48 & Yes & \cite{Boehm2014}\\
Bruno	 & 	  16	& 46 & Yes & \cite{Bruno16}\\
Raia	 	 & 	  39 & 205 &  No & \cite{Raia2011}\\
Schwen	 & 	  30	& 292 & No &  \cite{Schwen15}\\
Swameye	 & 	  16	& 46 & No &  \cite{Swameye2003}\\
Toensing-School	 & 	  5	& 15 & Yes & \cite{Toensing17}\\
Toensing-Zika	 & 	  17	& 57 & No  &\cite{Toensing17}\\ \hline
\end{tabular}
\caption{Overview about the investigated models. $^*$ denotes simulated data.\label{Tab:models}}
\end{table}

Two small illustration models as well as nine application models with real measurements were used
to demonstrate the applicability and capabilities of the presented approach in this chapter.
Table~\ref{Tab:models} provides an overview about the models
which have between 3 and 113 estimated parameters and between 11 and 542 data points.
Five models are structurally identifiable, six are structurally non-identifiable as shown in the Supplementary Information
where the mostly cited approach \citep{Raue2009} which is at the same also applicable for all investigated models 
has been applied as a reference.


\subsection{Illustration models}
\begin{figure*}[!tpb]
\includegraphics[width=\linewidth]{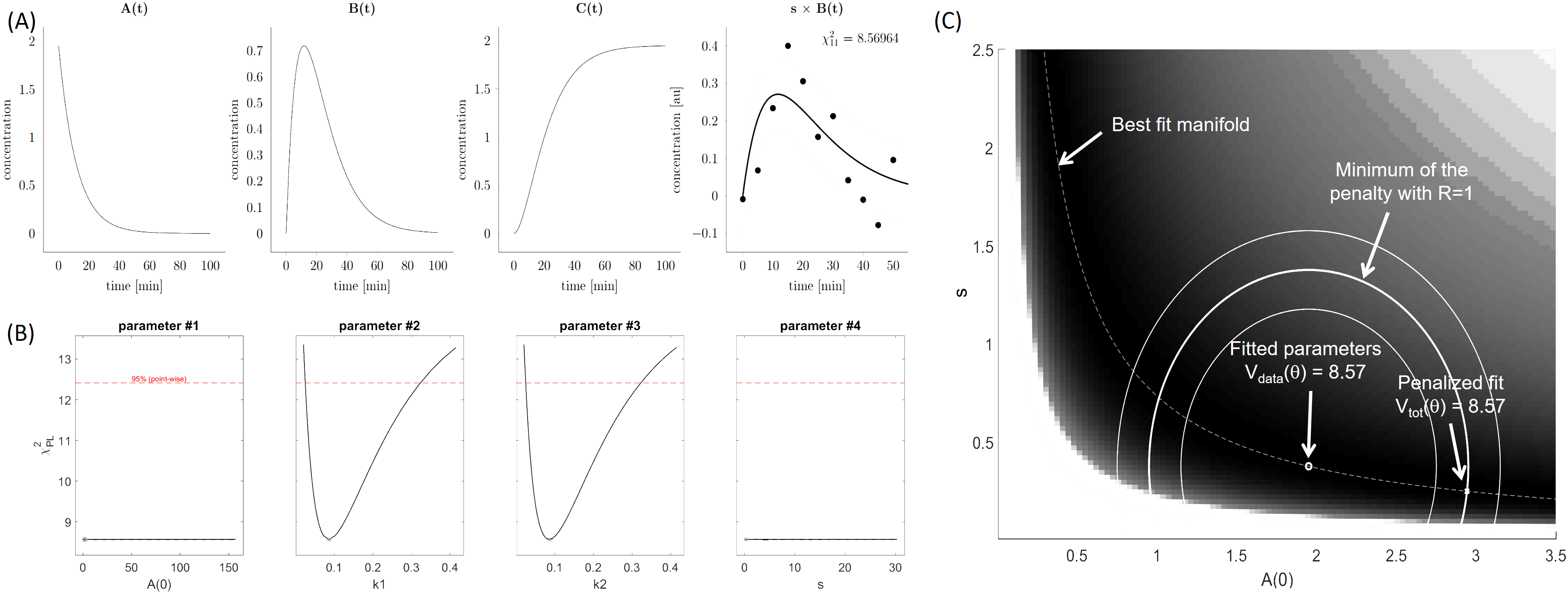}
\caption{Panel (A) shows the dynamics of the non-identifiable illustration model \emph{ABC\_rel} as well as the data. 
Gray shading again indicates the size of the measurement errors.
The likelihood profiles shown in panel (B) are flat for $A(0)$ and $s$ indicating non-identifiability.
The same outcome is obtained by the penalization-based \emph{identifiability test} as shown in panel (C).
Shading indicates the dependency of agreement between model and data from the two parameters. The remaining parameters $k_1$ and $k_2$ were optimized for all combinations.
Penalized fitting moves the estimated parameters to perfectly satisfy the penalty. 
Thereby, the resulting objective function does not increase, i.e.~$V_\text{tot}=V_\text{data}$ indicating non-identifiability. \label{fig:illu2}}
\end{figure*}

A small and illustrative model of two consecutive reactions
\begin{equation}
     A\: \overset{k_1}{\rightarrow} \:B\: \overset{k_2}{\rightarrow} \:C
\end{equation}
with rates $\theta_1\equiv k_1=0.1, \theta_2\equiv k_2=0.1$ and initial conditions $\theta_3\equiv A(0)=1, B(0)=0, C(0)=0$ is utilized to illustrate the \emph{ITRP}. 
For the simulated measurements, normally distributed noise with $\sigma=0.1$ has been assumed 
which corresponds to a typical signal-to-noise ratio for applications in molecular biology of around $10\%$. 
For an identifiable setting, $B(t)$ is assumed to be measured at $t=0,5,\dots,50$.
Although parameter log-transformation is reasonable for fitting ODE models \citep{Kreutz16}, we omit the log-transformation 
for the illustration models in the figures to keep the setting as simple as possible.
The identifiable model has a unique minimum and is termed \emph{ABC}, see first row in Table \ref{Tab:models}.

For the simulated data shown in panel (A) in Fig.~\ref{fig:illu1}, the maximum likelihood estimate is $\hat \theta = [1.27,   0.11,    0.11]$.
The profile likelihood for all parameters shown in panel (B) exhibit unique minima.
Panel (C) shows $\max_{k_2}V(A(0),k_1)$, i.e.~the dependency of the likelihood for given $A(0)$ and $k_1$ while optimizing $k_2$.
The maximum likelihood estimate is indicated by the circle.
If the penalty is added, the minimum shifts, but due to identifiability, the objective function increases: $V_\text{tot}(\theta^*) - V_\text{data}(\hat \theta) = 0.902$, i.e.~$\Delta V^R>0$ which correctly indicates identifiability.

For a non-identifiable setting, it is assumed that $B(t)$ is only measured on a relative scale, i.e.~the observation function is 
\begin{equation}
	g_i = s \times B(t_i)
	\end{equation}
with scaling parameter $s$.
This model has four fitted parameters and is termed \emph{ABC\_rel} (2nd row in Table \ref{Tab:models}).
Panel (A) in Fig.~\ref{fig:illu2} shows the dynamics and the measurements $ s \times B(t)$.
The profile likelihood of parameters $A(0)$ and $s$ shown in panel (B) are flat indicating a non-unique minimum and non-identifiability.
In the two-dimensional representation shown in panel (C), the parameters $k_1$ and $k_2$ are optimized for different combinations of $s$ and $A(0)$.
The flat best-fit manifold is indicated by the dashed line.

If the \emph{ITRP} is applied, the parameters are shifted due to the penalty but the objective function does not increase, 
i.e.~$V_\text{tot}(\theta^*) - V_\text{data}(\hat \theta) = 0$ which correctly indicates non-identifiability.

\subsection{Dependency on penalty location and parameter relationships} \label{sec:cont}
In \citep{Raman17}, radial constraints have been used to define \emph{multi-scale sloppiness} 
and combined with an integration-based approach to uncover non-identifiabilities 
and its corresponding parameter relationships.
Such a procedure can be interpreted as calculation of a prediction profile likelihood 
\begin{equation} \label{eq:ppl}
	\text{PPL}(R) = \min_{\{\theta | F(\theta)=R\}} V_\text{data}(\theta)
\end{equation}
as suggested in \citep{Kreutz2013}
for predicting the radial, euclidean distance 
\begin{equation} \label{eq:F}
	F(\theta) = || \theta - \hat \theta ||_2
\end{equation} to the estimated parameters.

In \citep{Kreutz2013}, is was shown that penalized optimization
\begin{equation} \label{eq:vpl}
	\text{V}(R) = \min_{\theta} \left( V_\text{data}(\theta) +  \lambda || F(\theta) - R ||_2 \right)
\end{equation}
can be used to calculate the solution of the constrained optimization in ($\ref{eq:ppl}$) in a numerically more robust manner
for any kind of prediction $F$.
In our setting ($\ref{eq:F}$), penalties are exactly satisfied in the case of non-identifiability and therefore 
both profiles ($\ref{eq:ppl}$) and ($\ref{eq:vpl}$) coincide.

The \emph{ITRP}, in turn, is equivalent to equation ($\ref{eq:vpl}$) for a single radius $R$.
For testing of non-identifiability, only one point at the profile ($\ref{eq:vpl}$) for the radial distance is sufficient.
Here, the exact choice of $R$ specifies the definition of ``local'' according to the definition ($\ref{eq:idMath}$).
Nevertheless, the function ($\ref{eq:vpl}$) can be used to obtain a more comprehensive picture 
and for assessing parameter relationship(s) as discussed in \citep{Raue2009,Raman17}.

\begin{figure}[!tpb]
\includegraphics[width=\linewidth]{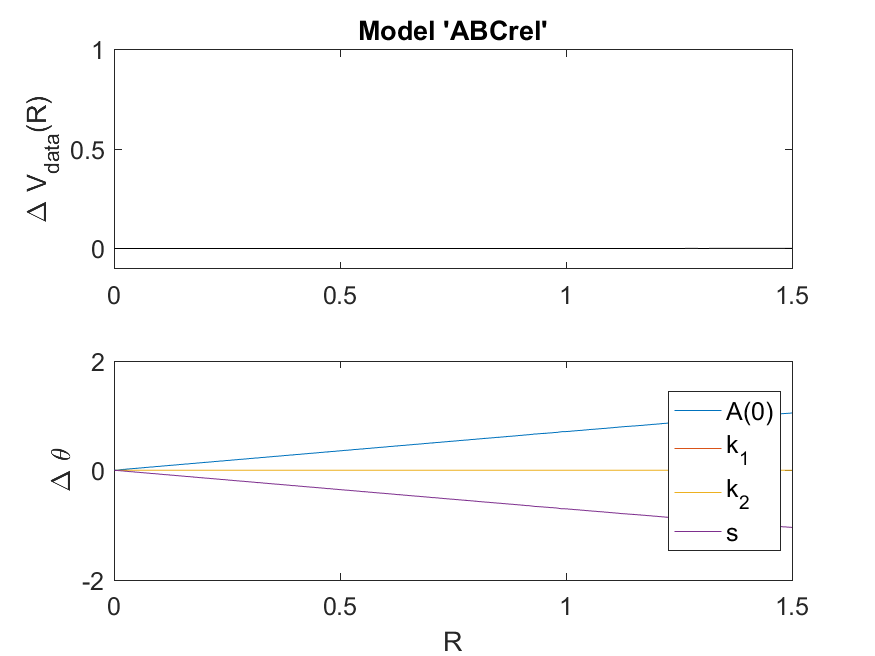}
\caption{\label{fig:ppl}The upper panel shows the fitted penalized objective function $V$ as a function of the radial distance of the penalty $R$.
The flat shape indicates non-identifiability.
The identifiabilty-test corresponds to a single point on this curve.
In the lower panel, the changes of the parameters while increasing the radius $R$ is depicted. 
The two non-identifiable parameters $A(0)$ and the scaling parameter $s$ are adjusted to satisfy the penalty.
$k_1$ and $k_2$ do not change and the curves are on top of each other.
$\Delta R$ as well as the euclidean distance are analyzed on the log10-scale.}
\end{figure}
Fig.~\ref{fig:ppl} shows the profile $V(R)$ for the non-identifiable model \emph{ABC\_rel}. 
Flatness of the profile in the upper panel indicates non-uniqueness of the parameters, i.e.~non-identifiability.
The lower panel in Fig.~\ref{fig:ppl} indicates that parameters $s$ and $A(0)$ have to be adjusted for increasing penalties $R>0$
and are therefore in the flat, redundant manifold.

\subsection{Application models} \label{sec:applications}
\begin{table}[!t]
\begin{tabular}{@{}lcc@{}}\hline
Name	 &   Identifiability 	& Computation time\\
&	 correct? &  (rel. to profile likelihood) \\ \hline
ABC  &  yes  & 0.22 sec (0.96\%) \\ 
ABC\_rel  &  yes  & 1.24 sec (2.47\%) \\ 
Bachmann  &  yes  & 9.05 sec (0.02\%) \\ 
Becker  &  yes  & 0.24 sec (0.25\%) \\ 
Boehm  &  yes  & 2.19 sec (2.48\%) \\ 
Bruno  &  yes  & 0.09 sec (0.38\%) \\ 
Raia  &  yes  & 4.50 sec (0.10\%) \\ 
Schwen  &  yes  & 8.40 sec (0.09\%) \\ 
Swameye  &  yes  & 0.66 sec (0.11\%) \\ 
Toensing-School  &  yes  & 0.12 sec (0.56\%) \\ 
Toensing-Zika  &  yes  & 0.80 sec (0.09\%) \\ \hline
\end{tabular}
\caption{Results of the identifiability analysis. The percentages in brackets
show the reduction of computation times relative to the standard approach (profile likelihood).\label{Tab:results}}
\end{table}
As application examples, nine published models were analyzed as summarized in Table \ref{Tab:models}, rows 3-11.
As in the original publication, the parameters were analyzed at the log10-scale.
The outcome of the identifiability test is summarized in Table \ref{Tab:results}. 
As a reference, the profile likelihood approach \citep{Raue2009} was used.
Our new approach correctly assesses identifiability for all models and requires less than 1\% computation times.
The total computation time, i.e.~the sum over all nine application models, was 26.1 seconds using five fits with five different initial guesses 
for each model
but 1005.5 minutes for the profile likelihood approach, and 17.0 seconds vs. 252.7 minutes if the computationally most demanding model (Bachmann) is excluded.
The command-line output of the implementation in Data2Dynamics modelling toolbox 
as well as the likelihood profiles are provided in detail in the Supplementary Information.

\begin{figure*}[!tpb]
\includegraphics[width=\linewidth]{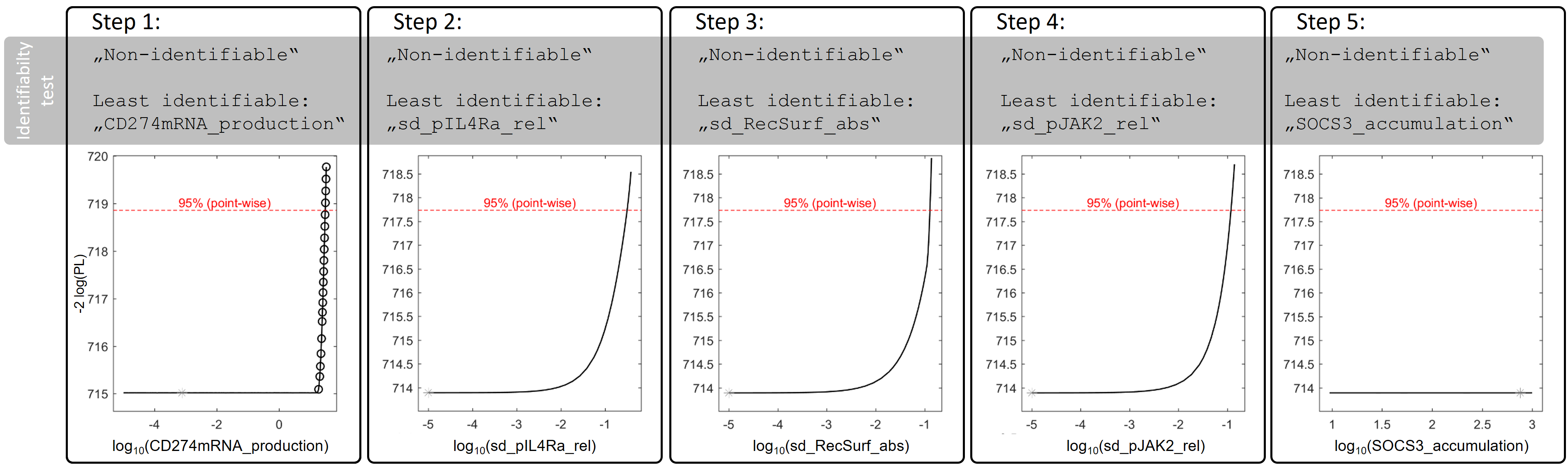}
\caption{In the first step, the identifiability test detects {\tt CD274mRNA\_production} as non-identifiable. 
In fact, the profile likelihood shown in the upper left panel is flat. 
In the second step, the relative error parameters {\tt sd\_pIL4Ra\_rel} is found as non-identifiable an the result of the identifiability 
test is confirmed by locally a flat profile likelihood.
Altogether, five non-identifiabilities are found before the model is completely identifiable.
The difference of the objective function between step 1 and 2 originates from the so-called Bessel-correction which depends on the number of parameters while not counting error parameters. 
The circles in the lower left panel indicate that another parameter is at the bounds of the predifined parameter space.
\label{fig:raia}}
\end{figure*}
As representative example, the model published in \citep{Raia2011} is shown in Fig.~\ref{fig:raia}.
For the published model, the identifiability test indicates non-identifiability and the least identifiable parameter ($\ref{eq:least}$)
is {\tt CD274mRNA\_production}.
The profile likelihood for this parameter shown in the left panel confirms that this parameter 
can be altered without loss of agreement with the data.

Next, the approach is iteratively applied as discussed in section \ref{sec:recursive}.
For this purpose, {\tt CD274mRNA\_production} is first fixed and the \emph{ITRP} is applied for this setup.
The model is still non-identifiable and next parameter {\tt sd\_pIL4Ra\_rel} is moved mostly for minimizing $V_\text{tot}$.
The profile likelihood confirms non-identifiability in this setup.
Repeating this procedure next detects {\tt sd\_pJAK2\_rel} as non-identifiable parameters, 
then {\tt sd\_RecSurf\_abs}, and finally {\tt SOCS3\_accumulation}. 
Fixing those five parameters yields a completely identifiable model.

\section{Conclusion and summary}
In this manuscript, the \emph{Identifiability-Test by Radial Penalization (ITRP)} 
for testing identifiability is applicable in any setting where the model can be fitted by optimization.
There is no restriction in terms of nonlinearity or size of the models and all systems biology models which are e.g.~covered
by \emph{SBML (Systems Biology Markup Langage)} model definitions can be analyzed.
The suggested \emph{ITRP} is based on comparison of the objective function of common fitting with a penalized fit pulling the
parameter vector away from the first estimate.
If this is feasible without worsen the objective function, non-uniqueness of the estimates is indicates which corresponds to non-identifiability.

The presented approach is more than 100 times faster than the profile likelihood approach which is according to citations 
currently the mostly frequently applied approach in systems biology and, 
at least according to our knowledge, the only approach with the same general applicability.
Feasibility and performance of the suggested method has been demonstrated using 11 ODE models.
Moreover, some extended analyses were introduced like investigation of parameter dependencies or analysis of mathematical identifiability.



\section*{Funding}
This work was supported by the German Ministry of Education and Research by grant \emph{EA:Sys [FKZ 031L0080]}.
\vspace*{-12pt}
 



\section*{Supplementary material}

# Supplementary Information:
# An easy and efficient approach for testing identifiability of parameters

C. Kreutz

August 11, 2017

**Analysis has been performed using the Data2Dynamics modelling toolbox [8, 9]**

**Website:** http://www.data2dynamics.org

# Contents







# 1 Result of the identifiability-tests

The command line output of the implementation of the identifiability-test in the Data2Dynamics toolbox is provided in sections 6.5 and 7.5 for the two illustration models. In 8.7 the result is shown the Becker model, in 9.5 for Boehm, in 10.5 for Raia, for Swameye in 11.4, for the Bachmann model in 12.17, for School in 13.4, for the Zika model in 14.5, and for Schwen in 15.9. for each individual model.

# 2 Number of fits/initial guesses

In principle, a single penalized fit is sufficient to check identifiability if optimization works reliably and there are no local optima in the local parameters space given by the radius $R$. In this paper, the *identifiability-test* was performed by using five fits with different initial guesses for penalized optimization. Our implementation of the approach in the *Data2Dynamics* modelling toolbox automatically checks whether the same objective function is repeatedly found. The command-line output shows this agreement (see sections 6.5, 7.5, 8.7, 9.5, 10.5, 11.4, 12.17, 13.4, 14.5, 15.9). To obtain a reliable result which well guarantees convergence, a rather stringent threshold equals to 0.001 for comparing different $\Delta V^R$ is used.

For the first intial guess we always choose $\hat{\theta}$, i.e. the minimum of $V_{\text{data}}$. This ensures that the result of the penalized fit is at least as good as $\hat{\theta}$. Additional intial guesses were randomly drawn within the sphere with

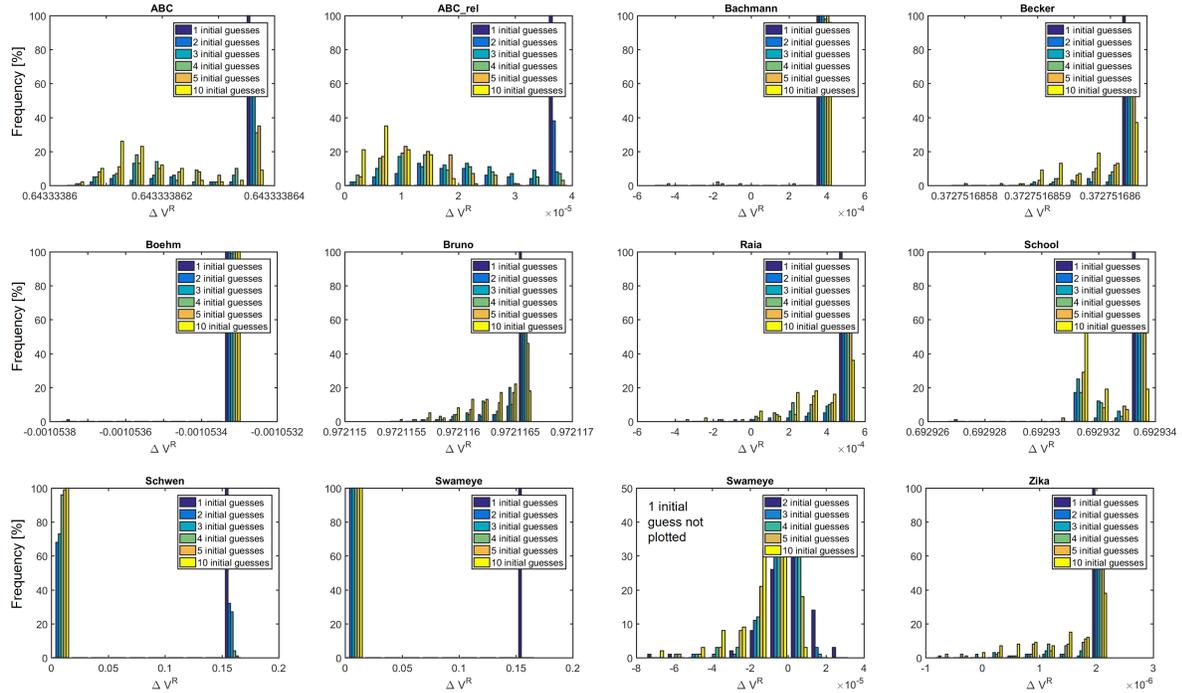

**Figure 1:** Comparison of the outcomes for increasing number of fits with different intial guesses for all 11 models. For nine models (ABC, ABS_rel, Bachmann, Becker, Bruno, Boehm, Raia, School, and Zika), we see only a very small dependency of the outcome. Here, the horizontal axis span only a small range indicating that $\Delta V^R$ is almost the same. For the Schwen and the Swameye model, a single fit using $\hat{\theta}$ as initial guess does not work. The fits then always end up in a local optimum. Here, adding a few random initial guesses yields reliable detection of the global optimum. For Swameye, the 2nd plot shows the result for the subset $n_{fit} = 2, 3, \ldots, 10$. For all plots, the histogram was made with 10 bins of equal size.

radius $R$. Since a uniform distribution in high-dimensions has the vast majority of its mass at the surface, we first draw the distance from a uniform distribution and then the direction of the initial guess. Thereby, a uniform distribution of the distance to $\hat{\theta}$ could be guaranteed.

Figure 1 shows how $\Delta V^R$ depends on the number of initial guesses. The different colors correspond to $n_{\text{fit}} = 1, 2, 3, 4, 5, 10$. The identifiability test has been performed for each model and each number $n_{\text{fits}}$ of fits 100 times with different random initial guesses. As discussed in the figure caption, for 9 out of the 11 models, there is only a very small dependency on the number of intial guesses and a single initial guess already yields desired outcomes. For the Swameye and the Schwen models a few random intial guesses already leads to reliable detection of the global optimum.

Note that the reproducibility of penalized optimization does not only depend on the optimization algorithm. It also depends on the accuracy of numerical integration of the ODEs. Therefore, adjusting the termination tolerances of optimization as well as improving the ODE integration tolerances might be a solution if optimization is not reproducible in other applications.

## 3 Investigating mathematical identifiability

As already summarized in the main text, parameter redundancy in a noise-free setting with continuous observables can be tested by investigating whether the minimum of Bellman's objective function

$$V_{\text{Bellman}}(\theta) = \int \left(y(t) - g(t, \theta)\right)^2 dt \ . \tag{1}$$

has a unique minimum. Since the integral can only be evaluated numerically for realistic systems biology models, we approximate the integral by a sum and use

$$V_{\text{data}}(\theta) = \sum_i \frac{1}{\sigma_i^2} \left( y_i - g_i(t_i, \theta) \right)^2 \quad (2)$$

with $N_{\text{sim}}$ points on the integrated trajectories and

$$\sigma_i = N_{\text{sim}} \left( atol + rtol \times x_i \right) \quad . \quad (3)$$

Then the *identifiability-test* is applied to investigate uniqueness of the parameters minimizing $V_{\text{data}}$.

In the following, we used the Swameye model to discuss the results of the suggested approach The outcomes were confirmed by comparison with the profile likelihood approach, although the plots are not shown since it would require a lot of space.

## 3.1 Swameye model

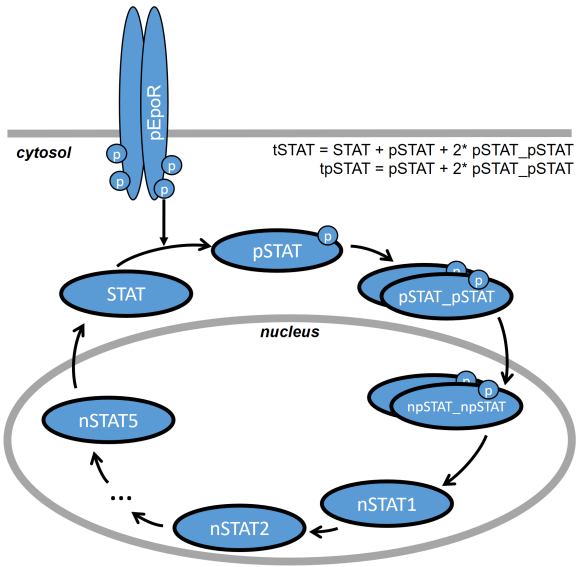

**Figure 2:** Investigation of identifiability for an (almost) noise-free and continuous setup yields an identifiable setting, if the input pEpoR is measured in combination with a dynamic variable. Measurements of a single dynamic variable are sufficient since all states are "linked" due to recycling of the STAT proteins.

The model structure is explained in detail in chapter 11. Here, only the results of the analysis of mathematical identifiability is discussed. For the analyses, all dynamic states and inputs are evaluated as observables without extra parameters in the observation functions, i.e. scaling and offset parameters were fixed to its estimated numbers. In addition to the dynamic states and the input, also the observables `tSTAT` and `tpSTAT` denoting the total cytosolic STAT as well as the total phosphorylated STAT in the cytosol are treated as potential observations (see chapter 11). Both observables are sums of dynamic states (see figure 2). Therefore, 12 "states" were analyzed as possible observables.

The *identifiability test* was applied as described in the main text with $R = 1$, $\lambda = 1$ and threshold $\delta = 0.001$. As a first result, the *identifiabilility-test* predicts parameters identifiability, if all states are observed. This outcome is expected since no redundant parameters occur at the right hand side of the ODEs and also the profile likelihood indicates non-identifiability.

7

If only a single state is observed, the model is predicted to be non-identifiable for all 12 scenarios, i.e. independently on the choice of the observed "state". This is confirmed by the profile likelihood approach resulting in flat profiles for the phosphorylation rate $p_1$ and the spline parameters $sp_1, \ldots, sp_5$ specifying the input pEpoR. The reason is that in the model only products $p_1 \times$pEpoR occur in the model and therefore a change in $p_1$ can be compensated by the spline parameters specifying the input pEpoR.

If all states except a single one is observed (again 12 scenarios), then the model is predicted to be identifiable for all choices for the unobserved state with one exception. If the phosphorylated receptors pEpoR which act as input in the pathway model are not observed, the model becomes non-identifiable because again the STAT phosphorylation rate $p_1$ and the spline parameters $sp_1, \ldots, sp_5$ are not identifiable.

Finally, it was investigated whether observation of the input in combination with another observable yields identifiable parameters (11 scenarios). Here, the result was that a combination with a dynamic state leads to identifiability. The reason for this is that all dynamic states are dynamically connected in the pathway because STAT can shuttle through the nucleus and is thereby converted into its original, unphosphorylated state. In contrast, combing the observation of the input pEpoR with tSTAT or tpSTAT yields to non-identifiable parameters. Figure 2 shows a pathway map illustrating that observation of the input and a single dynamic state in the cycle provides enough information for estimating all dynamic parameters.

## 3.2 Discussion

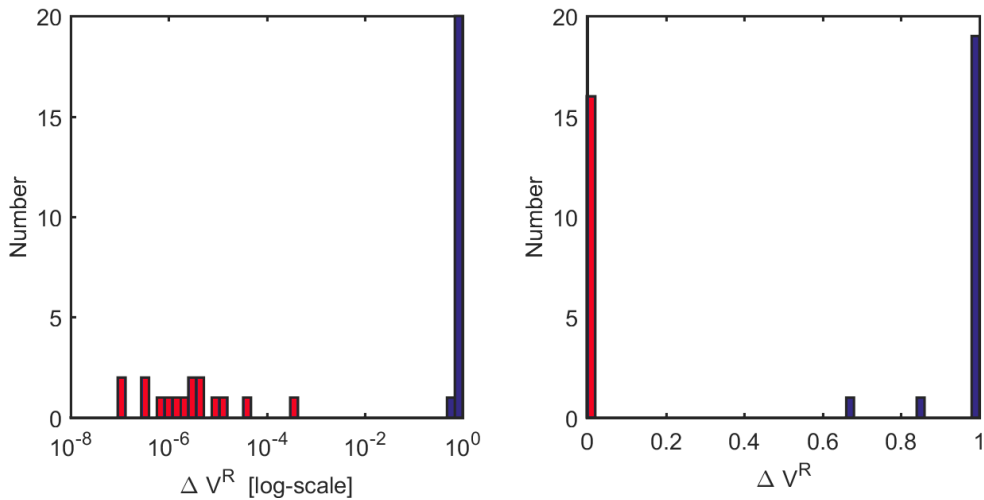

**Figure 3:** Outcome of the *identifiability-test* in terms of increase of the objective function $\Delta V^R$ due to penalization. There is a clear separation between the identifiable results (blue) and the non-identifiable ones shown as red bars.

Since the major focus of this paper was investigation of identifiability in the inverse setting, the focus has been put on the statistical setup for identifiability analysis. Therefore only some first analyses for the mathematical identifiability setting are provided without the aim of a comprehensive evaluation.

A major difference between both setups is the order of magnitude of the objective function. In an application setting with real measurements, there might be around 10-1000 data points. Since the contribution of each data point in (2) is weighted with the observation error $1/\sigma$ which roughly has the same size than $y_i - g_i$, the order of magnitude of $V_\text{data}$ is again 10-1000 and some orders of magnitude larger if the parameters are not fitted.

In contrast, in the continuous and (almost) noise-free setting used to investigate mathematical identifiability, the number of time points required to approximately sample a single continuous observation function has

8

to be around $100 - 10'000$. If there are several observation function, there might be up to $10^6$ (or even more) data points. The weights for every point in the objective function has to be chosen as 1 over the integration error which yields typical weights between $10^4$ - $10^{10}$. Therefore, a small mismatch of the parameters leads to much larger impairment of the objective function than in the setting with real data.

Therefore, it might be that the suggested choice of the threshold $\delta$ or of the penalization strength $\lambda$ has to be adapted for a general applicability in the case of investigation of mathematical identifiability, although we obtained reasonable results for our models. The number of fits used to circumvent convergence to local optima could also be increased to improve robustness of the results. Moreover, an essential aspect in our analysis was to use minimal changes of the objective function as termination threshold for optimization, i.e. `TolFun` in Matlab notation, as it was already suggested in the main text.

# 4 Implementation in Data2Dynamics

The *identifiability-test by radial penalization (ITRP)* is implemented in the Data2Dynamics modelling toolbox [8] as function `arIdentifiabilityTest`. The method can be applied to a model using the following Matlab commands:

```
>> cd [D2D-folder] % switch to the D2D basic directory

>> cd Examples/Becker_Science2010 % switch to folder of a model, here the Becker model
in the D2D examples

>> Setup % load the model, in the first call, the model is compiled automatically

>> arFit % fit the parameters

>> arIdentifiablityTest % apply the identidifability-test by radial penalization
```

The command line output of the implementation of the identifiability-test in the Data2Dynamics toolbox is provided in sections 6.5 and 7.5 for the two illustration models. In 8.7 the result is shown the Becker model, in 9.5 for Boehm, in 10.5 for Raia, for Swameye in 11.4, for the Bachmann model in 12.17, for School in 13.4, for the Zika model in 14.5, and for Schwen in 15.9. for each individual model.

# 5 Terms used in the following model definition chapters

*Derived variables* are variables defined for simplifying equations or because the are used as important outcomes in the model. A simple example is a sum of two dynamic states. In contrast, to observables, there us no experimental data available. *Initial states* $x(t = 0)$ are denoted by the prefix "init_".

# 6 Model 1 ("ABC")

## 6.1 Model definition of the identifiable illustration model

### 6.1.1 Description

Identifiable illustration model for identifiability test. The model consists of two consecutive reactions $A \to B \to C$, B is measured, B(0)=0, C(0)=0.

### 6.1.2 Dynamic variables

The model contains 3 dynamic variables. The dynamics of those variables evolve according to a system of ordinary differential equations (ODE) as will be defined in the following. The following list indicates the unique variable names and their initial conditions.

- **Dynamic variable 1:** A_state

    $$[\text{A\_state}](t = 0) = \text{init\_A\_state}$$

- **Dynamic variable 2:** B_state

    $$[\text{B\_state}](t = 0) = \text{init\_B\_state}$$

- **Dynamic variable 3:** C_state

    $$[\text{C\_state}](t = 0) = \text{init\_C\_state}$$

### 6.1.3 Reactions

The model contains 2 reactions. Reactions define interactions between dynamics variables and build up the ODE systems. The following list indicates the reaction laws and their corresponding reaction rate equations. Promoting rate modifiers are indicated in black above the rate law arrow. Inhibitory rate modifiers are indicated in red below the rate law arrow. In the reaction rate equations dynamic and input variables are indicated by square brackets. The remaining variables are model parameters that remain constant over time.

- **Reaction 1:**

    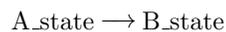

    $$v_1 = [\text{A\_state}] \cdot p1$$

- **Reaction 2:**

    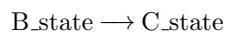

    $$v_2 = [\text{B\_state}] \cdot p2$$

### 6.1.4 ODE system

The specified reation laws and rate equations $v$ determine an ODE system. The time evolution of the dynamical variables is calculated by solving this equation system.

$$d[A\_state]/dt = -v_1$$
$$d[B\_state]/dt = +v_1 - v_2$$
$$d[C\_state]/dt = +v_2$$

Substituting the reaction rates $v_i$ yields:

$$d[A\_state]/dt = - [A\_state] \cdot p1$$

$$d[B\_state]/dt = [A\_state] \cdot p1 - [B\_state] \cdot p2$$

$$d[C\_state]/dt = [B\_state] \cdot p2$$

The ODE system was solved by a parallelized implementation of the CVODES algorithm [4]. It also supplies the parameter sensitivities utilized for parameter estimation.

### 6.1.5 Observables

The model contains only a single observable. Observables are calculated after the ODE system was solved and derived variables are calculated. Dynamic, input and derived variables are indicated by square brackets. The remaining variables are model parameters that remain constant over time. In additon to the equation for the observable, also their corresponding error model $\sigma$ is indicated.

- **Observable 1:** B_au

$$B\_au(t) = [B\_state]$$
$$\sigma\{B\_au\}(t) = sd\_B\_au$$

### 6.1.6 Conditions

Conditions modify the model according to replacement rules. New model parameters can be introduced or relations between existing model parameters can be implemented. The following list are default conditions that can be replace my experiment specific conditions defined seperately for each data set.

$$init\_B\_state \rightarrow 0$$
$$init\_C\_state \rightarrow 0$$

## 6.2 Simulated data of model "ABC"

Experiment specific conditions To evaluate the model for this experiment the following conditions are applied. Experimental data and model fit The model observables and the experimental data is shown in Figure 4. The agreement of the model observables and the experimental data, given in Table 1, yields a value of the objective function $\chi^2 = 14.3333$ for 11 data points in this data set. The trajectories of the input, dynamic and derived variables that correspond to the experimental conditions in this experiment are shown in Figure 5.

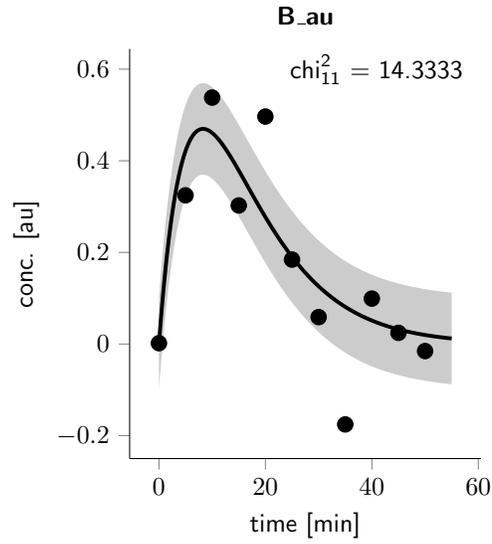

**Figure 4: Observables and simulated data.** The observables are displayed as solid lines. The error model that describes the measurement noise is indicated by shades.

|  | B_au |
| ---: | ---: |
| time [min] | conc. [au] |
| 0 | 0.00184943 |
| 5 | 0.324623 |
| 10 | 0.537945 |
| 15 | 0.302491 |
| 20 | 0.496547 |
| 25 | 0.184309 |
| 30 | 0.0586591 |
| 35 | -0.175556 |
| 40 | 0.0991418 |
| 45 | 0.0243198 |
| 50 | -0.0154171 |

**Table 1: Simulated data for the model "ABC"**

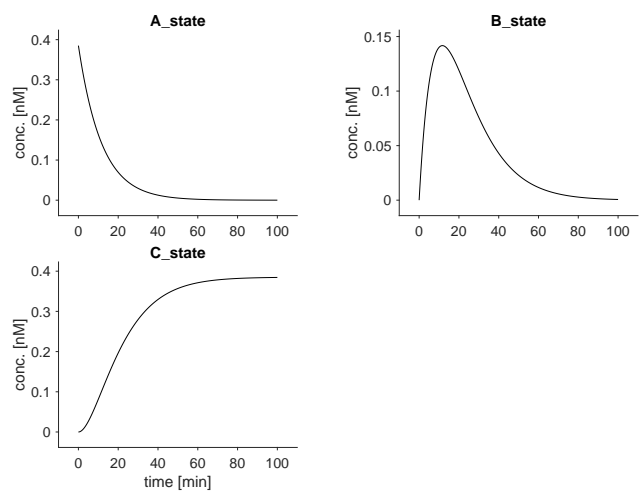

**Figure 5: Trajectories of the dynamic variables.** The dynamical behaviour is determined by integrating the ODE system defined in Section 6.1.4.

13

| | name | $\theta_{min}$ | $\hat{\theta}$ | $\theta_{max}$ | log | non-log $\hat{\theta}$ |
|---|---|---|---|---|---|---|
| 1 | init_A_state | -5 | +0.1646 | +3 | 1 | $+1.46 \cdot 10^{+00}$ |
| 2 | p1 | -5 | -0.9755 | +3 | 1 | $+1.06 \cdot 10^{-01}$ |
| 3 | p2 | -5 | -0.8633 | +3 | 1 | $+1.37 \cdot 10^{-01}$ |

**Table 2: Estimated parameter values**
$\hat{\theta}$ indicates the estimated value of the parameters. $\theta_{min}$ and $\theta_{max}$ indicate the upper and lower bounds for the parameters. The log-column indicates if the value of a parameter was log-transformed. If log $\equiv$ 1 the non-log-column indicates the non-logarithmic value of the estimate.

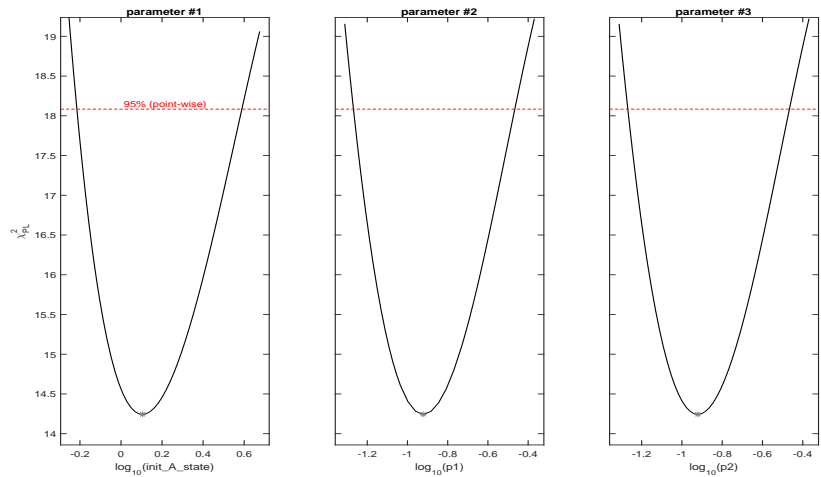

**Figure 6: Overview of the profile likelihood of the model parameters**
The solid lines indicate the profile likelihood. The broken lines indicate the threshold to assess confidence intervals. The asterisks indicate the optimal parameter values.

## 6.3 Estimated model parameters

In total 3 parameters are estimated from the experimental data, yielding a value of the objective function $\chi^2 = 14.3333$ for a total of 11 data points. The model parameters were estimated by maximum likelihood estimation. In Table 2 the estimated parameter values are given. Parameters highlighted in red color indicate parameter values close to their bounds. The parameter name prefix init_ indicates the initial value of a dynamic variable.

## 6.4 Profile likelihood of model parameters

As a classical approach, identifiability of the model parameters was assessed using the profile likelihood [7]. An overview is displayed in Figure 6.

## 6.5 Identifiability-test

Applying the new approach for investigating identifiability yielded the following outcome.

```
>> arIdentifiabilityTest
Identifiability-test started ...
... Identifiability-test finished.

Identifiability-test was performed with radius = 1 and penalty-SD = 1.

All 5 optimization runs are in the chi2-range 4.62478e-08.

Calculations took 0.22 seconds.
[Compared to 22.44 seconds required for calcuating the likelihood profiles.]

1.0000 (increase of merit by penalty, before fitting)
0.3567 (decrease of merit by fitting)
0.1837 (movement of parameters by penalized fitting)
0.6433 (total increase of merit by penalty) PRIMARY CRITERION
Model is identifiable.
```

# 7 Model 2 ("ABC_rel")

## 7.1 Model definition of the non-identifiable illustration model

### 7.1.1 Description

Non-identifiable illustration model for identifiability test. The model consists of two consecutive reactions $A \to B \to C$, B is measured, B(0)=0, C(0)=0. The non-identifiability occurs because measurements are only available on a relative scale.

### 7.1.2 Dynamic variables

The model contains 3 dynamic variables. The dynamics of those variables evolve according to a system of ordinary differential equations (ODE) as will be defined in the following. The following list indicates the unique variable names and their initial conditions.

- **Dynamic variable 1:** A_state

    $[\text{A\_state}](t = 0) = \text{init\_A\_state}$

- **Dynamic variable 2:** B_state

    $[\text{B\_state}](t = 0) = \text{init\_B\_state}$

- **Dynamic variable 3:** C_state

    $[\text{C\_state}](t = 0) = \text{init\_C\_state}$

### 7.1.3 Reactions

The model contains 2 reactions. Reactions define interactions between dynamics variables and build up the ODE systems. The following list indicates the reaction laws and their corresponding reaction rate equations. Promoting rate modifiers are indicated in black above the rate law arrow. Inhibitory rate modifiers are indicated in red below the rate law arrow. In the reaction rate equations dynamic and input variables are indicated by square brackets. The remaining variables are model parameters that remain constant over time.

- **Reaction 1:**

    A_state ⟶ B_state

    $v_1 = [\text{A\_state}] \cdot \text{p1}$

- **Reaction 2:**

    B_state ⟶ C_state

    $v_2 = [\text{B\_state}] \cdot \text{p2}$

### 7.1.4 ODE system

The specified reation laws and rate equations $v$ determine an ODE system. The time evolution of the dynamical variables is calculated by solving this equation system.

$\text{d}[\text{A\_state}]/\text{dt} = -v_1$
$\text{d}[\text{B\_state}]/\text{dt} = +v_1 - v_2$
$\text{d}[\text{C\_state}]/\text{dt} = +v_2$

Substituting the reaction rates $v_i$ yields:

$\text{d}[\text{A\_state}]/\text{dt} = -[\text{A\_state}] \cdot \text{p1}$

$\text{d}[\text{B\_state}]/\text{dt} = [\text{A\_state}] \cdot \text{p1} - [\text{B\_state}] \cdot \text{p2}$

$\text{d}[\text{C\_state}]/\text{dt} = [\text{B\_state}] \cdot \text{p2}$

The ODE system was solved by a parallelized implementation of the CVODES algorithm [4]. It also supplies the parameter sensitivities utilized for parameter estimation.

### 7.1.5 Observables

The model contains only a single observable. Observables are calculated after the ODE system was solved and derived variables are calculated. Dynamic, input and derived variables are indicated by square brackets. The remaining variables are model parameters that remain constant over time. In additon to the equation for the observable, also their corresponding error model $\sigma$ is indicated.

- **Observable 1:** B_au

    $\text{B\_au}(t) = [\text{B\_state}] \cdot \text{scale}$
    $\sigma\{\text{B\_au}\}(t) = \text{sd\_B\_au}$

| time [min] | B_au conc. [au] |
|---|---|
| 0 | -0.00870753 |
| 5 | 0.0679168 |
| 10 | 0.233758 |
| 15 | 0.399967 |
| 20 | 0.305549 |
| 25 | 0.157266 |
| 30 | 0.212539 |
| 35 | 0.041672 |
| 40 | -0.0102211 |
| 45 | -0.0779158 |
| 50 | 0.095701 |

Table 3: Simulated data

### 7.1.6 Conditions

Conditions modify the model according to replacement rules. New model parameters can be introduced or relations between existing model parameters can be implemented. The following list are default conditions that can be replace my experiment specific conditions defined seperately for each data set.

$\text{init\_B\_state} \to 0$

$\text{init\_C\_state} \to 0$

## 7.2 Experiment: ABC_data_Aobs

### 7.2.1 Experiment specific conditions

To evaluate the model for this experiment the following conditions are applied.

- **Local condition #1 (global condition #1):**

### 7.2.2 Simulated data and model fit

The agreement of the model observables and the experimental data, given in Table 3, yields a value of the objective function $\chi^2 = 8.56964$ for 11 data points in this data set.

## 7.3 Estimated model parameters

In total 4 parameters are estimated from the experimental data, yielding a value of the objective function $\chi^2 = 8.56964$ for a total of 12 data points. The model parameters were estimated by maximum likelihood estimation. In Table 4 the estimated parameter values are given. Parameters highlighted in red color indicate parameter values close to their bounds. The parameter name prefix init_ indicates the initial value of a dynamic variable.

## 7.4 Profile likelihood of model parameters

As a classical approach, identifiability of the model parameters was assessed using the profile likelihood [7]. An overview is displayed in Figure 7.

|   | name | $\theta_{min}$ | $\hat{\theta}$ | $\theta_{max}$ | log | non-log $\hat{\theta}$ |
|---|------|---|---|---|---|---|
| 1 | init_A_state | -5 | -0.4144 | +3 | 1 | $+3.85 \cdot 10^{-01}$ |
| 2 | p1 | -5 | -1.0676 | +3 | 1 | $+8.56 \cdot 10^{-02}$ |
| 3 | p2 | -5 | -1.0676 | +3 | 1 | $+8.56 \cdot 10^{-02}$ |
| 4 | scale | -5 | +0.2813 | +3 | 1 | $+1.91 \cdot 10^{+00}$ |

**Table 4: Estimated parameter values**
$\hat{\theta}$ indicates the estimated value of the parameters. $\theta_{min}$ and $\theta_{max}$ indicate the upper and lower bounds for the parameters. The log-column indicates if the value of a parameter was log-transformed. If log $\equiv$ 1 the non-log-column indicates the non-logarithmic value of the estimate.

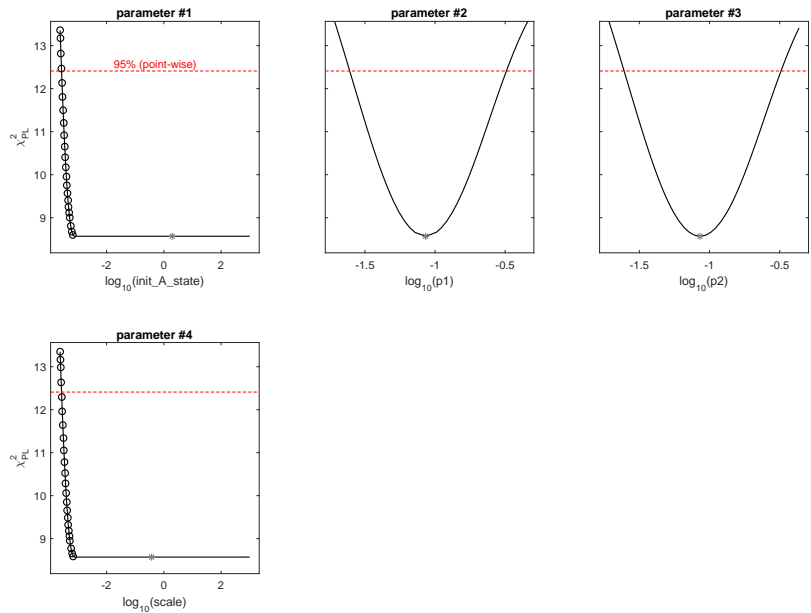

**Figure 7: Overview of the profile likelihood of the model parameters**
The solid lines indicate the profile likelihood. The broken lines indicate the threshold to assess confidence intervals. The asterisks indicate the optimal parameter values.

## 7.5 Identifiability-test

Applying the new approach for investigating identifiability yielded the following outcome.

```
>> arIdentifiabilityTest
Identifiability-test started ...
... Identifiability-test finished.

Identifiability-test was performed with radius = 1 and penalty-SD = 1.

All 5 optimization runs are in the chi2-range 1.65692e-05.

Calculations took 1.24 seconds.
[Compared to 50.26 seconds required for calcuating the likelihood profiles.]

1.0000 (increase of merit by penalty, before fitting)
1.0000 (decrease of merit by fitting)
0.9949 (movement of parameters by penalized fitting)
0.0000 (total increase of merit by penalty) PRIMARY CRITERION
Model is structurally non-identifiable.
```

# 8 Model 3 ("Becker")

## 8.1 Model definition

This model has been published in [2].

### 8.1.1 Description

The model describes EPO receptor binding and internalization after stimulation with erythropoetin (EPO). The model has been used to understand why EPO signalling works for a large range of ligand concentrations.

### 8.1.2 Dynamic variables

The model contains 6 dynamic variables. The dynamics of those variables evolve according to a system of ordinary differential equations (ODE) as will be defined in the following. The following list indicates the unique variable names and their initial conditions.

- **Dynamic variable 1:** Epo

  $[\text{Epo}](t = 0) = \text{init\_Epo}$

- **Dynamic variable 2:** EpoR

  $[\text{EpoR}](t = 0) = \text{init\_EpoR}$

- **Dynamic variable 3:** Epo_EpoR

  $[\text{Epo\_EpoR}](t = 0) = \text{init\_Epo\_EpoR}$

- **Dynamic variable 4:** Epo_EpoR_i

    [Epo_EpoR_i](t = 0) = init_Epo_EpoR_i

- **Dynamic variable 5:** dEpo_i

    [dEpo_i](t = 0) = init_dEpo_i

- **Dynamic variable 6:** dEpo_e

    [dEpo_e](t = 0) = init_dEpo_e

### 8.1.3 Reactions

The model contains 8 reactions. Reactions define interactions between dynamics variables and build up the ODE systems. The following list indicates the reaction laws and their corresponding reaction rate equations. Promoting rate modifiers are indicated in black above the rate law arrow. Inhibitory rate modifiers are indicated in red below the rate law arrow. In the reaction rate equations dynamic and input variables are indicated by square brackets. The remaining variables are model parameters that remain constant over time.

- **Reaction 1:**

    $$\mathrm{Epo} + \mathrm{EpoR} \longrightarrow \mathrm{Epo\_EpoR}$$

    $$v_1 = [\mathrm{Epo}] \cdot [\mathrm{EpoR}] \cdot \mathrm{kon}$$

- **Reaction 2:**

    $$\mathrm{Epo\_EpoR} \longrightarrow \mathrm{Epo} + \mathrm{EpoR}$$

    $$v_2 = [\mathrm{Epo\_EpoR}] \cdot \mathrm{koff}$$

- **Reaction 3:**

    $$\varnothing \longrightarrow \mathrm{EpoR}$$

    $$v_3 = \mathrm{init\_EpoR} \cdot \mathrm{kt}$$

- **Reaction 4:**

    $$\mathrm{EpoR} \longrightarrow \varnothing$$

    $$v_4 = [\mathrm{EpoR}] \cdot \mathrm{kt}$$

- **Reaction 5:**

    $$\mathrm{Epo\_EpoR} \longrightarrow \mathrm{Epo\_EpoR\_i}$$

    $$v_5 = [\mathrm{Epo\_EpoR}] \cdot \mathrm{ke}$$

- **Reaction 6:**

$$\text{Epo\_EpoR\_i} \longrightarrow \text{Epo} + \text{EpoR}$$

$$v_6 = [\text{Epo\_EpoR\_i}] \cdot \text{kex}$$

- **Reaction 7:**

$$\text{Epo\_EpoR\_i} \longrightarrow \text{dEpo\_i}$$

$$v_7 = [\text{Epo\_EpoR\_i}] \cdot \text{kdi}$$

- **Reaction 8:**

$$\text{Epo\_EpoR\_i} \longrightarrow \text{dEpo\_e}$$

$$v_8 = [\text{Epo\_EpoR\_i}] \cdot \text{kde}$$

### 8.1.4 ODE system

The specified reation laws and rate equations $v$ determine an ODE system. The time evolution of the dynamical variables is calculated by solving this equation system.

$$\begin{aligned}
d[\text{Epo}]/dt &= -v_1 + v_2 + v_6 \\
d[\text{EpoR}]/dt &= -v_1 + v_2 + v_3 - v_4 + v_6 \\
d[\text{Epo\_EpoR}]/dt &= +v_1 - v_2 - v_5 \\
d[\text{Epo\_EpoR\_i}]/dt &= +v_5 - v_6 - v_7 - v_8 \\
d[\text{dEpo\_i}]/dt &= +v_7 \\
d[\text{dEpo\_e}]/dt &= +v_8
\end{aligned}$$

Substituting the reaction rates $v_i$ yields:

$d[\text{Epo}]/dt = [\text{Epo\_EpoR}] \cdot \text{koff} + [\text{Epo\_EpoR\_i}] \cdot \text{kex} - [\text{Epo}] \cdot [\text{EpoR}] \cdot \text{kon}$

$d[\text{EpoR}]/dt = [\text{Epo\_EpoR}] \cdot \text{koff} + [\text{Epo\_EpoR\_i}] \cdot \text{kex} - [\text{EpoR}] \cdot \text{kt} + \text{init\_EpoR} \cdot \text{kt} - [\text{Epo}] \cdot [\text{EpoR}] \cdot \text{kon}$

$d[\text{Epo\_EpoR}]/dt = [\text{Epo}] \cdot [\text{EpoR}] \cdot \text{kon} - [\text{Epo\_EpoR}] \cdot \text{koff} - [\text{Epo\_EpoR}] \cdot \text{ke}$

$d[\text{Epo\_EpoR\_i}]/dt = [\text{Epo\_EpoR}] \cdot \text{ke} - [\text{Epo\_EpoR\_i}] \cdot \text{kde} - [\text{Epo\_EpoR\_i}] \cdot \text{kdi} - [\text{Epo\_EpoR\_i}] \cdot \text{kex}$

$d[\text{dEpo\_i}]/dt = [\text{Epo\_EpoR\_i}] \cdot \text{kdi}$

$d[\text{dEpo\_e}]/dt = [\text{Epo\_EpoR\_i}] \cdot \text{kde}$

The ODE system was solved by a parallelized implementation of the CVODES algorithm [4]. It also supplies the parameter sensitivities utilized for parameter estimation.

### 8.1.5 Derived variables

The model contains 2 derived variables. Derived variables are calculated after the ODE system was solved. Dynamic and input variables are indicated by square brackets. The remaining variables are model parameters that remain constant over time.

- **Derived variable 1:** Epo_ext

$$[\text{Epo\_ext}](t) = [\text{Epo}] + [\text{dEpo\_e}]$$

- **Derived variable 2:** Epo_int

$$[\text{Epo\_int}](t) = [\text{Epo\_EpoR\_i}] + [\text{dEpo\_i}]$$

### 8.1.6 Conditions

Conditions modify the model according to replacement rules. New model parameters can be introduced or relations between existing model parameters can be implemented. The following list are default conditions that can be replace my experiment specific conditions defined seperately for each data set.

$$\text{init\_EpoR} \to \text{init\_Epo} \cdot \text{init\_EpoR\_rel}$$
$$\text{init\_Epo\_EpoR} \to 0$$
$$\text{init\_Epo\_EpoR\_i} \to 0$$
$$\text{init\_dEpo\_e} \to 0$$
$$\text{init\_dEpo\_i} \to 0$$
$$\text{kon} \to \frac{\text{kon}}{\text{init\_Epo}}$$

## 8.2 Experiment: Epo_alpha_BaF3_Exp1_cpm_rep

### 8.2.1 Description

Experimenter: Verena Becker
Cells: BaF3
Ligand: Epo
Date: 16.04.2007

### 8.2.2 Experiment specific conditions

To evaluate the model for this experiment the following conditions are applied.

- **Local condition #1 (global condition #1):**

$$\text{init\_EpoR} \to 4 \cdot \text{init\_Epo} \cdot \text{init\_EpoR\_rel}$$
$$\text{scale} \to \frac{\text{scale}}{\text{init\_Epo}}$$

## 8.3 Model definition

### 8.3.1 Description

Epo binding described by a Michealis-Menten function

### 8.3.2 Input variables

The model contains 1 external inputs variables. Those variables evolve according to a regular algebraic equation. They are calculated before the ODE systems is solved and can appear in reaction rate equations. The following list indicates the unique variable names and their corresponding equations.

- **Input variable 1:** epo_bound

$$[\text{epo\_bound}](\text{epo\_free}) = \frac{10^{\text{epo\_free}} \cdot \text{init\_EpoR}}{\text{kD} + 10^{\text{epo\_free}}}$$

### 8.3.3 Conditions

Conditions modify the model according to replacement rules. New model parameters can be introduced or relations between existing model parameters can be implemented. The following list are default conditions that can be replace my experiment specific conditions defined seperately for each data set.

$$\text{kD} \rightarrow \frac{\text{koff}}{\text{kon}}$$

## 8.4 Experiment: Epo_binding_rep

### 8.4.1 Description

Experimenter: Verena Becker
Cells: BaF3
Ligand: Epo

### 8.4.2 Experiment specific conditions

To evaluate the model for this experiment the following conditions are applied.

- **Local condition #1 (global condition #1):**

$$\text{init\_EpoR} \rightarrow \text{init\_Epo} \cdot \text{init\_EpoR\_rel}$$
$$\text{kD} \rightarrow \text{kD}$$

## 8.5 Estimated model parameters

In total 16 parameters are estimated from the experimental data. The best fit yields a value of the objective function $-2\log(L) = -349.073$ for a total of 86 data points. The model parameters were estimated by maximum likelihood estimation. In Table 5 the estimated parameter values are given. Parameters highlighted in red color indicate parameter values close to their bounds. The parameter name prefix init_ indicates the initial value of a dynamic variable.

## 8.6 Profile likelihood of model parameters

As a classical approach, identifiability of the model parameters was assessed using the profile likelihood [7]. An overview is displayed in Figure 8.

|    | name          | $\theta_{min}$ | $\hat{\theta}$ | $\theta_{max}$ | log | non-log $\hat{\theta}$ |
|----|---------------|--------|---------|--------|-----|------------------------|
| 1  | init_Epo      | -5     | +3.1329 | +4     | 1   | $+1.36 \cdot 10^{+03}$ |
| 2  | init_EpoR_rel | -5     | -1.0394 | +3     | 1   | $+9.13 \cdot 10^{-02}$ |
| 3  | kD            | -5     | +2.1519 | +3     | 1   | $+1.42 \cdot 10^{+02}$ |
| 4  | kde           | -5     | -1.9137 | +3     | 1   | $+1.22 \cdot 10^{-02}$ |
| 5  | kdi           | -5     | -2.8563 | +3     | 1   | $+1.39 \cdot 10^{-03}$ |
| 6  | ke            | -5     | -1.2482 | +3     | 1   | $+5.65 \cdot 10^{-02}$ |
| 7  | kex           | -5     | -3.0437 | +3     | 1   | $+9.04 \cdot 10^{-04}$ |
| 8  | koff          | -5     | -1.1078 | +3     | 1   | $+7.80 \cdot 10^{-02}$ |
| 9  | kon           | -5     | -0.8211 | +3     | 1   | $+1.51 \cdot 10^{-01}$ |
| 10 | kt            | -5     | -1.7830 | +3     | 1   | $+1.65 \cdot 10^{-02}$ |
| 11 | offset        | -5     | -5.0000 | +3     | 1   | $+1.00 \cdot 10^{-05}$ |
| 12 | scale         | -5     | -0.0089 | +3     | 1   | $+9.80 \cdot 10^{-01}$ |
| 13 | sd_Epo_bound  | -5     | -1.4007 | +3     | 1   | $+3.97 \cdot 10^{-02}$ |
| 14 | sd_Epo_ext    | -5     | -2.0779 | +3     | 1   | $+8.36 \cdot 10^{-03}$ |
| 15 | sd_Epo_int    | -5     | -1.2553 | +3     | 1   | $+5.56 \cdot 10^{-02}$ |
| 16 | sd_Epo_mem    | -5     | -1.3173 | +3     | 1   | $+4.82 \cdot 10^{-02}$ |

**Table 5: Estimated parameter values**
$\hat{\theta}$ indicates the estimated value of the parameters. $\theta_{min}$ and $\theta_{max}$ indicate the upper and lower bounds for the parameters. The log-column indicates if the value of a parameter was log-transformed. If log $\equiv 1$ the non-log-column indicates the non-logarithmic value of the estimate.

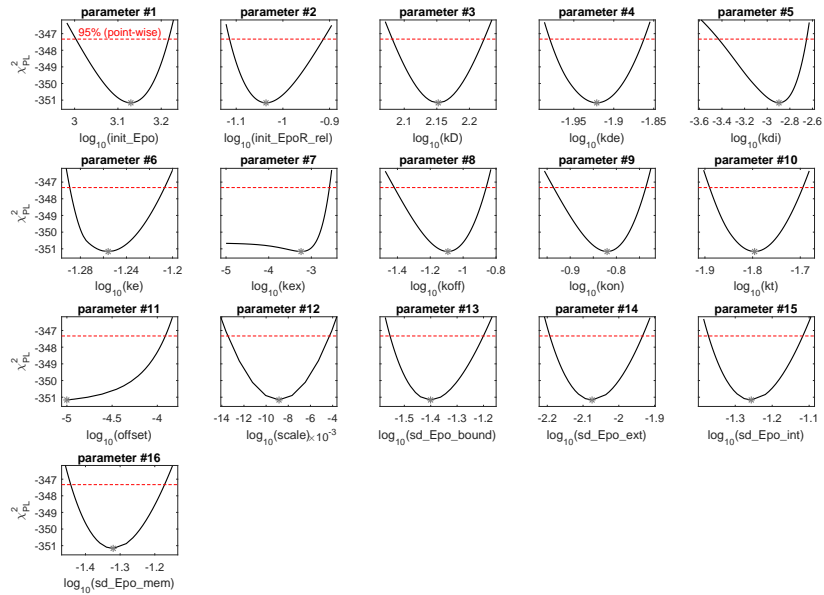

**Figure 8: Overview of the profile likelihood of the model parameters**
The solid lines indicate the profile likelihood. The broken lines indicate the threshold to assess confidence intervals. The asterisks indicate the optimal parameter values.

## 8.7 Identifiability-test

Applying the new approach for investigating identifiability yielded the following outcome.

```
>> arIdentifiabilityTest
Identifiability-test started ...
... Identifiability-test finished.

Identifiability-test was performed with radius = 1 and penalty-SD = 1.

All 5 optimization runs are in the chi2-range 1.83376e-08.

Calculations took 0.24 seconds.
[Compared to 95.98 seconds required for calcuating the likelihood profiles.]

1.0000 (increase of merit by penalty, before fitting)
0.8201 (decrease of merit by fitting)
0.7652 (movement of parameters by penalized fitting)
0.1799 (total increase of merit by penalty) PRIMARY CRITERION
Model is identifiable.
```

# 9 Model 4 ("Boehm")

## 9.1 Model definition

This model has been published in [3].

### 9.1.1 Description

STAT5AB phosphorylation and dimerization measured by mass spectrometry data.

### 9.1.2 Dynamic variables

The model contains 8 dynamic variables. The dynamics of those variables evolve according to a system of ordinary differential equations (ODE) as will be defined in the following. The following list indicates the unique variable names and their initial conditions.

- **Dynamic variable 1:** STAT5A

  $[\text{STAT5A}](t = 0) = \text{init\_STAT5A}$

- **Dynamic variable 2:** STAT5B

  $[\text{STAT5B}](t = 0) = \text{init\_STAT5B}$

- **Dynamic variable 3:** pApB

  $[\text{pApB}](t = 0) = \text{init\_pApB}$

- **Dynamic variable 4:** pApA

  $[\text{pApA}](t = 0) = \text{init\_pApA}$

- **Dynamic variable 5:** pBpB

  $[\text{pBpB}](t = 0) = \text{init\_pBpB}$

- **Dynamic variable 6:** nucpApA

  $[\text{nucpApA}](t = 0) = \text{init\_nucpApA}$

- **Dynamic variable 7:** nucpApB

  $[\text{nucpApB}](t = 0) = \text{init\_nucpApB}$

- **Dynamic variable 8:** nucpBpB

  $[\text{nucpBpB}](t = 0) = \text{init\_nucpBpB}$

### 9.1.3 Input variables

The model contains 1 external inputs variables. Those variables evolve according to a regular algebraic equation. They are calculated before the ODE systems is solved and can appear in reaction rate equations. The following list indicates the unique variable names and their corresponding equations.

- **Input variable 1:** BaF3_Epo

  $[\text{BaF3\_Epo}](t) = \text{epo\_level} \cdot e^{-\text{Epo\_degradation\_BaF3} \cdot t}$

### 9.1.4 Reactions

The model contains 9 reactions. Reactions define interactions between dynamics variables and build up the ODE systems. The following list indicates the reaction laws and their corresponding reaction rate equations. Promoting rate modifiers are indicated in black above the rate law arrow. Inhibitory rate modifiers are indicated in red below the rate law arrow. In the reaction rate equations dynamic and input variables are indicated by square brackets. The remaining variables are model parameters that remain constant over time.

- **Reaction 1:**

  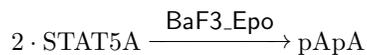

  $v_1 = [\text{BaF3\_Epo}] \cdot [\text{STAT5A}]^2 \cdot \text{k\_phos}$

- **Reaction 2:**

  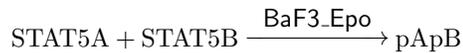

  $v_2 = [\text{BaF3\_Epo}] \cdot [\text{STAT5A}] \cdot [\text{STAT5B}] \cdot \text{k\_phos}$

- **Reaction 3:**

  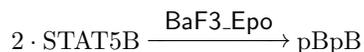

  $v_3 = [\text{BaF3\_Epo}] \cdot [\text{STAT5B}]^2 \cdot \text{k\_phos}$

- **Reaction 4:**

    pApA ⟶ nucpApA

    $v_4 = \text{k\_imp\_homo} \cdot [\text{pApA}]$

- **Reaction 5:**

    pApB ⟶ nucpApB

    $v_5 = \text{k\_imp\_hetero} \cdot [\text{pApB}]$

- **Reaction 6:**

    pBpB ⟶ nucpBpB

    $v_6 = \text{k\_imp\_homo} \cdot [\text{pBpB}]$

- **Reaction 7:**

    nucpApA ⟶ 2 · STAT5A

    $v_7 = \text{k\_exp\_homo} \cdot [\text{nucpApA}]$

- **Reaction 8:**

    nucpApB ⟶ STAT5A + STAT5B

    $v_8 = \text{k\_exp\_hetero} \cdot [\text{nucpApB}]$

- **Reaction 9:**

    nucpBpB ⟶ 2 · STAT5B

    $v_9 = \text{k\_exp\_homo} \cdot [\text{nucpBpB}]$

### 9.1.5 ODE system

The specified reation laws and rate equations $v$ determine an ODE system. The time evolution of the dynamical variables is calculated by solving this equation system.

$$d[\text{STAT5A}]/dt = -2 \cdot v_1 - v_2 + 2 \cdot v_7 \cdot \frac{0.45}{1.4} + v_8 \cdot \frac{0.45}{1.4}$$

$$d[\text{STAT5B}]/dt = -v_2 - 2 \cdot v_3 + v_8 \cdot \frac{0.45}{1.4} + 2 \cdot v_9 \cdot \frac{0.45}{1.4}$$

$$d[\text{pApB}]/dt = +v_2 - v_5$$

$$d[\text{pApA}]/dt = +v_1 - v_4$$

$$d[\text{pBpB}]/dt = +v_3 - v_6$$

$$d[\text{nucpApA}]/dt = +v_4 \cdot \frac{1.4}{0.45} - v_7$$

$$d[\text{nucpApB}]/dt = +v_5 \cdot \frac{1.4}{0.45} - v_8$$

$$d[\text{nucpBpB}]/dt = +v_6 \cdot \frac{1.4}{0.45} - v_9$$

Substituting the reaction rates $v_i$ yields:

$d[\text{STAT5A}]/dt = -2 \cdot [\text{BaF3\_Epo}] \cdot \text{k\_phos} \cdot [\text{STAT5A}]^2 - 1 \cdot [\text{BaF3\_Epo}] \cdot [\text{STAT5B}] \cdot \text{k\_phos} \cdot [\text{STAT5A}] + 0.321428571428571428571428571428571 \cdot \text{k\_exp\_hetero} \cdot [\text{nucpApB}] + 0.642857142857142857142857142857142857142857142857142857142857142857142857142857142857142857142857142857142857142857142857142857142857142857142857142857142857142857142857142857142857142857142857142857142857142857142857142857142857142857142857142857142857142857142857142857142857142857142857142857142857142857142857142857142857142857142857142857142857142857142857142857142857142857142857142857142857142857142857142857142857142857142857142857142857142857142857142857142857142857142857142857142857142857142857142857142857142857142857142857142857142857142857142857142857142857142857142857142857142857142857142857142857142857142857142857142857142857142857142857142857142857142857142857142857142857142857142857142857142857142857142857142857142857142857142857142857142857142857142857142857142857142857142857142857142857142857142857142857142857142857142857142857142857142857142857142857142857142857142857142857142857142857142857142857142857142857142857142857142857142857142857142857142857142857142857142857142857142857142857142857142857142857142857142857142857142857142857142857142857142857142857142857142857142857142857142857142857142857142857142857142857142857142857142857142857142857142857142857142857142857142857142857142857142857142857142857142857142857142857142857142857142857142857142857142857142857142857142857142857142857142857142857142857142857142857142857142857142857142857142857142857142857142857142857142857142857142857142857142857142857142857142857142857142857142857142857142857142857142857142857142857142857142857142857142857142857142857142857142857142857142857142857142857142857142857142857142857142857142857142857142857142857142857142857142857142857142857142857142857142857142857142857142857142857142857142857142857142857142857142857142857142857142857142857142857142857142857142857142857142857142857142857142857142857142857142857142857142857142857142857142857142857142857142857142857142857142857142857142857142857142857142857142857142857142857142857142857142857142857142857142857142857142857142857142857142857142857142857142857142857142857142857142857142857142857142857142857142857142857142857142857142857142857142857142857142857142857142857142857142857142857142857142857142857142857142857142857142857142857142857142857142857142857142857142857142857142857142857142857142857142857142857142857142857142857142857142857142857142857142857142857142857142857142857142857142857142857142857142857142857142857142857142857142857142857142857142857142857142857142857142857142857142857142857142857142857142857142857142857142857142857142857142857142857142857142857142857142857142857142857142857142857142857142857142857142857142857142857142857142857142857142857142857142857142857142857142857142857142857142857142857142857142857142857142857142857142857142857142857142857142857142857142857142857142857142857142857142857142857142857142857142857142857142857142857142857142857142857142857142857142857142857142857142857142857142857142857142857142857142857142857142857142857142857142857142857142857142857142857142857142857142857142857142857142857142857142857142857142857142857142857142857142857142857142857142857142857142857142857142857142857142857142857142857142857142857142857142857142857142857142857142857142857142857142857142857142857142857142857142857142857142857142857142857142857142857142857142857142857142857142857142857142857142857142857142857142857142857142857142857142857142857142857142857142857142857142857142857142857142857142857142857142857142857142857142857142857142857142857142857142857142857142857142857142857142857142857142857142857142857142857142857142857142857142857142857142857142857142857142857142857142857142857142857142857142857142857142857142857142857142857142857142857142857142857142857142857142857142857142857142857142857142857142857142857142857142857142857142857142857142857142857142857142857142857142857142857142857142857142857142857142857142857142857142857142857142857142857142857142857142857142857142857142857142857142857142857142857142857142857142857142857142857142857142857142857142857142857142857142857142857142857142857142857142857142857142857142857142857142857142857142857142857142857142857142857142857142857142857142857142857142857142857142857142857142857142857142857142857142857142857142857142857142857142857142857142857142857142857142857142857142857142857142857142857142857142857142857142857142857142857142857142857142857142857142857142857142857142857142857142857142857142857142857142857142857142857142857142857142857142857142857142857142857142857142857142857142857142857142857142857142857142857142857142857142857142857142857142857142857142857142857142857142857142857142857142857142857142857142857142857142857142857142857142857142857142857142857142857142857142857142857142857142857142857142857142857142857142857142857142857142857142857142857142857142857142857142857142857142857142857142857142857142857142857142857142857142857142857142857142857142857142857142857142857142857142857142857142857142857142857142857142857142857142857142857142857142857142857142857142857142857142857142857142857142857142857142857142857142857142857142857142857142857142857142857142857142857142857142857142857142857142857142857142857142857142857142857142857142857142857142857142857142857142857142857142857142857142857142857142857142857142857142857142857142857142857142857142857142857142857142857142857142857142857142857142857142857142857142857142857142857142857142857142857142857142857142857142857142857142857142857142857142857142857142857142857142857142857142857142857142857142857142857142857142857142857142857142857142857142857142857142857142857142857142857142857142857142857142857142857142857142857142857142857142857142857142857142857142857142857142857142857142857142857142857142857142857142857142857142857142857142857142857142857142857142857142857142857142857142857142857142857142857142857142857142857142857142857142857142857142857142857142857142857142857142857142857142857142857142857142857142857142857142857142857142857142857142857142857142857142857142857142857142857142857142857142857142857142857142857142857142857142857142857142857142857142857142857142857142857142857142857142857142857142857142857142857142857142857142857142857142857142857142857142857142857142857142857142857142857142857142857142857142857142857142857142857142857142857142857142857142857142857142857142857142857142857142857142857142857142857142857142857142857142857142857142857142857142857142857142857142857142857142857142857142857142857142857142857142857142857142857142857142857142857142857142857142857142857142857142857142857142857142857142857142857142857142857142857142857142857142857142857142857142857142857142857142857142857142857142857142857142857142857142857142857142857142857142857142857142857142857142857142857142857142857142857142857142857142857142857142857142857142857142857142857142857142857142857142857142857142857142857142857142857142857142857142857142857142857142857142857142857142857142857142857142857142857142857142857142857142857142857142857142857142857142857142857142857142857142857142857142857142857142857142857142857142857142857142857142857142857142857142857142857142857142857142857142857142857142857142857142857142857142857142857142857142857142857142857142857142857142857142857142857142857142857142857142857142857142857142857142857142857142857142857142857142857142857142857142857142857142857142857142857142857142857142857142857142857142857142857142857142857142857142857142857142857142857142857142857142857142857142857142857142857142857142857142857142857142857142857142857142857142857142857142857142857142857142857142857142857142857142857142857142857142857142857142857142857142857142857142857142857142857142857142857142857142857142857142857142857142857142857142857142857142857142857142857142857142857142857142857142857142857142857142857142857142857142857142857142857142857142857142857142857142857142857142857142857142857142857142857142857142857142857142857142857142857142857142857142857142857142857142857142857142857142857142857142857142857142857142857142857142857142857142857142857142857142857142857142857142857142857142857142857142857142857142857142857142857142857142857142857142857142857142857142857142857142857142857142857142857142857142857142857142857142857142857142857142857142857142857142857142857142857142857142857142857142857142857142857142857142857142857142857142857142857142857142857142857142857142857142857142857142857142857142857142857142857142857142857142857142857142857142857142857142857142857142857142857142857142857142857142857142857142857142857142857142857142857142857142857142857142857142857142857142857142857142857142857142857142857142857142857142857142857142857142857142857142857142857142857142857142857142857142857142857142857142857142857142857142857142857142857142857142857142857142857142857142857142857142857142857142857142857142857142857142857142857142857142857142857142857142857142857142857142857142857142857142857142857142857142857142857142857142857142857142857142857142857142857142857142857142857142857142857142857142857142857142857142857142857142857142857142857142857142857142857142857142857142857142857142857142857142857142857142857142857142857142857142857142857142857142857142857142857142857142857142857142857142857142857142857142857142857142857142857142857142857142857142857142857142857142857142857142857142857142857142857142857142857142857142857142857142857142857142857142857142857142857142857142857142857142857142857142857142857142857142857142857142857142857142857142857142857142857142857142857142857142857142857142857142857142857142857142857142857142857142857142857142857142857142857142857142857142857142857142857142857142857142857142857142857142857142857142857142857142857142857142857142857142857142857142857142857142857142857142857142857142857142857142857142857142857142857142857142857142857142857142857142857142857142857142857142857142857142857142857142857142857142857142857142857142857142857142857142857142857142857142857142857142857142857142857142857142857142857142857142857142857142857142857142857142857142857142857142857142857142857142857142857142857142857142857142857142857142857142857142857142857142857142857142857142857142857142857142857142857142857142857142857142857142857142857142857142857142857142857142857142857142857142857142857142857142857142857142857142857142857142857142857142857142857142857142857142857142857142857142857142857142857142857142857142857142857142857142857142857142857142857142857142857142857142857142857142857142857142857142857142857142857142857142857142857142857142857142857142857142857142857142857142857142857142857142857142857142857142857142857142857142857142857142857142857142857142857142857142857142857142857142857142857142857142857142857142857142857142857142857142857142857142857142857142857142857142857142857142857142857142857142857142857142857142857142857142857142857142857142857142857142857142857142857142857142857142857142857142857142857142857142857142857142857142857142857142857142857142857142857142857142857142857142857142857142857142857142857142857142857142857142857142857142857142857142857142857142857142857142857142857142857142857142857142857142857142857142857142857142857142857142857142857142857142857142857142857142857142857142857142857142857142857142857142857142857142857142857142857142857142857142857142857142857142857 \cdot \text{k\_exp\_homo} \cdot [\text{nucpApA}]$

$d[\text{STAT5B}]/dt = -2 \cdot [\text{BaF3\_Epo}] \cdot \text{k\_phos} \cdot [\text{STAT5B}]^2 - 1 \cdot [\text{BaF3\_Epo}] \cdot [\text{STAT5A}] \cdot \text{k\_phos} \cdot [\text{STAT5B}] + 0.321428571428571428571428571428571 \cdot \text{k\_exp\_hetero} \cdot [\text{nucpApB}] + 0.642857142857142857142857142857142857 \cdot \text{k\_exp\_homo} \cdot [\text{nucpBpB}]$

$d[\text{pApB}]/dt = [\text{BaF3\_Epo}] \cdot [\text{STAT5A}] \cdot [\text{STAT5B}] \cdot \text{k\_phos} - \text{k\_imp\_hetero} \cdot [\text{pApB}]$

$d[\text{pApA}]/dt = [\text{BaF3\_Epo}] \cdot [\text{STAT5A}]^2 \cdot \text{k\_phos} - \text{k\_imp\_homo} \cdot [\text{pApA}]$

$d[\text{pBpB}]/dt = [\text{BaF3\_Epo}] \cdot [\text{STAT5B}]^2 \cdot \text{k\_phos} - \text{k\_imp\_homo} \cdot [\text{pBpB}]$

$d[\text{nucpApA}]/dt = 3.11111111111111111111111111111111 \cdot \text{k\_imp\_homo} \cdot [\text{pApA}] - \text{k\_exp\_homo} \cdot [\text{nucpApA}]$

$d[\text{nucpApB}]/dt = 3.11111111111111111111111111111111 \cdot \text{k\_imp\_hetero} \cdot [\text{pApB}] - \text{k\_exp\_hetero} \cdot [\text{nucpApB}]$

$d[\text{nucpBpB}]/dt = 3.11111111111111111111111111111111 \cdot \text{k\_imp\_homo} \cdot [\text{pBpB}] - \text{k\_exp\_homo} \cdot [\text{nucpBpB}]$

The ODE system was solved by a parallelized implementation of the CVODES algorithm [4]. It also supplies the parameter sensitivities utilized for parameter estimation.

### 9.1.6 Observables

The model contains 3 standard observables. Observables are calculated after the ODE system was solved and derived variables are calculated. Dynamic, input and derived variables are indicated by square brackets. The remaining variables are model parameters that remain constant over time. In additon to the equation for the observable, also their corresponding error model $\sigma$ is indicated.

- **Observable 1:** pSTAT5A_rel

$$\text{pSTAT5A\_rel}(t) = \frac{100 \cdot [\text{pApB}] + 200 \cdot [\text{pApA}] \cdot \text{specC17}}{[\text{pApB}] + [\text{STAT5A}] \cdot \text{specC17} + 2 \cdot [\text{pApA}] \cdot \text{specC17}}$$

$$\sigma\{\text{pSTAT5A\_rel}\}(t) = \text{sd\_pSTAT5A\_rel}$$

- **Observable 2:** pSTAT5B_rel

$$\text{pSTAT5B\_rel}(t) = -\frac{100 \cdot [\text{pApB}] - 200 \cdot [\text{pBpB}] \cdot (\text{specC17} - 1)}{[\text{STAT5B}] \cdot (\text{specC17} - 1) - [\text{pApB}] + 2 \cdot [\text{pBpB}] \cdot (\text{specC17} - 1)}$$

$$\sigma\{\text{pSTAT5B\_rel}\}(t) = \text{sd\_pSTAT5B\_rel}$$

- **Observable 3:** rSTAT5A_rel

$$\text{rSTAT5A\_rel}(t) = \frac{100 \cdot [\text{pApB}] + 100 \cdot [\text{STAT5A}] \cdot \text{specC17} + 200 \cdot [\text{pApA}] \cdot \text{specC17}}{2 \cdot [\text{pApB}] + [\text{STAT5A}] \cdot \text{specC17} + 2 \cdot [\text{pApA}] \cdot \text{specC17} - [\text{STAT5B}] \cdot (\text{specC17} - 1) - 2 \cdot [\text{pBpB}] \cdot (\text{specC17} -}$$

$$\sigma\{\text{rSTAT5A\_rel}\}(t) = \text{sd\_rSTAT5A\_rel}$$

### 9.1.7 Conditions

Conditions modify the model according to replacement rules. New model parameters can be introduced or relations between existing model parameters can be implemented. The following list are default conditions that can be replace my experiment specific conditions defined seperately for each data set.

$$\text{init\_STAT5A} \to 207.6 \cdot \text{ratio}$$
$$\text{init\_STAT5B} \to 207.6 - 207.6 \cdot \text{ratio}$$
$$\text{init\_nucpApA} \to 0$$
$$\text{init\_nucpApB} \to 0$$
$$\text{init\_nucpBpB} \to 0$$
$$\text{init\_pApA} \to 0$$
$$\text{init\_pApB} \to 0$$
$$\text{init\_pBpB} \to 0$$

## 9.2 Experiment: TimeCourseData

### 9.2.1 Description

Experimenter: Lorenz Adlung and Martin Boehm  
Technique: Mass spectrometry  
Cells: BaF3  
Ligand: Epo 5 units/ml

### 9.2.2 Experiment specific conditions

To evaluate the model for this experiment the following conditions are applied.

- **Local condition #1 (global condition #1):**

$$\text{epo\_level} \to 0.000000125$$

## 9.3 Estimated model parameters

In total 9 parameters are estimated from the experimental data. The best fit yields a value of the objective function $-2\log(L) = 284.691$ for a total of 49 data points. The model parameters were estimated by maximum likelihood estimation. In Table 6 the estimated parameter values are given. Parameters highlighted in red color indicate parameter values close to their bounds. The parameter name prefix init_ indicates the initial value of a dynamic variable.

|   | name | $\theta_{min}$ | $\hat{\theta}$ | $\theta_{max}$ | log | non-log $\hat{\theta}$ |
|---|------|---|---|---|---|---|
| 1 | Epo_degradation_BaF3 | -5 | -1.5691 | +5 | 1 | $+2.70 \cdot 10^{-02}$ |
| 2 | k_exp_hetero | -5 | -4.9995 | +5 | 1 | $+1.00 \cdot 10^{-05}$ |
| 3 | k_exp_homo | -5 | -2.2098 | +5 | 1 | $+6.17 \cdot 10^{-03}$ |
| 4 | k_imp_hetero | -5 | -1.7859 | +5 | 1 | $+1.64 \cdot 10^{-02}$ |
| 5 | k_imp_homo | -5 | +3.9941 | +5 | 1 | $+9.86 \cdot 10^{+03}$ |
| 6 | k_phos | -5 | +4.1977 | +5 | 1 | $+1.58 \cdot 10^{+04}$ |
| 8 | sd_pSTAT5A_rel | -5 | +0.6147 | +5 | 1 | $+4.12 \cdot 10^{+00}$ |
| 9 | sd_pSTAT5B_rel | -5 | +0.8478 | +5 | 1 | $+7.04 \cdot 10^{+00}$ |
| 10 | sd_rSTAT5A_rel | -5 | +0.5276 | +5 | 1 | $+3.37 \cdot 10^{+00}$ |

**Table 6: Estimated parameter values**
$\hat{\theta}$ indicates the estimated value of the parameters. $\theta_{min}$ and $\theta_{max}$ indicate the upper and lower bounds for the parameters. The log-column indicates if the value of a parameter was log-transformed. If log $\equiv$ 1 the non-log-column indicates the non-logarithmic value of the estimate.

### 9.4 Profile likelihood of model parameters

As a classical approach, identifiability of the model parameters was assessed using the profile likelihood [7]. An overview is displayed in Figure 9.

### 9.5 Identifiability-test

Applying the new approach for investigating identifiability yielded the following outcome.

```
>> arIdentifiabilityTest
Identifiability-test started ...
... Identifiability-test finished.

Identifiability-test was performed with radius = 1 and penalty-SD = 1.

Warning: Penalization force addional parameters to bounds.  Decreasing the
radius is suggested in this case.

4 optimization runs (80.00 percent) are in the chi2-range 0.001.

Calculations took 2.19 seconds.
[Compared to 88.65 seconds required for calcuating the likelihood profiles.]

1.0000 (increase of merit by penalty, before fitting)
1.0004 (decrease of merit by fitting)
1.0031 (movement of parameters by penalized fitting)
-0.0004 (total increase of merit by penalty) PRIMARY CRITERION
Model is structurally non-identifiable.
```

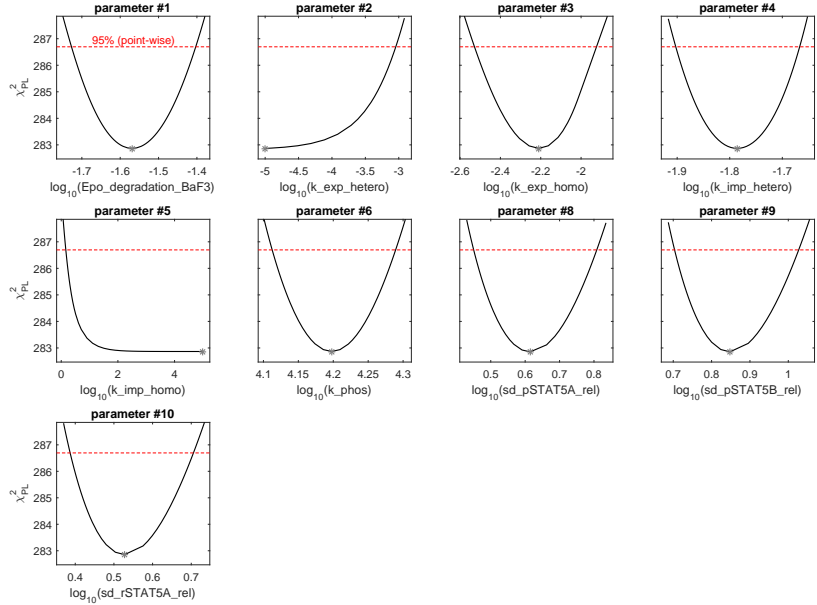

**Figure 9: Overview of the profile likelihood of the model parameters**
The solid lines indicate the profile likelihood. The broken lines indicate the threshold to assess confidence intervals. The asterisks indicate the optimal parameter values.

# 10 Model 6 ("Raia")

## 10.1 Model definition

This model has been published in [6].

### 10.1.1 Description

Model for JAK2-STAT5 signalling with feedbacks by SOCS3, CIS and SHP-1.

### 10.1.2 Dynamic variables

The model contains 14 dynamic variables. The dynamics of those variables evolve according to a system of ordinary differential equations (ODE) as will be defined in the following. The following list indicates the unique variable names and their initial conditions.

- **Dynamic variable 1:** Rec

  $[\text{Rec}](t = 0) = \text{init\_Rec}$

- **Dynamic variable 2:** Rec_i

  $[\text{Rec\_i}](t = 0) = \text{init\_Rec\_i}$

- **Dynamic variable 3:** IL13_Rec

  $[\text{IL13\_Rec}](t = 0) = \text{init\_IL13\_Rec}$

- **Dynamic variable 4:** p_IL13_Rec
  $[\text{p\_IL13\_Rec}](t = 0) = \text{init\_p\_IL13\_Rec}$
- **Dynamic variable 5:** p_IL13_Rec_i
  $[\text{p\_IL13\_Rec\_i}](t = 0) = \text{init\_p\_IL13\_Rec\_i}$
- **Dynamic variable 6:** JAK2
  $[\text{JAK2}](t = 0) = \text{init\_JAK2}$
- **Dynamic variable 7:** pJAK2
  $[\text{pJAK2}](t = 0) = \text{init\_pJAK2}$
- **Dynamic variable 8:** STAT5
  $[\text{STAT5}](t = 0) = \text{init\_STAT5}$
- **Dynamic variable 9:** pSTAT5
  $[\text{pSTAT5}](t = 0) = \text{init\_pSTAT5}$
- **Dynamic variable 10:** SOCS3mRNA
  $[\text{SOCS3mRNA}](t = 0) = \text{init\_SOCS3mRNA}$
- **Dynamic variable 11:** DecoyR
  $[\text{DecoyR}](t = 0) = \text{init\_DecoyR}$
- **Dynamic variable 12:** IL13_DecoyR
  $[\text{IL13\_DecoyR}](t = 0) = \text{init\_IL13\_DecoyR}$
- **Dynamic variable 13:** SOCS3
  $[\text{SOCS3}](t = 0) = \text{init\_SOCS3}$
- **Dynamic variable 14:** CD274mRNA
  $[\text{CD274mRNA}](t = 0) = \text{init\_CD274mRNA}$

### 10.1.3 Reactions

The model contains 16 reactions. Reactions define interactions between dynamics variables and build up the ODE systems. The following list indicates the reaction laws and their corresponding reaction rate equations. Promoting rate modifiers are indicated in black above the rate law arrow. Inhibitory rate modifiers are indicated in red below the rate law arrow. In the reaction rate equations dynamic and input variables are indicated by square brackets. The remaining variables are model parameters that remain constant over time.

- **Reaction 1:**

  $\text{Rec} \longrightarrow \text{IL13\_Rec}$

  $v_1 = 2.265 \cdot \text{Kon\_IL13Rec} \cdot [\text{Rec}] \cdot \text{il13\_level}$

- **Reaction 2:**

  $\text{Rec} \longrightarrow \text{Rec\_i}$

  $v_2 = [\text{Rec}] \cdot \text{Rec\_intern}$

32

- **Reaction 3:**

$$\text{Rec\_i} \longrightarrow \text{Rec}$$

$$v_3 = [\text{Rec\_i}] \cdot \text{Rec\_recycle}$$

- **Reaction 4:**

$$\text{IL13\_Rec} \xrightarrow{\text{pJAK2}} \text{p\_IL13\_Rec}$$

$$v_4 = [\text{IL13\_Rec}] \cdot \text{Rec\_phosphorylation} \cdot [\text{pJAK2}]$$

- **Reaction 5:**

$$\text{JAK2} \xrightarrow[\text{SOCS3}]{\text{IL13\_Rec}} \text{pJAK2}$$

$$v_5 = \frac{[\text{IL13\_Rec}] \cdot [\text{JAK2}] \cdot \text{JAK2\_phosphorylation}}{\text{JAK2\_p\_inhibition} \cdot [\text{SOCS3}] + 1}$$

- **Reaction 6:**

$$\text{JAK2} \xrightarrow[\text{SOCS3}]{\text{p\_IL13\_Rec}} \text{pJAK2}$$

$$v_6 = \frac{[\text{JAK2}] \cdot \text{JAK2\_phosphorylation} \cdot [\text{p\_IL13\_Rec}]}{\text{JAK2\_p\_inhibition} \cdot [\text{SOCS3}] + 1}$$

- **Reaction 7:**

$$\text{p\_IL13\_Rec} \longrightarrow \text{p\_IL13\_Rec\_i}$$

$$v_7 = \text{pRec\_intern} \cdot [\text{p\_IL13\_Rec}]$$

- **Reaction 8:**

$$\text{p\_IL13\_Rec\_i} \longrightarrow \varnothing$$

$$v_8 = \text{pRec\_degradation} \cdot [\text{p\_IL13\_Rec\_i}]$$

- **Reaction 9:**

$$\text{pJAK2} \longrightarrow \text{JAK2}$$

$$v_9 = \text{SHP1} \cdot [\text{pJAK2}] \cdot \text{pJAK2\_dephosphorylation}$$

- **Reaction 10:**

$$\text{STAT5} \xrightarrow{\text{pJAK2}} \text{pSTAT5}$$

$v_{10} = [\text{STAT5}] \cdot \text{STAT5\_phosphorylation} \cdot [\text{pJAK2}]$

- **Reaction 11:**

$$\text{pSTAT5} \longrightarrow \text{STAT5}$$

$v_{11} = \text{SHP1} \cdot [\text{pSTAT5}] \cdot \text{pSTAT5\_dephosphorylation}$

- **Reaction 12:**

$$\text{DecoyR} \longrightarrow \text{IL13\_DecoyR}$$

$v_{12} = 2.265 \cdot [\text{DecoyR}] \cdot \text{DecoyR\_binding} \cdot \text{il13\_level}$

- **Reaction 13:**

$$\varnothing \xrightarrow{\text{pSTAT5}} \text{SOCS3mRNA}$$

$v_{13} = \text{SOCS3mRNA\_production} \cdot [\text{pSTAT5}]$

- **Reaction 14:**

$$\varnothing \xrightarrow{\text{SOCS3mRNA}} \text{SOCS3}$$

$v_{14} = \dfrac{\text{SOCS3\_translation} \cdot [\text{SOCS3mRNA}]}{\text{SOCS3\_accumulation} + [\text{SOCS3mRNA}]}$

- **Reaction 15:**

$$\text{SOCS3} \longrightarrow \varnothing$$

$v_{15} = [\text{SOCS3}] \cdot \text{SOCS3\_degradation}$

- **Reaction 16:**

$$\varnothing \xrightarrow{\text{pSTAT5}} \text{CD274mRNA}$$

$v_{16} = \text{CD274mRNA\_production} \cdot [\text{pSTAT5}]$

### 10.1.4 ODE system

The specified reation laws and rate equations $v$ determine an ODE system. The time evolution of the dynamical variables is calculated by solving this equation system.

$$\begin{aligned}
\text{d[Rec]}/\text{dt} &= -v_1 - v_2 + v_3 \\
\text{d[Rec\_i]}/\text{dt} &= +v_2 - v_3 \\
\text{d[IL13\_Rec]}/\text{dt} &= +v_1 - v_4 \\
\text{d[p\_IL13\_Rec]}/\text{dt} &= +v_4 - v_7 \\
\text{d[p\_IL13\_Rec\_i]}/\text{dt} &= +v_7 - v_8 \\
\text{d[JAK2]}/\text{dt} &= -v_5 - v_6 + v_9 \\
\text{d[pJAK2]}/\text{dt} &= +v_5 + v_6 - v_9 \\
\text{d[STAT5]}/\text{dt} &= -v_{10} + v_{11} \\
\text{d[pSTAT5]}/\text{dt} &= +v_{10} - v_{11} \\
\text{d[SOCS3mRNA]}/\text{dt} &= +v_{13} \\
\text{d[DecoyR]}/\text{dt} &= -v_{12} \\
\text{d[IL13\_DecoyR]}/\text{dt} &= +v_{12} \\
\text{d[SOCS3]}/\text{dt} &= +v_{14} - v_{15} \\
\text{d[CD274mRNA]}/\text{dt} &= +v_{16}
\end{aligned}$$

Substituting the reaction rates $v_i$ yields:

d[Rec]/dt = [Rec\_i] · Rec\_recycle − [Rec] · Rec\_intern − 2.265 · Kon\_IL13Rec · [Rec] · il13\_level

d[Rec\_i]/dt = [Rec] · Rec\_intern − [Rec\_i] · Rec\_recycle

d[IL13\_Rec]/dt = 2.265 · Kon\_IL13Rec · [Rec] · il13\_level − [IL13\_Rec] · Rec\_phosphorylation · [pJAK2]

d[p\_IL13\_Rec]/dt = [IL13\_Rec] · Rec\_phosphorylation · [pJAK2] − pRec\_intern · [p\_IL13\_Rec]

d[p\_IL13\_Rec\_i]/dt = pRec\_intern · [p\_IL13\_Rec] − pRec\_degradation · [p\_IL13\_Rec\_i]

d[JAK2]/dt = SHP1·[pJAK2]·pJAK2\_dephosphorylation $- \frac{[\text{IL13\_Rec}]\cdot[\text{JAK2}]\cdot\text{JAK2\_phosphorylation}}{\text{JAK2\_p\_inhibition}\cdot[\text{SOCS3}]+1} - \frac{[\text{JAK2}]\cdot\text{JAK2\_phosphorylation}\cdot[\text{p\_IL13\_Rec}]}{\text{JAK2\_p\_inhibition}\cdot[\text{SOCS3}]+1}$

d[pJAK2]/dt = $\frac{[\text{IL13\_Rec}]\cdot[\text{JAK2}]\cdot\text{JAK2\_phosphorylation}}{\text{JAK2\_p\_inhibition}\cdot[\text{SOCS3}]+1}$ −SHP1·[pJAK2]·pJAK2\_dephosphorylation+ $\frac{[\text{JAK2}]\cdot\text{JAK2\_phosphorylation}\cdot[\text{p\_IL13\_Rec}]}{\text{JAK2\_p\_inhibition}\cdot[\text{SOCS3}]+1}$

d[STAT5]/dt = SHP1 · [pSTAT5] · pSTAT5\_dephosphorylation − [STAT5] · STAT5\_phosphorylation · [pJAK2]

d[pSTAT5]/dt = [STAT5] · STAT5\_phosphorylation · [pJAK2] − SHP1 · [pSTAT5] · pSTAT5\_dephosphorylation

d[SOCS3mRNA]/dt = SOCS3mRNA\_production · [pSTAT5]

d[DecoyR]/dt = −2.265 · [DecoyR] · DecoyR\_binding · il13\_level

d[IL13\_DecoyR]/dt = 2.265 · [DecoyR] · DecoyR\_binding · il13\_level

d[SOCS3]/dt = $\frac{\text{SOCS3\_translation}\cdot[\text{SOCS3mRNA}]}{\text{SOCS3\_accumulation}+[\text{SOCS3mRNA}]}$ − [SOCS3] · SOCS3\_degradation

d[CD274mRNA]/dt = CD274mRNA\_production · [pSTAT5]

The ODE system was solved by a parallelized implementation of the CVODES algorithm [4]. It also supplies the parameter sensitivities utilized for parameter estimation.

### 10.1.5 Derived variables

The model contains 2 derived variables. Derived variables are calculated after the ODE system was solved. Dynamic and input variables are indicated by square brackets. The remaining variables are model parameters that remain constant over time.

- **Derived variable 1:** IL13_cell

$$[\text{IL13\_cell}](t) = [\text{IL13\_DecoyR}] + [\text{IL13\_Rec}] + [\text{p\_IL13\_Rec}] + [\text{p\_IL13\_Rec\_i}]$$

- **Derived variable 2:** pIL4Ra

$$[\text{pIL4Ra}](t) = [\text{p\_IL13\_Rec}] + [\text{p\_IL13\_Rec\_i}]$$

### 10.1.6 Observables

The model contains 8 standard observables. Observables are calculated after the ODE system was solved and derived variables are calculated. Dynamic, input and derived variables are indicated by square brackets. The remaining variables are model parameters that remain constant over time. In additon to the equation for the observable, also their corresponding error model $\sigma$ is indicated.

- **Observable 1:** RecSurf_obs

$$\text{RecSurf\_obs}(t) = [\text{IL13\_Rec}] + [\text{Rec}] + [\text{p\_IL13\_Rec}]$$
$$\sigma\{\text{RecSurf\_obs}\}(t) = \text{sd\_RecSurf\_abs} + \text{RecSurf\_obs} \cdot \text{sd\_RecSurf\_rel}$$

- **Observable 2:** IL13_cell_obs

$$\text{IL13\_cell\_obs}(t) = [\text{IL13\_cell}] \cdot \text{scale\_IL13\_cell\_obs}$$
$$\sigma\{\text{IL13\_cell\_obs}\}(t) = \text{sd\_IL13\_cell\_abs} + \text{IL13\_cell\_obs} \cdot \text{sd\_IL13\_cell\_rel}$$

- **Observable 3:** pIL4Ra_obs

$$\text{pIL4Ra\_obs}(t) = [\text{pIL4Ra}] \cdot \text{scale\_pIL4Ra\_obs}$$
$$\sigma\{\text{pIL4Ra\_obs}\}(t) = \text{sd\_pIL4Ra\_abs} + \text{pIL4Ra\_obs} \cdot \text{sd\_pIL4Ra\_rel}$$

- **Observable 4:** pJAK2_obs

$$\text{pJAK2\_obs}(t) = [\text{pJAK2}] \cdot \text{scale\_pJAK2\_obs}$$
$$\sigma\{\text{pJAK2\_obs}\}(t) = \text{sd\_pJAK2\_abs} + \text{pJAK2\_obs} \cdot \text{sd\_pJAK2\_rel}$$

- **Observable 5:** SOCS3mRNA_obs

$$\text{SOCS3mRNA\_obs}(t) = [\text{SOCS3mRNA}] \cdot \text{scale\_SOCS3mRNA\_obs}$$
$$\sigma\{\text{SOCS3mRNA\_obs}\}(t) = \text{sd\_SOCS3mRNA\_abs} + \text{SOCS3mRNA\_obs} \cdot \text{sd\_SOCS3mRNA\_rel}$$

- **Observable 6:** CD274mRNA_obs

$$\text{CD274mRNA\_obs}(t) = [\text{CD274mRNA}] \cdot \text{scale\_CD274mRNA\_obs}$$
$$\sigma\{\text{CD274mRNA\_obs}\}(t) = \text{sd\_CD274mRNA\_abs} + \text{CD274mRNA\_obs} \cdot \text{sd\_CD274mRNA\_rel}$$

- **Observable 7:** SOCS3_obs

$$\text{SOCS3\_obs}(t) = [\text{SOCS3}]$$
$$\sigma\{\text{SOCS3\_obs}\}(t) = \text{sd\_SOCS3\_abs} + \text{SOCS3\_obs} \cdot \text{sd\_SOCS3\_rel}$$

- **Observable 8:** pSTAT5_obs

$$\text{pSTAT5\_obs}(t) = [\text{pSTAT5}]$$
$$\sigma\{\text{pSTAT5\_obs}\}(t) = \text{sd\_pSTAT5\_abs} + \text{pSTAT5\_obs} \cdot \text{sd\_pSTAT5\_rel}$$

### 10.1.7 Conditions

Conditions modify the model according to replacement rules. New model parameters can be introduced or relations between existing model parameters can be implemented. The following list are default conditions that can be replace my experiment specific conditions defined seperately for each data set.

$$SHP1 \to 91$$
$$il13\_level \to 1$$
$$init\_CD274mRNA \to 0$$
$$init\_DecoyR \to 0.34$$
$$init\_IL13\_DecoyR \to 0$$
$$init\_IL13\_Rec \to 0$$
$$init\_JAK2 \to 2.8$$
$$init\_Rec \to 1.3$$
$$init\_SOCS3 \to 0$$
$$init\_SOCS3mRNA \to 0$$
$$init\_STAT5 \to 165$$
$$init\_pJAK2 \to 0$$
$$init\_pSTAT5 \to 0$$
$$init\_p\_IL13\_Rec \to 0$$
$$init\_p\_IL13\_Rec\_i \to 0$$

## 10.2 Experiment: MedB1_real_data

### 10.2.1 Description

Experimenter: Valentina Raia

### 10.2.2 Experiment specific conditions

To evaluate the model for this experiment the following conditions are applied.

- **Local condition #1 (global condition #1):**

    $il13\_level \to 0$

- **Local condition #2 (global condition #2):**

    $il13\_level \to 20$

- **Local condition #3 (global condition #3):**

    $il13\_level \to 4$

- **Local condition #4 (global condition #4):**

    $il13\_level \to 80$

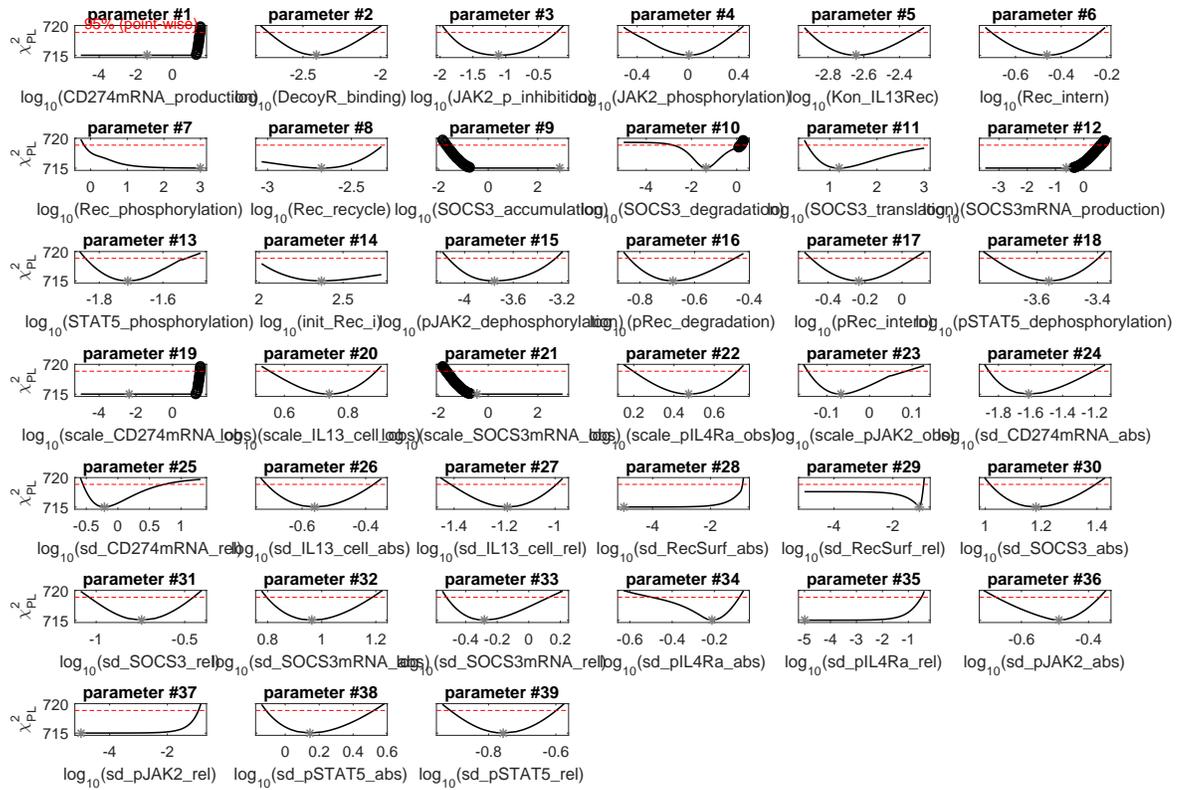

**Figure 10: Overview of the profile likelihood of the model parameters**
The solid lines indicate the profile likelihood. The broken lines indicate the threshold to assess confidence intervals. The asterisks indicate the optimal parameter values.

## 10.3 Estimated model parameters

In total 39 parameters are estimated from the experimental data. The best fit yields a value of the objective function $-2\log(L) = 716.853$ for a total of 206 data points. The model parameters were estimated by maximum likelihood estimation. In Table 7 the estimated parameter values are given. Parameters highlighted in red color indicate parameter values close to their bounds. The parameter name prefix init_ indicates the initial value of a dynamic variable.

## 10.4 Profile likelihood of model parameters

As a classical approach, identifiability of the model parameters was assessed using the profile likelihood [7]. An overview is displayed in Figure 10.

## 10.5 Identifiability-test

Applying the new approach for investigating identifiability yielded the following outcome.

|   | name | $\theta_{min}$ | $\hat{\theta}$ | $\theta_{max}$ | log | non-log $\hat{\theta}$ |
|---|---|---|---|---|---|---|
| 1 | CD274mRNA_production | -5 | -2.0667 | +3 | 1 | $+8.58 \cdot 10^{-03}$ |
| 2 | DecoyR_binding | -5 | -2.4144 | +3 | 1 | $+3.85 \cdot 10^{-03}$ |
| 3 | JAK2_p_inhibition | -5 | -1.1078 | +3 | 1 | $+7.80 \cdot 10^{-02}$ |
| 4 | JAK2_phosphorylation | -5 | +0.0075 | +3 | 1 | $+1.02 \cdot 10^{+00}$ |
| 5 | Kon_IL13Rec | -5 | -2.6424 | +3 | 1 | $+2.28 \cdot 10^{-03}$ |
| 6 | Rec_intern | -5 | -0.4632 | +3 | 1 | $+3.44 \cdot 10^{-01}$ |
| 7 | Rec_phosphorylation | -5 | +3.0000 | +3 | 1 | $+1.00 \cdot 10^{+03}$ |
| 8 | Rec_recycle | -5 | -2.6785 | +3 | 1 | $+2.10 \cdot 10^{-03}$ |
| 9 | SOCS3_accumulation | -5 | +2.7047 | +3 | 1 | $+5.07 \cdot 10^{+02}$ |
| 10 | SOCS3_degradation | -5 | -1.3623 | +3 | 1 | $+4.34 \cdot 10^{-02}$ |
| 11 | SOCS3_translation | -5 | +1.1970 | +3 | 1 | $+1.57 \cdot 10^{+01}$ |
| 12 | SOCS3mRNA_production | -5 | -0.7943 | +3 | 1 | $+1.61 \cdot 10^{-01}$ |
| 13 | STAT5_phosphorylation | -5 | -1.7104 | +3 | 1 | $+1.95 \cdot 10^{-02}$ |
| 14 | init_Rec_i | -5 | +2.3715 | +3 | 1 | $+2.35 \cdot 10^{+02}$ |
| 15 | pJAK2_dephosphorylation | -5 | -3.7545 | +3 | 1 | $+1.76 \cdot 10^{-04}$ |
| 16 | pRec_degradation | -5 | -0.6786 | +3 | 1 | $+2.10 \cdot 10^{-01}$ |
| 17 | pRec_intern | -5 | -0.2325 | +3 | 1 | $+5.85 \cdot 10^{-01}$ |
| 18 | pSTAT5_dephosphorylation | -5 | -3.5610 | +3 | 1 | $+2.75 \cdot 10^{-04}$ |
| 19 | scale_CD274mRNA_obs | -5 | -1.6954 | +3 | 1 | $+2.02 \cdot 10^{-02}$ |
| 20 | scale_IL13_cell_obs | -5 | +0.7405 | +3 | 1 | $+5.50 \cdot 10^{+00}$ |
| 21 | scale_SOCS3mRNA_obs | -5 | -0.2848 | +3 | 1 | $+5.19 \cdot 10^{-01}$ |
| 22 | scale_pIL4Ra_obs | -5 | +0.4736 | +3 | 1 | $+2.98 \cdot 10^{+00}$ |
| 23 | scale_pJAK2_obs | -5 | -0.0669 | +3 | 1 | $+8.57 \cdot 10^{-01}$ |
| 24 | sd_CD274mRNA_abs | -5 | -1.6112 | +3 | 1 | $+2.45 \cdot 10^{-02}$ |
| 25 | sd_CD274mRNA_rel | -5 | -0.2153 | +3 | 1 | $+6.09 \cdot 10^{-01}$ |
| 26 | sd_IL13_cell_abs | -5 | -0.5605 | +3 | 1 | $+2.75 \cdot 10^{-01}$ |
| 27 | sd_IL13_cell_rel | -5 | -1.1891 | +3 | 1 | $+6.47 \cdot 10^{-02}$ |
| 28 | sd_RecSurf_abs | -5 | -5.0000 | +3 | 1 | $+1.00 \cdot 10^{-05}$ |
| 29 | sd_RecSurf_rel | -5 | -1.1067 | +3 | 1 | $+7.82 \cdot 10^{-02}$ |
| 30 | sd_SOCS3_abs | -5 | +1.1809 | +3 | 1 | $+1.52 \cdot 10^{+01}$ |
| 31 | sd_SOCS3_rel | -5 | -0.7441 | +3 | 1 | $+1.80 \cdot 10^{-01}$ |
| 32 | sd_SOCS3mRNA_abs | -5 | +0.9632 | +3 | 1 | $+9.19 \cdot 10^{+00}$ |
| 33 | sd_SOCS3mRNA_rel | -5 | -0.2808 | +3 | 1 | $+5.24 \cdot 10^{-01}$ |
| 34 | sd_pIL4Ra_abs | -5 | -0.2104 | +3 | 1 | $+6.16 \cdot 10^{-01}$ |
| 35 | sd_pIL4Ra_rel | -5 | -5.0000 | +3 | 1 | $+1.00 \cdot 10^{-05}$ |
| 36 | sd_pJAK2_abs | -5 | -0.4889 | +3 | 1 | $+3.24 \cdot 10^{-01}$ |
| 37 | sd_pJAK2_rel | -5 | -5.0000 | +3 | 1 | $+1.00 \cdot 10^{-05}$ |
| 38 | sd_pSTAT5_abs | -5 | +0.1467 | +3 | 1 | $+1.40 \cdot 10^{+00}$ |
| 39 | sd_pSTAT5_rel | -5 | -0.7562 | +3 | 1 | $+1.75 \cdot 10^{-01}$ |

**Table 7: Estimated parameter values**
$\hat{\theta}$ indicates the estimated value of the parameters. $\theta_{min}$ and $\theta_{max}$ indicate the upper and lower bounds for the parameters. The log-column indicates if the value of a parameter was log-transformed. If log $\equiv 1$ the non-log-column indicates the non-logarithmic value of the estimate.

```
>> arIdentifiabilityTest
Identifiability-test started ...
...  Identifiability-test finished.


Identifiability-test was performed with radius = 1 and penalty-SD = 1.


All 5 optimization runs are in the chi2-range 2.81407e-05.


Calculations took 4.50 seconds.
[Compared to 4496.81 seconds required for calcuating the likelihood profiles.]


1.0000 (increase of merit by penalty, before fitting)
1.0000 (decrease of merit by fitting)
1.0002 (movement of parameters by penalized fitting)
0.0000 (total increase of merit by penalty) PRIMARY CRITERION
Model is structurally non-identifiable.
```

# 11 Model 7 ("Swameye")

## 11.1 Model definition

This model has been published in [11].

### 11.1.1 Description

One of the first models of JAK-STAT signalling used to show that cycling of STAT to and from the nucleus occurs.

### 11.1.2 Dynamic variables

The model contains 9 dynamic variables. The dynamics of those variables evolve according to a system of ordinary differential equations (ODE) as will be defined in the following. The following list indicates the unique variable names and their initial conditions.

- **Dynamic variable 1:** STAT

    $[\text{STAT}](t=0) = \text{init\_STAT}$

- **Dynamic variable 2:** pSTAT

    $[\text{pSTAT}](t=0) = \text{init\_pSTAT}$

- **Dynamic variable 3:** pSTAT_pSTAT

    $[\text{pSTAT\_pSTAT}](t=0) = \text{init\_pSTAT\_pSTAT}$

- **Dynamic variable 4:** npSTAT_npSTAT

    $[\text{npSTAT\_npSTAT}](t=0) = \text{init\_npSTAT\_npSTAT}$

- **Dynamic variable 5:** nSTAT1

    $[\text{nSTAT1}](t=0) = \text{init\_nSTAT1}$

- **Dynamic variable 6:** nSTAT2

    $[\text{nSTAT2}](t=0) = \text{init\_nSTAT2}$

- **Dynamic variable 7:** nSTAT3
  $$[\text{nSTAT3}](t=0) = \text{init\_nSTAT3}$$

- **Dynamic variable 8:** nSTAT4
  $$[\text{nSTAT4}](t=0) = \text{init\_nSTAT4}$$

- **Dynamic variable 9:** nSTAT5
  $$[\text{nSTAT5}](t=0) = \text{init\_nSTAT5}$$

### 11.1.3 Input variables

The model contains 1 external inputs variables. Those variables evolve according to a regular algebraic equation. They are calculated before the ODE systems is solved and can appear in reaction rate equations. The following list indicates the unique variable names and their corresponding equations.

- **Input variable 1:** pEpoR
  $$[\text{pEpoR}](t) = \text{spline\_pos5}(t, 0.0, \text{sp1}, 5.0, \text{sp2}, 10.0, \text{sp3}, 20.0, \text{sp4}, 60.0, \text{sp5}, 0, 0.0)$$

### 11.1.4 Reactions

The model contains 9 reactions. Reactions define interactions between dynamics variables and build up the ODE systems. The following list indicates the reaction laws and their corresponding reaction rate equations. Promoting rate modifiers are indicated in black above the rate law arrow. Inhibitory rate modifiers are indicated in red below the rate law arrow. In the reaction rate equations dynamic and input variables are indicated by square brackets. The remaining variables are model parameters that remain constant over time.

- **Reaction 1:**

  $$\text{STAT} \xrightarrow{\text{pEpoR}} \text{pSTAT}$$

  $$v_1 = [\text{STAT}] \cdot \text{p1} \cdot [\text{pEpoR}]$$

- **Reaction 2:**

  $$2 \cdot \text{pSTAT} \longrightarrow \text{pSTAT\_pSTAT}$$

  $$v_2 = \text{p2} \cdot [\text{pSTAT}]^2$$

- **Reaction 3:**

  $$\text{pSTAT\_pSTAT} \longrightarrow \text{npSTAT\_npSTAT}$$

  $$v_3 = \text{p3} \cdot [\text{pSTAT\_pSTAT}]$$

- **Reaction 4:**

    npSTAT_npSTAT $\longrightarrow 2 \cdot$ nSTAT1

    $v_4 = [\text{npSTAT\_npSTAT}] \cdot \text{p4}$

- **Reaction 5:**

    nSTAT1 $\longrightarrow$ nSTAT2

    $v_5 = [\text{nSTAT1}] \cdot \text{p4}$

- **Reaction 6:**

    nSTAT2 $\longrightarrow$ nSTAT3

    $v_6 = [\text{nSTAT2}] \cdot \text{p4}$

- **Reaction 7:**

    nSTAT3 $\longrightarrow$ nSTAT4

    $v_7 = [\text{nSTAT3}] \cdot \text{p4}$

- **Reaction 8:**

    nSTAT4 $\longrightarrow$ nSTAT5

    $v_8 = [\text{nSTAT4}] \cdot \text{p4}$

- **Reaction 9:**

    nSTAT5 $\longrightarrow$ STAT

    $v_9 = [\text{nSTAT5}] \cdot \text{p4}$

### 11.1.5 ODE system

The specified reation laws and rate equations $v$ determine an ODE system. The time evolution of the dynamical variables is calculated by solving this equation system.

$$\mathrm{d[STAT]/dt} = -v_1 + v_9 \cdot \frac{0.45}{1.4}$$
$$\mathrm{d[pSTAT]/dt} = +v_1 - 2 \cdot v_2$$
$$\mathrm{d[pSTAT\_pSTAT]/dt} = +v_2 - v_3$$
$$\mathrm{d[npSTAT\_npSTAT]/dt} = +v_3 \cdot \frac{1.4}{0.45} - v_4$$
$$\mathrm{d[nSTAT1]/dt} = +2 \cdot v_4 - v_5$$
$$\mathrm{d[nSTAT2]/dt} = +v_5 - v_6$$
$$\mathrm{d[nSTAT3]/dt} = +v_6 - v_7$$
$$\mathrm{d[nSTAT4]/dt} = +v_7 - v_8$$
$$\mathrm{d[nSTAT5]/dt} = +v_8 - v_9$$

Substituting the reaction rates $v_i$ yields:

$\mathrm{d[STAT]/dt} = 0.32142857142857142857142857142857 \cdot \mathrm{[nSTAT5]} \cdot \mathrm{p4} - \mathrm{[STAT]} \cdot \mathrm{p1} \cdot \mathrm{[pEpoR]}$

$\mathrm{d[pSTAT]/dt} = \mathrm{[STAT]} \cdot \mathrm{p1} \cdot \mathrm{[pEpoR]} - 2 \cdot \mathrm{p2} \cdot \mathrm{[pSTAT]}^2$

$\mathrm{d[pSTAT\_pSTAT]/dt} = \mathrm{p2} \cdot \mathrm{[pSTAT]}^2 - \mathrm{p3} \cdot \mathrm{[pSTAT\_pSTAT]}$

$\mathrm{d[npSTAT\_npSTAT]/dt} = 3.1111111111111111111111111111111 \cdot \mathrm{p3} \cdot \mathrm{[pSTAT\_pSTAT]} - \mathrm{[npSTAT\_npSTAT]} \cdot \mathrm{p4}$

$\mathrm{d[nSTAT1]/dt} = 2 \cdot \mathrm{[npSTAT\_npSTAT]} \cdot \mathrm{p4} - \mathrm{[nSTAT1]} \cdot \mathrm{p4}$

$\mathrm{d[nSTAT2]/dt} = \mathrm{[nSTAT1]} \cdot \mathrm{p4} - \mathrm{[nSTAT2]} \cdot \mathrm{p4}$

$\mathrm{d[nSTAT3]/dt} = \mathrm{[nSTAT2]} \cdot \mathrm{p4} - \mathrm{[nSTAT3]} \cdot \mathrm{p4}$

$\mathrm{d[nSTAT4]/dt} = \mathrm{[nSTAT3]} \cdot \mathrm{p4} - \mathrm{[nSTAT4]} \cdot \mathrm{p4}$

$\mathrm{d[nSTAT5]/dt} = \mathrm{[nSTAT4]} \cdot \mathrm{p4} - \mathrm{[nSTAT5]} \cdot \mathrm{p4}$

The ODE system was solved by a parallelized implementation of the CVODES algorithm [4]. It also supplies the parameter sensitivities utilized for parameter estimation.

### 11.1.6 Derived variables

The model contains 2 derived variables. Derived variables are calculated after the ODE system was solved. Dynamic and input variables are indicated by square brackets. The remaining variables are model parameters that remain constant over time.

- **Derived variable 1:** tSTAT

    $\mathrm{[tSTAT]}(t) = \mathrm{[STAT]} + \mathrm{[pSTAT]} + 2 \cdot \mathrm{[pSTAT\_pSTAT]}$

- **Derived variable 2:** tpSTAT

    $\mathrm{[tpSTAT]}(t) = \mathrm{[pSTAT]} + 2 \cdot \mathrm{[pSTAT\_pSTAT]}$

### 11.1.7 Observables

The model contains 3 standard observables. Observables are calculated after the ODE system was solved and derived variables are calculated. Dynamic, input and derived variables are indicated by square brackets. The remaining variables are model parameters that remain constant over time. In additon to the equation for the observable, also their corresponding error model $\sigma$ is indicated.

- **Observable 1:** tSTAT_au

$$\text{tSTAT\_au}(t) = \text{offset\_tSTAT} + \text{scale\_tSTAT} \cdot [\text{tSTAT}]$$
$$\sigma\{\text{tSTAT\_au}\}(t) = \text{sd\_tSTAT\_au}$$

- **Observable 2:** pSTAT_au

$$\text{pSTAT\_au}(t) = \text{offset\_pSTAT} + \text{scale\_pSTAT} \cdot [\text{tpSTAT}]$$
$$\sigma\{\text{pSTAT\_au}\}(t) = \text{sd\_pSTAT\_au}$$

- **Observable 3:** pEpoR_au

$$\text{pEpoR\_au}(t) = [\text{pEpoR}]$$
$$\sigma\{\text{pEpoR\_au}\}(t) = \text{sd\_pEpoR\_au}$$

### 11.1.8 Conditions

Conditions modify the model according to replacement rules. New model parameters can be introduced or relations between existing model parameters can be implemented. The following list are default conditions that can be replace my experiment specific conditions defined seperately for each data set.

$$\text{init\_nSTAT1} \to 0$$
$$\text{init\_nSTAT2} \to 0$$
$$\text{init\_nSTAT3} \to 0$$
$$\text{init\_nSTAT4} \to 0$$
$$\text{init\_nSTAT5} \to 0$$
$$\text{init\_npSTAT\_npSTAT} \to 0$$
$$\text{init\_pSTAT} \to 0$$
$$\text{init\_pSTAT\_pSTAT} \to 0$$
$$\text{p2} \to \frac{\text{p2}}{\text{init\_STAT}}$$
$$\text{scale\_pSTAT} \to \frac{\text{scale\_pSTAT}}{\text{init\_STAT}}$$
$$\text{scale\_tSTAT} \to \frac{\text{scale\_tSTAT}}{\text{init\_STAT}}$$

## 11.2 Estimated model parameters

In total 16 parameters are estimated from the experimental data. The best fit yields a value of the objective function $-2\log(L) = -123.782$ for a total of 47 data points. The model parameters were estimated by maximum likelihood estimation. In Table 8 the estimated parameter values are given. Parameters highlighted in red color indicate parameter values close to their bounds. The parameter name prefix init_ indicates the initial value of a dynamic variable.

## 11.3 Profile likelihood of model parameters

As a classical approach, identifiability of the model parameters was assessed using the profile likelihood [7]. An overview is displayed in Figure 11.

| | name | $\theta_{min}$ | $\hat{\theta}$ | $\theta_{max}$ | log | non-log $\hat{\theta}$ |
|---|---|---|---|---|---|---|
| 2 | offset_pSTAT | -5 | -0.6231 | +3 | 1 | $+2.38 \cdot 10^{-01}$ |
| 3 | offset_tSTAT | -5 | -0.5777 | +3 | 1 | $+2.64 \cdot 10^{-01}$ |
| 4 | p1 | -5 | +0.5391 | +3 | 1 | $+3.46 \cdot 10^{+00}$ |
| 5 | p2 | -5 | +3.0000 | +3 | 1 | $+1.00 \cdot 10^{+03}$ |
| 6 | p3 | -5 | -0.9157 | +3 | 1 | $+1.21 \cdot 10^{-01}$ |
| 7 | p4 | -5 | -0.0141 | +3 | 1 | $+9.68 \cdot 10^{-01}$ |
| 8 | scale_pSTAT | -5 | +0.0392 | +3 | 1 | $+1.09 \cdot 10^{+00}$ |
| 9 | scale_tSTAT | -5 | -0.1705 | +3 | 1 | $+6.75 \cdot 10^{-01}$ |
| 10 | sd_pEpoR_au | -5 | -1.0913 | +3 | 1 | $+8.10 \cdot 10^{-02}$ |
| 11 | sd_pSTAT_au | -5 | +2.2157 | +3 | 1 | $+1.64 \cdot 10^{+02}$ |
| 12 | sd_tSTAT_au | -5 | +2.2054 | +3 | 1 | $+1.60 \cdot 10^{+02}$ |
| 13 | sp1 | -5 | -2.6737 | +3 | 1 | $+2.12 \cdot 10^{-03}$ |
| 14 | sp2 | -5 | -0.2513 | +3 | 1 | $+5.61 \cdot 10^{-01}$ |
| 15 | sp3 | -5 | -0.0720 | +3 | 1 | $+8.47 \cdot 10^{-01}$ |
| 16 | sp4 | -5 | -0.4266 | +3 | 1 | $+3.74 \cdot 10^{-01}$ |
| 17 | sp5 | -5 | -4.7601 | +3 | 1 | $+1.74 \cdot 10^{-05}$ |

**Table 8: Estimated parameter values**
$\hat{\theta}$ indicates the estimated value of the parameters. $\theta_{min}$ and $\theta_{max}$ indicate the upper and lower bounds for the parameters. The log-column indicates if the value of a parameter was log-transformed. If log $\equiv$ 1 the non-log-column indicates the non-logarithmic value of the estimate.

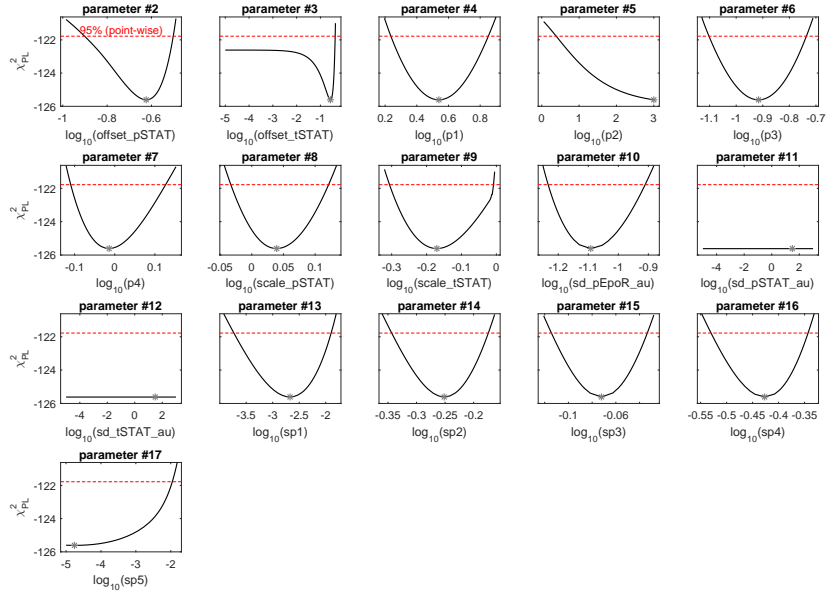

**Figure 11: Overview of the profile likelihood of the model parameters**
The solid lines indicate the profile likelihood. The broken lines indicate the threshold to assess confidence intervals. The asterisks indicate the optimal parameter values.

## 11.4 Identifiability-test

Applying the new approach for investigating identifiability yielded the following outcome.

```
>> arIdentifiabilityTest
Identifiability-test started ...
... Identifiability-test finished.

Identifiability-test was performed with radius = 1 and penalty-SD = 1.

All 4 optimization runs with random intial guesses are in the chi2-range
5.02689e-07.

Calculations took 0.66 seconds.
[Compared to 584.43 seconds required for calcuating the likelihood profiles.]


1.0000 (increase of merit by penalty, before fitting)
1.0000 (decrease of merit by fitting)
1.0000 (movement of parameters by penalized fitting)
0.0000 (total increase of merit by penalty) PRIMARY CRITERION
Model is structurally non-identifiable.
```

# 12 Model 8 ("Bachmann")

## 12.1 Model definition

This model has been published in [1].

### 12.1.1 Description

Epo induced JAK2-STAT5 signaling
with negative feedback by SOCS3, CIS and SHP1

### 12.1.2 Dynamic variables

The model contains 25 dynamic variables. The dynamics of those variables evolve according to a system of ordinary differential equations (ODE) as will be defined in the following. The following list indicates the unique variable names and their initial conditions.

- **Dynamic variable 1:** EpoRJAK2

    $[\text{EpoRJAK2}](t=0) = \text{init\_EpoRJAK2}$

- **Dynamic variable 2:** EpoRpJAK2

    $[\text{EpoRpJAK2}](t=0) = \text{init\_EpoRpJAK2}$

- **Dynamic variable 3:** p1EpoRpJAK2
  $[\text{p1EpoRpJAK2}](t=0) = \text{init\_p1EpoRpJAK2}$
- **Dynamic variable 4:** p2EpoRpJAK2
  $[\text{p2EpoRpJAK2}](t=0) = \text{init\_p2EpoRpJAK2}$
- **Dynamic variable 5:** p12EpoRpJAK2
  $[\text{p12EpoRpJAK2}](t=0) = \text{init\_p12EpoRpJAK2}$
- **Dynamic variable 6:** EpoRJAK2_CIS
  $[\text{EpoRJAK2\_CIS}](t=0) = \text{init\_EpoRJAK2\_CIS}$
- **Dynamic variable 7:** SHP1
  $[\text{SHP1}](t=0) = \text{init\_SHP1}$
- **Dynamic variable 8:** SHP1Act
  $[\text{SHP1Act}](t=0) = \text{init\_SHP1Act}$
- **Dynamic variable 9:** STAT5
  $[\text{STAT5}](t=0) = \text{init\_STAT5}$
- **Dynamic variable 10:** pSTAT5
  $[\text{pSTAT5}](t=0) = \text{init\_pSTAT5}$
- **Dynamic variable 11:** npSTAT5
  $[\text{npSTAT5}](t=0) = \text{init\_npSTAT5}$
- **Dynamic variable 12:** CISnRNA1
  $[\text{CISnRNA1}](t=0) = \text{init\_CISnRNA1}$
- **Dynamic variable 13:** CISnRNA2
  $[\text{CISnRNA2}](t=0) = \text{init\_CISnRNA2}$
- **Dynamic variable 14:** CISnRNA3
  $[\text{CISnRNA3}](t=0) = \text{init\_CISnRNA3}$
- **Dynamic variable 15:** CISnRNA4
  $[\text{CISnRNA4}](t=0) = \text{init\_CISnRNA4}$
- **Dynamic variable 16:** CISnRNA5
  $[\text{CISnRNA5}](t=0) = \text{init\_CISnRNA5}$
- **Dynamic variable 17:** CISRNA
  $[\text{CISRNA}](t=0) = \text{init\_CISRNA}$
- **Dynamic variable 18:** CIS
  $[\text{CIS}](t=0) = \text{init\_CIS}$
- **Dynamic variable 19:** SOCS3nRNA1
  $[\text{SOCS3nRNA1}](t=0) = \text{init\_SOCS3nRNA1}$
- **Dynamic variable 20:** SOCS3nRNA2
  $[\text{SOCS3nRNA2}](t=0) = \text{init\_SOCS3nRNA2}$
- **Dynamic variable 21:** SOCS3nRNA3
  $[\text{SOCS3nRNA3}](t=0) = \text{init\_SOCS3nRNA3}$
- **Dynamic variable 22:** SOCS3nRNA4
  $[\text{SOCS3nRNA4}](t=0) = \text{init\_SOCS3nRNA4}$
- **Dynamic variable 23:** SOCS3nRNA5
  $[\text{SOCS3nRNA5}](t=0) = \text{init\_SOCS3nRNA5}$
- **Dynamic variable 24:** SOCS3RNA
  $[\text{SOCS3RNA}](t=0) = \text{init\_SOCS3RNA}$
- **Dynamic variable 25:** SOCS3
  $[\text{SOCS3}](t=0) = \text{init\_SOCS3}$

### 12.1.3 Input variables

The model contains 1 external inputs variables. Those variables evolve according to a regular algebraic equation. They are calculated before the ODE systems is solved and can appear in reaction rate equations. The following list indicates the unique variable names and their corresponding equations.

- **Input variable 1:** Epo

  $[\text{Epo}](t) = \text{epo\_level}$

### 12.1.4 Reactions

The model contains 36 reactions. Reactions define interactions between dynamics variables and build up the ODE systems. The following list indicates the reaction laws and their corresponding reaction rate equations. Promoting rate modifiers are indicated in black above the rate law arrow. Inhibitory rate modifiers are indicated in red below the rate law arrow. In the reaction rate equations dynamic and input variables are indicated by square brackets. The remaining variables are model parameters that remain constant over time.

- **Reaction 1:**

  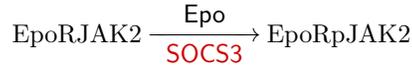

  $$v_1 = \frac{[\text{Epo}] \cdot [\text{EpoRJAK2}] \cdot \text{JAK2ActEpo}}{[\text{SOCS3}] \cdot \text{SOCS3Inh} + 1}$$

- **Reaction 2:**

  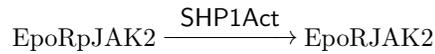

  $$v_2 = [\text{EpoRpJAK2}] \cdot \text{JAK2EpoRDeaSHP1} \cdot [\text{SHP1Act}]$$

- **Reaction 3:**

  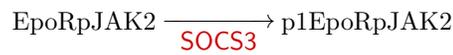

  $$v_3 = \frac{\text{EpoRActJAK2} \cdot [\text{EpoRpJAK2}]}{[\text{SOCS3}] \cdot \text{SOCS3Inh} + 1}$$

- **Reaction 4:**

  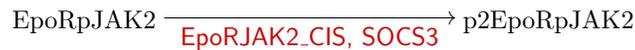

  $$v_4 = \frac{3 \cdot \text{EpoRActJAK2} \cdot [\text{EpoRpJAK2}]}{(\text{EpoRCISInh} \cdot [\text{EpoRJAK2\_CIS}] + 1) \cdot ([\text{SOCS3}] \cdot \text{SOCS3Inh} + 1)}$$

- **Reaction 5:**

  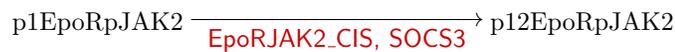

  $$v_5 = \frac{3 \cdot \text{EpoRActJAK2} \cdot [\text{p1EpoRpJAK2}]}{(\text{EpoRCISInh} \cdot [\text{EpoRJAK2\_CIS}] + 1) \cdot ([\text{SOCS3}] \cdot \text{SOCS3Inh} + 1)}$$

- **Reaction 6:**

$$\text{p2EpoRpJAK2} \xrightarrow{\text{\color{red}SOCS3}} \text{p12EpoRpJAK2}$$

$$v_6 = \frac{\text{EpoRActJAK2} \cdot [\text{p2EpoRpJAK2}]}{[\text{SOCS3}] \cdot \text{SOCS3Inh} + 1}$$

- **Reaction 7:**

$$\text{p1EpoRpJAK2} \xrightarrow{\text{SHP1Act}} \text{EpoRJAK2}$$

$$v_7 = \text{JAK2EpoRDeaSHP1} \cdot [\text{SHP1Act}] \cdot [\text{p1EpoRpJAK2}]$$

- **Reaction 8:**

$$\text{p2EpoRpJAK2} \xrightarrow{\text{SHP1Act}} \text{EpoRJAK2}$$

$$v_8 = \text{JAK2EpoRDeaSHP1} \cdot [\text{SHP1Act}] \cdot [\text{p2EpoRpJAK2}]$$

- **Reaction 9:**

$$\text{p12EpoRpJAK2} \xrightarrow{\text{SHP1Act}} \text{EpoRJAK2}$$

$$v_9 = \text{JAK2EpoRDeaSHP1} \cdot [\text{SHP1Act}] \cdot [\text{p12EpoRpJAK2}]$$

- **Reaction 10:**

$$\text{EpoRJAK2\_CIS} \xrightarrow{\text{p1EpoRpJAK2, p12EpoRpJAK2}} \varnothing$$

$$v_{10} = \text{EpoRCISRemove} \cdot [\text{EpoRJAK2\_CIS}] \cdot ([\text{p12EpoRpJAK2}] + [\text{p1EpoRpJAK2}])$$

- **Reaction 11:**

$$\text{SHP1} \xrightarrow{\text{EpoRpJAK2, p1EpoRpJAK2, p2EpoRpJAK2, p12EpoRpJAK2}} \text{SHP1Act}$$

$$v_{11} = [\text{SHP1}] \cdot \text{SHP1ActEpoR} \cdot ([\text{EpoRpJAK2}] + [\text{p12EpoRpJAK2}] + [\text{p1EpoRpJAK2}] + [\text{p2EpoRpJAK2}])$$

- **Reaction 12:**

$$\text{SHP1Act} \longrightarrow \text{SHP1}$$

$$v_{12} = [\text{SHP1Act}] \cdot \text{SHP1Dea}$$

- **Reaction 13:**

$$\text{STAT5} \xrightarrow[\text{SOCS3}]{\text{EpoRpJAK2, p1EpoRpJAK2, p2EpoRpJAK2, p12EpoRpJAK2}} \text{pSTAT5}$$

$$v_{13} = \frac{[\text{STAT5}] \cdot \text{STAT5ActJAK2} \cdot ([\text{EpoRpJAK2}] + [\text{p12EpoRpJAK2}] + [\text{p1EpoRpJAK2}] + [\text{p2EpoRpJAK2}])}{[\text{SOCS3}] \cdot \text{SOCS3Inh} + 1}$$

- **Reaction 14:**

$$\text{STAT5} \xrightarrow[\text{CIS, SOCS3}]{\text{p1EpoRpJAK2, p12EpoRpJAK2}} \text{pSTAT5}$$

$$v_{14} = \frac{[\text{STAT5}] \cdot \text{STAT5ActEpoR} \cdot ([\text{p12EpoRpJAK2}] + [\text{p1EpoRpJAK2}])^2}{([\text{CIS}] \cdot \text{CISInh} + 1) \cdot ([\text{SOCS3}] \cdot \text{SOCS3Inh} + 1)}$$

- **Reaction 15:**

$$\text{pSTAT5} \longrightarrow \text{npSTAT5}$$

$$v_{15} = \text{STAT5Imp} \cdot [\text{pSTAT5}]$$

- **Reaction 16:**

$$\text{npSTAT5} \longrightarrow \text{STAT5}$$

$$v_{16} = \text{STAT5Exp} \cdot [\text{npSTAT5}]$$

- **Reaction 17:**

$$\varnothing \xrightarrow{\text{npSTAT5}} \text{CISnRNA1}$$

$$v_{17} = -\text{CISRNAEqc} \cdot \text{CISRNATurn} \cdot [\text{npSTAT5}] \cdot (\text{ActD} - 1)$$

- **Reaction 18:**

$$\text{CISnRNA1} \longrightarrow \text{CISnRNA2}$$

$$v_{18} = \text{CISRNADelay} \cdot [\text{CISnRNA1}]$$

- **Reaction 19:**

$$\text{CISnRNA2} \longrightarrow \text{CISnRNA3}$$

$$v_{19} = \text{CISRNADelay} \cdot [\text{CISnRNA2}]$$

- **Reaction 20:**

    CISnRNA3 $\longrightarrow$ CISnRNA4

    $v_{20} = \text{CISRNADelay} \cdot [\text{CISnRNA3}]$

- **Reaction 21:**

    CISnRNA4 $\longrightarrow$ CISnRNA5

    $v_{21} = \text{CISRNADelay} \cdot [\text{CISnRNA4}]$

- **Reaction 22:**

    CISnRNA5 $\longrightarrow$ CISRNA

    $v_{22} = \text{CISRNADelay} \cdot [\text{CISnRNA5}]$

- **Reaction 23:**

    CISRNA $\longrightarrow \varnothing$

    $v_{23} = [\text{CISRNA}] \cdot \text{CISRNATurn}$

- **Reaction 24:**

    $\varnothing \xrightarrow{\text{CISRNA}} \text{CIS}$

    $v_{24} = \text{CISEqc} \cdot [\text{CISRNA}] \cdot \text{CISTurn}$

- **Reaction 25:**

    CIS $\longrightarrow \varnothing$

    $v_{25} = [\text{CIS}] \cdot \text{CISTurn}$

- **Reaction 26:**

    $\varnothing \longrightarrow \text{CIS}$

    $v_{26} = \text{CISEqcOE} \cdot \text{CISTurn} \cdot \text{CISoe}$

- **Reaction 27:**

$$\varnothing \xrightarrow{\text{npSTAT5}} \text{SOCS3nRNA1}$$

$v_{27} = -\text{SOCS3RNAEqc} \cdot \text{SOCS3RNATurn} \cdot [\text{npSTAT5}] \cdot (\text{ActD} - 1)$

- **Reaction 28:**

$$\text{SOCS3nRNA1} \longrightarrow \text{SOCS3nRNA2}$$

$v_{28} = \text{SOCS3RNADelay} \cdot [\text{SOCS3nRNA1}]$

- **Reaction 29:**

$$\text{SOCS3nRNA2} \longrightarrow \text{SOCS3nRNA3}$$

$v_{29} = \text{SOCS3RNADelay} \cdot [\text{SOCS3nRNA2}]$

- **Reaction 30:**

$$\text{SOCS3nRNA3} \longrightarrow \text{SOCS3nRNA4}$$

$v_{30} = \text{SOCS3RNADelay} \cdot [\text{SOCS3nRNA3}]$

- **Reaction 31:**

$$\text{SOCS3nRNA4} \longrightarrow \text{SOCS3nRNA5}$$

$v_{31} = \text{SOCS3RNADelay} \cdot [\text{SOCS3nRNA4}]$

- **Reaction 32:**

$$\text{SOCS3nRNA5} \longrightarrow \text{SOCS3RNA}$$

$v_{32} = \text{SOCS3RNADelay} \cdot [\text{SOCS3nRNA5}]$

- **Reaction 33:**

$$\text{SOCS3RNA} \longrightarrow \varnothing$$

$v_{33} = [\text{SOCS3RNA}] \cdot \text{SOCS3RNATurn}$

- **Reaction 34:**

$$\varnothing \xrightarrow{\text{SOCS3RNA}} \text{SOCS3}$$

$$v_{34} = \text{SOCS3Eqc} \cdot [\text{SOCS3RNA}] \cdot \text{SOCS3Turn}$$

- **Reaction 35:**

$$\text{SOCS3} \longrightarrow \varnothing$$

$$v_{35} = [\text{SOCS3}] \cdot \text{SOCS3Turn}$$

- **Reaction 36:**

$$\varnothing \longrightarrow \text{SOCS3}$$

$$v_{36} = \text{SOCS3EqcOE} \cdot \text{SOCS3Turn} \cdot \text{SOCS3oe}$$

### 12.1.5 ODE system

The specified reation laws and rate equations $v$ determine an ODE system. The time evolution of the dynamical variables is calculated by solving this equation system.

$$\begin{aligned}
d[\text{EpoRJAK2}]/dt &= -v_1 + v_2 + v_7 + v_8 + v_9 \\
d[\text{EpoRpJAK2}]/dt &= +v_1 - v_2 - v_3 - v_4 \\
d[\text{p1EpoRpJAK2}]/dt &= +v_3 - v_5 - v_7 \\
d[\text{p2EpoRpJAK2}]/dt &= +v_4 - v_6 - v_8 \\
d[\text{p12EpoRpJAK2}]/dt &= +v_5 + v_6 - v_9 \\
d[\text{EpoRJAK2\_CIS}]/dt &= -v_{10} \\
d[\text{SHP1}]/dt &= -v_{11} + v_{12} \\
d[\text{SHP1Act}]/dt &= +v_{11} - v_{12} \\
d[\text{STAT5}]/dt &= -v_{13} - v_{14} + v_{16} \cdot \frac{0.275}{0.4} \\
d[\text{pSTAT5}]/dt &= +v_{13} + v_{14} - v_{15} \\
d[\text{npSTAT5}]/dt &= +v_{15} \cdot \frac{0.4}{0.275} - v_{16} \\
d[\text{CISnRNA1}]/dt &= +v_{17} - v_{18} \\
d[\text{CISnRNA2}]/dt &= +v_{18} - v_{19} \\
d[\text{CISnRNA3}]/dt &= +v_{19} - v_{20} \\
d[\text{CISnRNA4}]/dt &= +v_{20} - v_{21} \\
d[\text{CISnRNA5}]/dt &= +v_{21} - v_{22} \\
d[\text{CISRNA}]/dt &= +v_{22} \cdot \frac{0.275}{0.4} - v_{23} \\
d[\text{CIS}]/dt &= +v_{24} - v_{25} + v_{26} \\
d[\text{SOCS3nRNA1}]/dt &= +v_{27} - v_{28} \\
d[\text{SOCS3nRNA2}]/dt &= +v_{28} - v_{29} \\
d[\text{SOCS3nRNA3}]/dt &= +v_{29} - v_{30}
\end{aligned}$$

53

$$\mathrm{d[SOCS3nRNA4]}/\mathrm{dt} = +v_{30} - v_{31}$$
$$\mathrm{d[SOCS3nRNA5]}/\mathrm{dt} = +v_{31} - v_{32}$$
$$\mathrm{d[SOCS3RNA]}/\mathrm{dt} = +v_{32} \cdot \frac{0.275}{0.4} - v_{33}$$
$$\mathrm{d[SOCS3]}/\mathrm{dt} = +v_{34} - v_{35} + v_{36}$$

Substituting the reaction rates $v_i$ yields:

$\mathrm{d[EpoRJAK2]/dt} = \mathrm{[EpoRpJAK2] \cdot JAK2EpoRDeaSHP1 \cdot [SHP1Act]} + \mathrm{JAK2EpoRDeaSHP1 \cdot [SHP1Act] \cdot [p12EpoRpJAK2]} + \mathrm{JAK2EpoRDeaSHP1 \cdot [SHP1Act]} \cdot$
$\mathrm{[p1EpoRpJAK2]} + \mathrm{JAK2EpoRDeaSHP1 \cdot [SHP1Act] \cdot [p2EpoRpJAK2]} - \frac{\mathrm{[Epo] \cdot [EpoRJAK2] \cdot JAK2ActEpo}}{\mathrm{[SOCS3] \cdot SOCS3Inh + 1}}$

$\mathrm{d[EpoRpJAK2]/dt} = \frac{\mathrm{[Epo] \cdot [EpoRJAK2] \cdot JAK2ActEpo}}{\mathrm{[SOCS3] \cdot SOCS3Inh + 1}} - \mathrm{[EpoRpJAK2] \cdot JAK2EpoRDeaSHP1 \cdot [SHP1Act]} - \frac{3 \cdot \mathrm{EpoRActJAK2 \cdot [EpoRpJAK2]}}{(\mathrm{EpoRCISInh \cdot [EpoRJAK2\_CIS] + 1}) \cdot (\mathrm{[SOCS3] \cdot SOCS3Inh + 1})} - \frac{\mathrm{EpoRActJAK2 \cdot [EpoRpJAK2]}}{\mathrm{[SOCS3] \cdot SOCS3Inh + 1}}$

$\mathrm{d[p1EpoRpJAK2]/dt} = \frac{\mathrm{EpoRActJAK2 \cdot [EpoRpJAK2]}}{\mathrm{[SOCS3] \cdot SOCS3Inh + 1}} - \mathrm{JAK2EpoRDeaSHP1 \cdot [SHP1Act] \cdot [p1EpoRpJAK2]} - \frac{3 \cdot \mathrm{EpoRActJAK2 \cdot [p1EpoRpJAK2]}}{(\mathrm{EpoRCISInh \cdot [EpoRJAK2\_CIS] + 1}) \cdot (\mathrm{[SOCS3] \cdot SOCS3Inh + 1})}$

$\mathrm{d[p2EpoRpJAK2]/dt} = \frac{3 \cdot \mathrm{EpoRActJAK2 \cdot [EpoRpJAK2]}}{(\mathrm{EpoRCISInh \cdot [EpoRJAK2\_CIS] + 1}) \cdot (\mathrm{[SOCS3] \cdot SOCS3Inh + 1})} - \mathrm{JAK2EpoRDeaSHP1 \cdot [SHP1Act] \cdot [p2EpoRpJAK2]} - \frac{\mathrm{EpoRActJAK2 \cdot [p2EpoRpJAK2]}}{\mathrm{[SOCS3] \cdot SOCS3Inh + 1}}$

$\mathrm{d[p12EpoRpJAK2]/dt} = \frac{\mathrm{EpoRActJAK2 \cdot [p2EpoRpJAK2]}}{\mathrm{[SOCS3] \cdot SOCS3Inh + 1}} - \mathrm{JAK2EpoRDeaSHP1 \cdot [SHP1Act] \cdot [p12EpoRpJAK2]} + \frac{3 \cdot \mathrm{EpoRActJAK2 \cdot [p1EpoRpJAK2]}}{(\mathrm{EpoRCISInh \cdot [EpoRJAK2\_CIS] + 1}) \cdot (\mathrm{[SOCS3] \cdot SOCS3Inh + 1})}$

$\mathrm{d[EpoRJAK2\_CIS]/dt} = -\mathrm{EpoRCISRemove \cdot [EpoRJAK2\_CIS] \cdot ([p12EpoRpJAK2] + [p1EpoRpJAK2])}$

$\mathrm{d[SHP1]/dt} = \mathrm{[SHP1Act] \cdot SHP1Dea} - \mathrm{[SHP1] \cdot SHP1ActEpoR \cdot ([EpoRpJAK2] + [p12EpoRpJAK2] + [p1EpoRpJAK2] + [p2EpoRpJAK2])}$

$\mathrm{d[SHP1Act]/dt} = \mathrm{[SHP1] \cdot SHP1ActEpoR \cdot ([EpoRpJAK2] + [p12EpoRpJAK2] + [p1EpoRpJAK2] + [p2EpoRpJAK2])} - \mathrm{[SHP1Act] \cdot SHP1Dea}$

$\mathrm{d[STAT5]/dt} = 0.6875 \cdot \mathrm{STAT5Exp \cdot [npSTAT5]} - \frac{\mathrm{[STAT5] \cdot STAT5ActJAK2 \cdot ([EpoRpJAK2] + [p12EpoRpJAK2] + [p1EpoRpJAK2] + [p2EpoRpJAK2])}}{\mathrm{[SOCS3] \cdot SOCS3Inh + 1}} - \frac{\mathrm{[STAT5] \cdot STAT5ActEpoR \cdot ([p12EpoRpJAK2] + [p1EpoRpJAK2])}^2}{(\mathrm{[CIS] \cdot CISInh + 1}) \cdot (\mathrm{[SOCS3] \cdot SOCS3Inh + 1})}$

$\mathrm{d[pSTAT5]/dt} = \frac{\mathrm{[STAT5] \cdot STAT5ActJAK2 \cdot ([EpoRpJAK2] + [p12EpoRpJAK2] + [p1EpoRpJAK2] + [p2EpoRpJAK2])}}{\mathrm{[SOCS3] \cdot SOCS3Inh + 1}} - \mathrm{STAT5Imp \cdot [pSTAT5]} + \frac{\mathrm{[STAT5] \cdot STAT5ActEpoR \cdot ([p12EpoRpJA}}{(\mathrm{[CIS] \cdot CISInh + 1}) \cdot (\mathrm{[SOCS3] \cdot S}}$

$\mathrm{d[npSTAT5]/dt} = 1.4545454545454545454545454545455 \cdot \mathrm{STAT5Imp \cdot [pSTAT5]} - \mathrm{STAT5Exp \cdot [npSTAT5]}$

$\mathrm{d[CISnRNA1]/dt} = -\mathrm{CISRNADelay \cdot [CISnRNA1]} - \mathrm{CISRNAEqc \cdot CISRNATurn \cdot [npSTAT5] \cdot (ActD - 1)}$

$\mathrm{d[CISnRNA2]/dt} = \mathrm{CISRNADelay \cdot [CISnRNA1]} - \mathrm{CISRNADelay \cdot [CISnRNA2]}$

$\mathrm{d[CISnRNA3]/dt} = \mathrm{CISRNADelay \cdot [CISnRNA2]} - \mathrm{CISRNADelay \cdot [CISnRNA3]}$

$\mathrm{d[CISnRNA4]/dt} = \mathrm{CISRNADelay \cdot [CISnRNA3]} - \mathrm{CISRNADelay \cdot [CISnRNA4]}$

$\mathrm{d[CISnRNA5]/dt} = \mathrm{CISRNADelay \cdot [CISnRNA4]} - \mathrm{CISRNADelay \cdot [CISnRNA5]}$

$\mathrm{d[CISRNA]/dt} = 0.6875 \cdot \mathrm{CISRNADelay \cdot [CISnRNA5]} - \mathrm{[CISRNA] \cdot CISRNATurn}$

$\mathrm{d[CIS]/dt} = \mathrm{CISEqc \cdot [CISRNA] \cdot CISTurn} - \mathrm{[CIS] \cdot CISTurn} + \mathrm{CISEqcOE \cdot CISTurn \cdot CISoe}$

$\mathrm{d[SOCS3nRNA1]/dt} = -\mathrm{SOCS3RNADelay \cdot [SOCS3nRNA1]} - \mathrm{SOCS3RNAEqc \cdot SOCS3RNATurn \cdot [npSTAT5] \cdot (ActD - 1)}$

$\mathrm{d[SOCS3nRNA2]/dt} = \mathrm{SOCS3RNADelay \cdot [SOCS3nRNA1]} - \mathrm{SOCS3RNADelay \cdot [SOCS3nRNA2]}$

$\mathrm{d[SOCS3nRNA3]/dt} = \mathrm{SOCS3RNADelay \cdot [SOCS3nRNA2]} - \mathrm{SOCS3RNADelay \cdot [SOCS3nRNA3]}$

$$d[\text{SOCS3nRNA4}]/dt = \text{SOCS3RNADelay} \cdot [\text{SOCS3nRNA3}] - \text{SOCS3RNADelay} \cdot [\text{SOCS3nRNA4}]$$

$$d[\text{SOCS3nRNA5}]/dt = \text{SOCS3RNADelay} \cdot [\text{SOCS3nRNA4}] - \text{SOCS3RNADelay} \cdot [\text{SOCS3nRNA5}]$$

$$d[\text{SOCS3RNA}]/dt = 0.6875 \cdot \text{SOCS3RNADelay} \cdot [\text{SOCS3nRNA5}] - [\text{SOCS3RNA}] \cdot \text{SOCS3RNATurn}$$

$$d[\text{SOCS3}]/dt = \text{SOCS3Eqc} \cdot [\text{SOCS3RNA}] \cdot \text{SOCS3Turn} - [\text{SOCS3}] \cdot \text{SOCS3Turn} + \text{SOCS3EqcOE} \cdot \text{SOCS3Turn} \cdot \text{SOCS3oe}$$

The ODE system was solved by a parallelized implementation of the CVODES algorithm [4]. It also supplies the parameter sensitivities utilized for parameter estimation.

### 12.1.6 Derived variables

The model contains 3 derived variables. Derived variables are calculated after the ODE system was solved. Dynamic and input variables are indicated by square brackets. The remaining variables are model parameters that remain constant over time.

- **Derived variable 1: pJAK2**

$$[\text{pJAK2}](t) = 2 \cdot [\text{EpoRpJAK2}] + 2 \cdot [\text{p12EpoRpJAK2}] + 2 \cdot [\text{p1EpoRpJAK2}] + 2 \cdot [\text{p2EpoRpJAK2}]$$

- **Derived variable 2: pEpoR**

$$[\text{pEpoR}](t) = 16 \cdot [\text{p12EpoRpJAK2}] + 16 \cdot [\text{p1EpoRpJAK2}] + 16 \cdot [\text{p2EpoRpJAK2}]$$

- **Derived variable 3: tSTAT5**

$$[\text{tSTAT5}](t) = [\text{STAT5}] + [\text{pSTAT5}]$$

### 12.1.7 Conditions

Conditions modify the model according to replacement rules. New model parameters can be introduced or relations between existing model parameters can be implemented. The following list are default conditions that can be replace my experiment specific conditions defined seperately for each data set.

$$\text{CISEqc} \rightarrow \frac{\text{CISEqc}}{\text{CISRNAEqc}}$$

$$\text{CISEqcOE} \rightarrow \text{CISEqc} \cdot \text{CISEqcOE}$$

$$\text{CISInh} \rightarrow \frac{\text{CISInh}}{\text{CISEqc}}$$

$$\text{CISRNAEqc} \rightarrow \frac{\text{CISRNAEqc}}{\text{init\_STAT5}}$$

$$\text{EpoRCISRemove} \rightarrow \frac{\text{EpoRCISRemove}}{\text{init\_EpoRJAK2}}$$

$$\text{JAK2EpoRDeaSHP1} \rightarrow \frac{\text{JAK2EpoRDeaSHP1}}{\text{init\_SHP1}}$$

$$\text{SHP1ActEpoR} \rightarrow \frac{\text{SHP1ActEpoR}}{\text{init\_EpoRJAK2}}$$

$$\text{SOCS3Eqc} \rightarrow \frac{\text{SOCS3Eqc}}{\text{SOCS3RNAEqc}}$$

$$\text{SOCS3EqcOE} \rightarrow \text{SOCS3Eqc} \cdot \text{SOCS3EqcOE}$$

$$\text{SOCS3Inh} \rightarrow \frac{\text{SOCS3Inh}}{\text{SOCS3Eqc}}$$

$$\text{SOCS3RNAEqc} \rightarrow \frac{\text{SOCS3RNAEqc}}{\text{init\_STAT5}}$$

$$\begin{aligned}
\text{STAT5ActEpoR} &\to \frac{\text{STAT5ActEpoR}}{\text{init\_EpoRJAK2}^2} \\
\text{STAT5ActJAK2} &\to \frac{\text{STAT5ActJAK2}}{\text{init\_EpoRJAK2}} \\
\text{init\_CIS} &\to 0 \\
\text{init\_CISRNA} &\to 0 \\
\text{init\_CISnRNA1} &\to 0 \\
\text{init\_CISnRNA2} &\to 0 \\
\text{init\_CISnRNA3} &\to 0 \\
\text{init\_CISnRNA4} &\to 0 \\
\text{init\_CISnRNA5} &\to 0 \\
\text{init\_EpoRJAK2\_CIS} &\to 0 \\
\text{init\_EpoRpJAK2} &\to 0 \\
\text{init\_SHP1Act} &\to 0 \\
\text{init\_SOCS3} &\to 0 \\
\text{init\_SOCS3RNA} &\to 0 \\
\text{init\_SOCS3nRNA1} &\to 0 \\
\text{init\_SOCS3nRNA2} &\to 0 \\
\text{init\_SOCS3nRNA3} &\to 0 \\
\text{init\_SOCS3nRNA4} &\to 0 \\
\text{init\_SOCS3nRNA5} &\to 0 \\
\text{init\_npSTAT5} &\to 0 \\
\text{init\_p12EpoRpJAK2} &\to 0 \\
\text{init\_p1EpoRpJAK2} &\to 0 \\
\text{init\_p2EpoRpJAK2} &\to 0 \\
\text{init\_pSTAT5} &\to 0
\end{aligned}$$

## 12.2 Experiment: CFUE_Long

### 12.2.1 Description

Experimenter: Julie Bachmann
Cells: CFU-E
Ligand: Epo 5 units/ml (1.25e-7 units/cell)

### 12.2.2 Experiment specific conditions

To evaluate the model for this experiment the following conditions are applied.

- **Local condition #1 (global condition #1):**

$$\begin{aligned}
\text{ActD} &\to 0 \\
\text{CISoe} &\to 0 \\
\text{SOCS3oe} &\to 0 \\
\text{epo\_level} &\to 0.000000125
\end{aligned}$$

## 12.3 Experiment: CFUE_Concentrations

### 12.3.1 Description

Experimenter: Julie Bachmann
Cells: CFU-E
Ligand: Epo 5 units/ml (1.25e-7 units/cell)

### 12.3.2 Experiment specific conditions

To evaluate the model for this experiment the following conditions are applied.

- **Local condition #2 (global condition #1):**

$$\begin{aligned} \text{ActD} &\to 0 \\ \text{CISoe} &\to 0 \\ \text{SOCS3oe} &\to 0 \\ \text{epo\_level} &\to 0.000000125 \end{aligned}$$

## 12.4 Experiment: CFUE_RNA

### 12.4.1 Description

Experimenter: Julie Bachmann
Cells: CFU-E
Ligand: Epo 5 units/ml (1.25e-7 units/cell)

### 12.4.2 Experiment specific conditions

To evaluate the model for this experiment the following conditions are applied.

- **Local condition #3 (global condition #1):**

$$\begin{aligned} \text{ActD} &\to 0 \\ \text{CISoe} &\to 0 \\ \text{SOCS3oe} &\to 0 \\ \text{epo\_level} &\to 0.000000125 \end{aligned}$$

## 12.5 Experiment: CFUE_ActD

### 12.5.1 Description

Experimenter: Julie Bachmann
Cells: CFU-E
Ligand: Epo 5 units/ml (1.25e-7 units/cell) + ActD 1 mug/ml

### 12.5.2 Experiment specific conditions

To evaluate the model for this experiment the following conditions are applied.

- **Local condition #4 (global condition #1):**

    $$\begin{aligned} \text{ActD} &\to 0 \\ \text{CISoe} &\to 0 \\ \text{SOCS3oe} &\to 0 \\ \text{epo\_level} &\to 0.000000125 \end{aligned}$$

- **Local condition #5 (global condition #2):**

    $$\begin{aligned} \text{ActD} &\to 1 \\ \text{CISoe} &\to 0 \\ \text{SOCS3oe} &\to 0 \\ \text{epo\_level} &\to 0.000000125 \end{aligned}$$

## 12.6 Experiment: CFUE_Fine

### 12.6.1 Description

Experimenter: Marcel Schilling
Cells: CFU-E
Ligand: Epo 50 units/ml (1.25e-6 units/cell)

### 12.6.2 Experiment specific conditions

To evaluate the model for this experiment the following conditions are applied.

- **Local condition #6 (global condition #3):**

    $$\begin{aligned} \text{ActD} &\to 0 \\ \text{CISoe} &\to 0 \\ \text{SOCS3oe} &\to 0 \\ \text{epo\_level} &\to 0.00000125 \end{aligned}$$

## 12.7 Experiment: CFUE_CISoe

### 12.7.1 Description

Experimenter: Julie Bachmann
Cells: CFU-E
Ligand: Epo 5 units/ml (1.25e-7 units/cell) + Cis 10-fold OE

### 12.7.2 Experiment specific conditions

To evaluate the model for this experiment the following conditions are applied.

- **Local condition #7 (global condition #4):**

$$\begin{aligned}
\text{ActD} &\to 0 \\
\text{CISoe} &\to 0 \\
\text{SOCS3oe} &\to 0 \\
\text{epo\_level} &\to 0.000000125 \\
\text{init\_CIS} &\to 0 \\
\text{init\_EpoRJAK2\_CIS} &\to 0
\end{aligned}$$

- **Local condition #8 (global condition #5):**

$$\begin{aligned}
\text{ActD} &\to 0 \\
\text{CISoe} &\to 1 \\
\text{SOCS3oe} &\to 0 \\
\text{epo\_level} &\to 0.000000125 \\
\text{init\_CIS} &\to \text{CISEqc} \cdot \text{CISEqcOE} \\
\text{init\_EpoRJAK2\_CIS} &\to 1
\end{aligned}$$

## 12.8 Experiment: CFUE_CISoe_pEpoR

### 12.8.1 Description

Experimenter: Julie Bachmann
Cells: CFU-E
Ligand: Epo 5 units/ml (1.25e-7 units/cell) + Cis 10-fold OE

### 12.8.2 Experiment specific conditions

To evaluate the model for this experiment the following conditions are applied.

- **Local condition #9 (global condition #4):**

$$\begin{aligned}
\text{ActD} &\to 0 \\
\text{CISoe} &\to 0 \\
\text{SOCS3oe} &\to 0 \\
\text{epo\_level} &\to 0.000000125 \\
\text{init\_CIS} &\to 0 \\
\text{init\_EpoRJAK2\_CIS} &\to 0
\end{aligned}$$

- **Local condition #10 (global condition #5):**

$$\begin{aligned}
\text{ActD} &\to 0 \\
\text{CISoe} &\to 1 \\
\text{SOCS3oe} &\to 0 \\
\text{epo\_level} &\to 0.000000125
\end{aligned}$$

$$\text{init\_CIS} \to \text{CISEqc} \cdot \text{CISEqcOE}$$
$$\text{init\_EpoRJAK2\_CIS} \to 1$$

## 12.9 Experiment: CFUE_SOCS3oe

### 12.9.1 Description

Experimenter: Julie Bachmann
Cells: CFU-E
Ligand: Epo 5 units/ml (1.25e-7 units/cell) + Socs3 10-fold OE

### 12.9.2 Experiment specific conditions

To evaluate the model for this experiment the following conditions are applied.

- **Local condition #11 (global condition #6):**

$$\text{ActD} \to 0$$
$$\text{CISoe} \to 0$$
$$\text{SOCS3oe} \to 0$$
$$\text{epo\_level} \to 0.000000125$$
$$\text{init\_SOCS3} \to 0$$

- **Local condition #12 (global condition #7):**

$$\text{ActD} \to 0$$
$$\text{CISoe} \to 0$$
$$\text{SOCS3oe} \to 1$$
$$\text{epo\_level} \to 0.000000125$$
$$\text{init\_SOCS3} \to \text{SOCS3Eqc} \cdot \text{SOCS3EqcOE}$$

## 12.10 Experiment: CFUE_SHP1oe

### 12.10.1 Description

Experimenter: Julie Bachmann
Cells: CFU-E
Ligand: Epo 5 units/ml (1.25e-7 units/cell) + Shp1 3.5-fold OE

### 12.10.2 Experiment specific conditions

To evaluate the model for this experiment the following conditions are applied.

- **Local condition #13 (global condition #8):**

$$\text{ActD} \to 0$$
$$\text{CISoe} \to 0$$
$$\text{SHP1oe} \to 0$$

SOCS3oe → 0
epo_level → 0.000000125

- **Local condition #14 (global condition #9):**

    ActD → 0
    CISoe → 0
    SHP1oe → 1
    SOCS3oe → 0
    epo_level → 0.000000125

## 12.11 Experiment: CFUE_DoseResp_7min

### 12.11.1 Description

Experimenter: Julie Bachmann
Cells: CFU-E
Ligand: Epo dose response at 7 minutes

### 12.11.2 Experiment specific conditions

To evaluate the model for this experiment the following conditions are applied.

- **Local condition #15 (global condition #10):**

    ActD → 0
    CISoe → 0
    SOCS3oe → 0
    epo_level → 0.000025

- **Local condition #16 (global condition #11):**

    ActD → 0
    CISoe → 0
    SOCS3oe → 0
    epo_level → 0.0000025

- **Local condition #17 (global condition #12):**

    ActD → 0
    CISoe → 0
    SOCS3oe → 0
    epo_level → 0.00000025

- **Local condition #18 (global condition #13):**

  $\text{ActD} \to 0$
  $\text{CISoe} \to 0$
  $\text{SOCS3oe} \to 0$
  $\text{epo\_level} \to 0.000000025$

- **Local condition #19 (global condition #14):**

  $\text{ActD} \to 0$
  $\text{CISoe} \to 0$
  $\text{SOCS3oe} \to 0$
  $\text{epo\_level} \to 0.0000000025$

## 12.12 Experiment: CFUE_DoseResp_30min

### 12.12.1 Description

Experimenter: Julie Bachmann
Cells: CFU-E
Ligand: Epo dose response at 30 minutes

### 12.12.2 Experiment specific conditions

To evaluate the model for this experiment the following conditions are applied.

- **Local condition #20 (global condition #15):**

  $\text{ActD} \to 0$
  $\text{CISoe} \to 0$
  $\text{SOCS3oe} \to 0$
  $\text{epo\_level} \to 0.00000125$

- **Local condition #21 (global condition #16):**

  $\text{ActD} \to 0$
  $\text{CISoe} \to 0$
  $\text{SOCS3oe} \to 0$
  $\text{epo\_level} \to 0.000000125$

- **Local condition #22 (global condition #11):**

  $\text{ActD} \to 0$
  $\text{CISoe} \to 0$
  $\text{SOCS3oe} \to 0$
  $\text{epo\_level} \to 0.0000025$

- **Local condition #23 (global condition #12):**

    $\text{ActD} \to 0$
    $\text{CISoe} \to 0$
    $\text{SOCS3oe} \to 0$
    $\text{epo\_level} \to 0.00000025$

- **Local condition #24 (global condition #13):**

    $\text{ActD} \to 0$
    $\text{CISoe} \to 0$
    $\text{SOCS3oe} \to 0$
    $\text{epo\_level} \to 0.000000025$

- **Local condition #25 (global condition #14):**

    $\text{ActD} \to 0$
    $\text{CISoe} \to 0$
    $\text{SOCS3oe} \to 0$
    $\text{epo\_level} \to 0.0000000025$

## 12.13 Experiment: CFUE_DoseResp_pSTAT5_10min_fine

### 12.13.1 Description

Experimenter: Julie Bachmann
Cells: CFU-E
Ligand: Epo dose response at 10 minutes

### 12.13.2 Experiment specific conditions

To evaluate the model for this experiment the following conditions are applied.

- **Local condition #26 (global condition #16):**

    $\text{ActD} \to 0$
    $\text{CISoe} \to 0$
    $\text{SOCS3oe} \to 0$
    $\text{epo\_level} \to 0.000000125$

- **Local condition #27 (global condition #17):**

    $\text{ActD} \to 0$
    $\text{CISoe} \to 0$
    $\text{SOCS3oe} \to 0$
    $\text{epo\_level} \to 0.0000000125$

- **Local condition #28 (global condition #18):**

    $\text{ActD} \to 0$
    $\text{CISoe} \to 0$
    $\text{SOCS3oe} \to 0$
    $\text{epo\_level} \to 0.0000000175$

- **Local condition #29 (global condition #19):**

    $\text{ActD} \to 0$
    $\text{CISoe} \to 0$
    $\text{SOCS3oe} \to 0$
    $\text{epo\_level} \to 0.00000017675$

- **Local condition #30 (global condition #11):**

    $\text{ActD} \to 0$
    $\text{CISoe} \to 0$
    $\text{SOCS3oe} \to 0$
    $\text{epo\_level} \to 0.0000025$

- **Local condition #31 (global condition #12):**

    $\text{ActD} \to 0$
    $\text{CISoe} \to 0$
    $\text{SOCS3oe} \to 0$
    $\text{epo\_level} \to 0.00000025$

- **Local condition #32 (global condition #13):**

    $\text{ActD} \to 0$
    $\text{CISoe} \to 0$
    $\text{SOCS3oe} \to 0$
    $\text{epo\_level} \to 0.000000025$

- **Local condition #33 (global condition #14):**

    $\text{ActD} \to 0$
    $\text{CISoe} \to 0$
    $\text{SOCS3oe} \to 0$
    $\text{epo\_level} \to 0.0000000025$

- **Local condition #34 (global condition #20):**

    $ActD \rightarrow 0$
    $CISoe \rightarrow 0$
    $SOCS3oe \rightarrow 0$
    $epo\_level \rightarrow 0.0000000395$

- **Local condition #35 (global condition #21):**

    $ActD \rightarrow 0$
    $CISoe \rightarrow 0$
    $SOCS3oe \rightarrow 0$
    $epo\_level \rightarrow 0.0000007905$

- **Local condition #36 (global condition #22):**

    $ActD \rightarrow 0$
    $CISoe \rightarrow 0$
    $SOCS3oe \rightarrow 0$
    $epo\_level \rightarrow 0.000000007905$

- **Local condition #37 (global condition #23):**

    $ActD \rightarrow 0$
    $CISoe \rightarrow 0$
    $SOCS3oe \rightarrow 0$
    $epo\_level \rightarrow 0.000000079$

## 12.14 Experiment: CFUE_DoseResp_CIS_90min

### 12.14.1 Description

Experimenter: Julie Bachmann
Cells: CFU-E
Ligand: Epo dose response at 90 minutes

### 12.14.2 Experiment specific conditions

To evaluate the model for this experiment the following conditions are applied.

- **Local condition #38 (global condition #16):**

    $ActD \rightarrow 0$
    $CISoe \rightarrow 0$
    $SOCS3oe \rightarrow 0$
    $epo\_level \rightarrow 0.000000125$

- **Local condition #39 (global condition #11):**

$$\text{ActD} \to 0$$
$$\text{CISoe} \to 0$$
$$\text{SOCS3oe} \to 0$$
$$\text{epo\_level} \to 0.0000025$$

- **Local condition #40 (global condition #12):**

$$\text{ActD} \to 0$$
$$\text{CISoe} \to 0$$
$$\text{SOCS3oe} \to 0$$
$$\text{epo\_level} \to 0.00000025$$

- **Local condition #41 (global condition #13):**

$$\text{ActD} \to 0$$
$$\text{CISoe} \to 0$$
$$\text{SOCS3oe} \to 0$$
$$\text{epo\_level} \to 0.000000025$$

- **Local condition #42 (global condition #14):**

$$\text{ActD} \to 0$$
$$\text{CISoe} \to 0$$
$$\text{SOCS3oe} \to 0$$
$$\text{epo\_level} \to 0.0000000025$$

## 12.15 Estimated model parameters

In total 113 parameters are estimated from the experimental data. The best fit yields a value of the objective function $-2\log(L) = -364.827$ for a total of 543 data points. The model parameters were estimated by maximum likelihood estimation. In Table 9 – 11 the estimated parameter values are given. Parameters highlighted in red color indicate parameter values close to their bounds. The parameter name prefix init_ indicates the initial value of a dynamic variable.

## 12.16 Profile likelihood of model parameters

As a classical approach, identifiability of the model parameters was assessed using the profile likelihood [7]. An overview is displayed in Figure 12.

## 12.17 Identifiability-test

Applying the new approach for investigating identifiability yielded the following outcome.

|    | name | $\theta_{min}$ | $\hat{\theta}$ | $\theta_{max}$ | log | non-log $\hat{\theta}$ |
|----|------|------|------|------|-----|------|
| 1  | CISEqc | -3 | +2.6364 | +4 | 1 | $+4.33 \cdot 10^{+02}$ |
| 2  | CISEqcOE | -3 | -0.2755 | +3 | 1 | $+5.30 \cdot 10^{-01}$ |
| 3  | CISInh | -3 | +8.8949 | +1e+01 | 1 | $+7.85 \cdot 10^{+08}$ |
| 4  | CISRNADelay | -3 | -0.8393 | +3 | 1 | $+1.45 \cdot 10^{-01}$ |
| 6  | CISRNATurn | -3 | +3.0000 | +3 | 1 | $+1.00 \cdot 10^{+03}$ |
| 7  | CISTurn | -3 | -2.0758 | +3 | 1 | $+8.40 \cdot 10^{-03}$ |
| 8  | EpoRActJAK2 | -3 | -0.5730 | +4 | 1 | $+2.67 \cdot 10^{-01}$ |
| 9  | EpoRCISInh | -3 | +6.0000 | +6 | 1 | $+1.00 \cdot 10^{+06}$ |
| 10 | EpoRCISRemove | -3 | +0.7348 | +3 | 1 | $+5.43 \cdot 10^{+00}$ |
| 11 | JAK2ActEpo | -3 | +5.8015 | +9 | 1 | $+6.33 \cdot 10^{+05}$ |
| 12 | JAK2EpoRDeaSHP1 | -3 | +2.1545 | +4 | 1 | $+1.43 \cdot 10^{+02}$ |
| 13 | SHP1ActEpoR | -3 | -3.0000 | +3 | 1 | $+1.00 \cdot 10^{-03}$ |
| 14 | SHP1Dea | -3 | -2.0882 | +3 | 1 | $+8.16 \cdot 10^{-03}$ |
| 15 | SHP1ProOE | -3 | +0.4511 | +3 | 1 | $+2.83 \cdot 10^{+00}$ |
| 16 | SOCS3Eqc | -3 | +2.2397 | +3 | 1 | $+1.74 \cdot 10^{+02}$ |
| 17 | SOCS3EqcOE | -3 | -0.1680 | +3 | 1 | $+6.79 \cdot 10^{-01}$ |
| 18 | SOCS3Inh | -3 | +1.0174 | +3 | 1 | $+1.04 \cdot 10^{+01}$ |
| 19 | SOCS3RNADelay | -3 | +0.0272 | +3 | 1 | $+1.06 \cdot 10^{+00}$ |
| 21 | SOCS3RNATurn | -3 | -2.0805 | +3 | 1 | $+8.31 \cdot 10^{-03}$ |
| 22 | SOCS3Turn | -3 | +3.0006 | +4 | 1 | $+1.00 \cdot 10^{+03}$ |
| 23 | STAT5ActEpoR | -3 | +1.5910 | +3 | 1 | $+3.90 \cdot 10^{+01}$ |
| 24 | STAT5ActJAK2 | -3 | -1.1073 | +3 | 1 | $+7.81 \cdot 10^{-02}$ |
| 25 | STAT5Exp | -3 | -1.1278 | +3 | 1 | $+7.45 \cdot 10^{-02}$ |
| 26 | STAT5Imp | -3 | -1.5705 | +3 | 1 | $+2.69 \cdot 10^{-02}$ |
| 27 | init_EpoRJAK2 | -3 | +0.5995 | +3 | 1 | $+3.98 \cdot 10^{+00}$ |
| 28 | init_SHP1 | -3 | +1.4269 | +3 | 1 | $+2.67 \cdot 10^{+01}$ |
| 29 | init_STAT5 | -3 | +1.9017 | +3 | 1 | $+7.98 \cdot 10^{+01}$ |
| 30 | offset_CIS_actd | -3 | -2.0274 | +3 | 1 | $+9.39 \cdot 10^{-03}$ |
| 31 | offset_CIS_cisoe | -3 | -1.5227 | +3 | 1 | $+3.00 \cdot 10^{-02}$ |
| 32 | offset_CIS_long | -3 | -1.5774 | +3 | 1 | $+2.65 \cdot 10^{-02}$ |
| 33 | offset_CIS_shp1oe | -3 | -1.2307 | +3 | 1 | $+5.88 \cdot 10^{-02}$ |
| 34 | offset_CIS_socs3oe | -3 | -1.0434 | +3 | 1 | $+9.05 \cdot 10^{-02}$ |
| 35 | offset_SOCS3_cisoe | -3 | -0.5555 | +3 | 1 | $+2.78 \cdot 10^{-01}$ |
| 36 | offset_SOCS3_long | -3 | -0.9413 | +3 | 1 | $+1.14 \cdot 10^{-01}$ |
| 37 | offset_SOCS3_socs3oe | -3 | -1.5920 | +3 | 1 | $+2.56 \cdot 10^{-02}$ |
| 38 | offset_pEpoR_actd | -3 | -1.7249 | +3 | 1 | $+1.88 \cdot 10^{-02}$ |
| 39 | offset_pEpoR_cisoe | -3 | -1.5198 | +3 | 1 | $+3.02 \cdot 10^{-02}$ |
| 40 | offset_pEpoR_cisoe_pepor | -3 | -0.8826 | +3 | 1 | $+1.31 \cdot 10^{-01}$ |
| 41 | offset_pEpoR_dr30 | -3 | -3.0000 | +3 | 1 | $+1.00 \cdot 10^{-03}$ |
| 42 | offset_pEpoR_dr7 | -3 | -1.5466 | +3 | 1 | $+2.84 \cdot 10^{-02}$ |
| 43 | offset_pEpoR_fine | -3 | -1.1901 | +3 | 1 | $+6.46 \cdot 10^{-02}$ |
| 44 | offset_pEpoR_long | -3 | -2.3574 | +3 | 1 | $+4.39 \cdot 10^{-03}$ |
| 45 | offset_pEpoR_shp1oe | -3 | -1.5041 | +3 | 1 | $+3.13 \cdot 10^{-02}$ |
| 46 | offset_pEpoR_socs3oe | -3 | -1.2463 | +3 | 1 | $+5.67 \cdot 10^{-02}$ |
| 47 | offset_pJAK2_actd | -3 | -1.7680 | +3 | 1 | $+1.71 \cdot 10^{-02}$ |
| 48 | offset_pJAK2_cisoe | -3 | -1.6590 | +3 | 1 | $+2.19 \cdot 10^{-02}$ |
| 49 | offset_pJAK2_dr30 | -3 | -1.5237 | +3 | 1 | $+2.99 \cdot 10^{-02}$ |

**Table 9: Estimated parameter values**
$\hat{\theta}$ indicates the estimated value of the parameters. $\theta_{min}$ and $\theta_{max}$ indicate the upper and lower bounds for the parameters. The log-column indicates if the value of a parameter was log-transformed. If log ≡ 1 the non-log-column indicates the non-logarithmic value of the estimate.

| | name | $\theta_{min}$ | $\hat{\theta}$ | $\theta_{max}$ | log | non-log $\hat{\theta}$ | estimated |
|---|---|---|---|---|---|---|---|
| 50 | offset_pJAK2_dr7 | -3 | -1.3248 | +3 | 1 | $+4.73 \cdot 10^{-02}$ | |
| 51 | offset_pJAK2_fine | -3 | -1.6644 | +3 | 1 | $+2.17 \cdot 10^{-02}$ | |
| 52 | offset_pJAK2_long | -3 | -2.0235 | +3 | 1 | $+9.47 \cdot 10^{-03}$ | |
| 53 | offset_pJAK2_shp1oe | -3 | -1.5740 | +3 | 1 | $+2.67 \cdot 10^{-02}$ | |
| 54 | offset_pJAK2_socs3oe | -3 | -1.2236 | +3 | 1 | $+5.98 \cdot 10^{-02}$ | |
| 55 | offset_pSTAT5_actd | -3 | -2.7284 | +3 | 1 | $+1.87 \cdot 10^{-03}$ | |
| 56 | offset_pSTAT5_cisoe | -3 | -1.1500 | +3 | 1 | $+7.08 \cdot 10^{-02}$ | |
| 57 | offset_pSTAT5_conc | -3 | -0.6255 | +3 | 1 | $+2.37 \cdot 10^{-01}$ | |
| 58 | offset_pSTAT5_long | -3 | -2.9614 | +3 | 1 | $+1.09 \cdot 10^{-03}$ | |
| 59 | offset_pSTAT5_shp1oe | -3 | -1.2601 | +3 | 1 | $+5.49 \cdot 10^{-02}$ | |
| 60 | offset_pSTAT5_socs3oe | -3 | -2.2508 | +3 | 1 | $+5.61 \cdot 10^{-03}$ | |
| 61 | scale1_CIS_dr90 | -3 | +1.2446 | +3 | 1 | $+1.76 \cdot 10^{+01}$ | |
| 62 | scale2_CIS_dr90 | -3 | +1.2074 | +3 | 1 | $+1.61 \cdot 10^{+01}$ | |
| 63 | scale_CISRNA_foldA | -3 | +1.5222 | +3 | 1 | $+3.33 \cdot 10^{+01}$ | |
| 64 | scale_CISRNA_foldB | -3 | +1.4916 | +3 | 1 | $+3.10 \cdot 10^{+01}$ | |
| 65 | scale_CISRNA_foldC | -3 | +1.2927 | +3 | 1 | $+1.96 \cdot 10^{+01}$ | |
| 66 | scale_CIS_actd | -3 | +1.1628 | +3 | 1 | $+1.45 \cdot 10^{+01}$ | |
| 67 | scale_CIS_cisoe | -3 | +0.1370 | +3 | 1 | $+1.37 \cdot 10^{+00}$ | |
| 68 | scale_CIS_long | -3 | +1.2150 | +3 | 1 | $+1.64 \cdot 10^{+01}$ | |
| 69 | scale_CIS_shp1oe | -3 | +1.7136 | +3 | 1 | $+5.17 \cdot 10^{+01}$ | |
| 70 | scale_CIS_socs3oe | -3 | +1.3253 | +3 | 1 | $+2.12 \cdot 10^{+01}$ | |
| 71 | scale_SHP1_shp1oe | -3 | -0.6468 | +3 | 1 | $+2.26 \cdot 10^{-01}$ | |
| 72 | scale_SOCS3RNA_foldA | -3 | +1.7556 | +3 | 1 | $+5.70 \cdot 10^{+01}$ | |
| 73 | scale_SOCS3RNA_foldB | -3 | +1.6909 | +3 | 1 | $+4.91 \cdot 10^{+01}$ | |
| 74 | scale_SOCS3RNA_foldC | -3 | +1.9071 | +3 | 1 | $+8.07 \cdot 10^{+01}$ | |
| 75 | scale_SOCS3_cisoe | -3 | +1.0633 | +3 | 1 | $+1.16 \cdot 10^{+01}$ | |
| 76 | scale_SOCS3_long | -3 | +1.1899 | +3 | 1 | $+1.55 \cdot 10^{+01}$ | |
| 77 | scale_SOCS3_socs3oe | -3 | +0.0809 | +3 | 1 | $+1.20 \cdot 10^{+00}$ | |
| 78 | scale_pEpoR_actd | -3 | -0.6662 | +3 | 1 | $+2.16 \cdot 10^{-01}$ | |
| 79 | scale_pEpoR_cisoe | -3 | -0.5628 | +3 | 1 | $+2.74 \cdot 10^{-01}$ | |
| 80 | scale_pEpoR_cisoe_pepor | -3 | -0.8052 | +3 | 1 | $+1.57 \cdot 10^{-01}$ | |
| 81 | scale_pEpoR_dr30 | -3 | -0.2697 | +3 | 1 | $+5.37 \cdot 10^{-01}$ | |
| 82 | scale_pEpoR_dr7 | -3 | -0.9938 | +3 | 1 | $+1.01 \cdot 10^{-01}$ | |
| 83 | scale_pEpoR_fine | -3 | -1.0955 | +3 | 1 | $+8.03 \cdot 10^{-02}$ | |
| 84 | scale_pEpoR_long | -3 | -0.5909 | +3 | 1 | $+2.57 \cdot 10^{-01}$ | |
| 85 | scale_pEpoR_shp1oe | -3 | -0.6175 | +3 | 1 | $+2.41 \cdot 10^{-01}$ | |
| 86 | scale_pEpoR_socs3oe | -3 | -0.1957 | +3 | 1 | $+6.37 \cdot 10^{-01}$ | |
| 87 | scale_pJAK2_actd | -3 | -0.0918 | +3 | 1 | $+8.09 \cdot 10^{-01}$ | |
| 88 | scale_pJAK2_cisoe | -3 | +0.2682 | +3 | 1 | $+1.85 \cdot 10^{+00}$ | |
| 89 | scale_pJAK2_dr30 | -3 | +0.2493 | +3 | 1 | $+1.78 \cdot 10^{+00}$ | |
| 90 | scale_pJAK2_dr7 | -3 | -0.2956 | +3 | 1 | $+5.06 \cdot 10^{-01}$ | |
| 91 | scale_pJAK2_fine | -3 | -0.3971 | +3 | 1 | $+4.01 \cdot 10^{-01}$ | |
| 92 | scale_pJAK2_long | -3 | -0.0542 | +3 | 1 | $+8.83 \cdot 10^{-01}$ | |
| 93 | scale_pJAK2_shp1oe | -3 | +0.3636 | +3 | 1 | $+2.31 \cdot 10^{+00}$ | |
| 94 | scale_pJAK2_socs3oe | -3 | +0.1026 | +3 | 1 | $+1.27 \cdot 10^{+00}$ | |
| 95 | scale_pSTAT5_actd | -3 | +0.0404 | +3 | 1 | $+1.10 \cdot 10^{+00}$ | |
| 96 | scale_pSTAT5_cisoe | -3 | +0.3836 | +3 | 1 | $+2.42 \cdot 10^{+00}$ | |
| 97 | scale_pSTAT5_dr10 | -3 | +0.0041 | +3 | 1 | $+1.01 \cdot 10^{+00}$ | |
| 98 | scale_pSTAT5_long | -3 | +0.1684 | +3 | 1 | $+1.47 \cdot 10^{+00}$ | |
| 99 | scale_pSTAT5_shp1oe | -3 | +0.0613 | +3 | 1 | $+1.15 \cdot 10^{+00}$ | |

**Table 10: Estimated parameter values**
$\hat{\theta}$ indicates the estimated value of the parameters. $\theta_{min}$ and $\theta_{max}$ indicate the upper and lower bounds for the parameters. The log-column indicates if the value of a parameter was log-transformed. If log $\equiv$ 1 the non-log-column indicates the non-logarithmic value of the estimate.

**Figure 12: Overview of the profile likelihood of the model parameters**
The solid lines indicate the profile likelihood. The broken lines indicate the threshold to assess confidence intervals. The asterisks indicate the optimal parameter values.

69

|  | name | $\theta_{min}$ | $\hat{\theta}$ | $\theta_{max}$ | log | non-log $\hat{\theta}$ | estimated |
|---|---|---|---|---|---|---|---|
| 100 | scale_pSTAT5_socs3oe | -3 | +0.2456 | +3 | 1 | $+1.76 \cdot 10^{+00}$ | |
| 101 | scale_tSTAT5_actd | -3 | -0.0887 | +3 | 1 | $+8.15 \cdot 10^{-01}$ | |
| 102 | scale_tSTAT5_long | -3 | -0.1180 | +3 | 1 | $+7.62 \cdot 10^{-01}$ | |
| 103 | scale_tSTAT5_shp1oe | -3 | -0.1668 | +3 | 1 | $+6.81 \cdot 10^{-01}$ | |
| 104 | sd_CIS_abs | -3 | -1.0998 | +3 | 1 | $+7.95 \cdot 10^{-02}$ | |
| 105 | sd_CIS_au | -3 | -0.8279 | +3 | 1 | $+1.49 \cdot 10^{-01}$ | |
| 106 | sd_JAK2EpoR_au | -3 | -0.6722 | +3 | 1 | $+2.13 \cdot 10^{-01}$ | |
| 107 | sd_RNA_fold | -3 | -0.9420 | +3 | 1 | $+1.14 \cdot 10^{-01}$ | |
| 108 | sd_SHP1_abs | -3 | -1.1428 | +3 | 1 | $+7.20 \cdot 10^{-02}$ | |
| 109 | sd_SHP1_au | -3 | -1.0662 | +3 | 1 | $+8.59 \cdot 10^{-02}$ | |
| 110 | sd_SOCS3_abs | -3 | -0.9435 | +3 | 1 | $+1.14 \cdot 10^{-01}$ | |
| 111 | sd_SOCS3_au | -3 | -1.0723 | +3 | 1 | $+8.47 \cdot 10^{-02}$ | |
| 112 | sd_STAT5_abs | -3 | -0.8777 | +3 | 1 | $+1.33 \cdot 10^{-01}$ | |
| 113 | sd_STAT5_au | -3 | -0.8634 | +3 | 1 | $+1.37 \cdot 10^{-01}$ | |
| 114 | sd_pSTAT5_rel | -3 | +0.4543 | +3 | 1 | $+2.85 \cdot 10^{+00}$ | |
| 115 | sd_pSTAT5_socs3oe | -3 | -0.1963 | +3 | 1 | $+6.36 \cdot 10^{-01}$ | |

**Table 11: Estimated parameter values**
$\hat{\theta}$ indicates the estimated value of the parameters. $\theta_{min}$ and $\theta_{max}$ indicate the upper and lower bounds for the parameters. The log-column indicates if the value of a parameter was log-transformed. If log $\equiv 1$ the non-log-column indicates the non-logarithmic value of the estimate.

```
>> arIdentifiabilityTest
Identifiability-test started ...
... Identifiability-test finished.

Identifiability-test was performed with radius = 1 and penalty-SD = 1.

Warning: Penalization force addional parameters to bounds.  Decreasing the
radius is suggested in this case.

Only 1 optimization runs (20.00 percent) are in the chi2-range 0.001.
Increasing the number of fits should be considerd.

Calculations took 9.05 seconds.
[Compared to 45168.14 seconds required for calcuating the likelihood
profiles.]

1.0000 (increase of merit by penalty, before fitting)
1.0005 (decrease of merit by fitting)
0.9994 (movement of parameters by penalized fitting)
-0.0005 (total increase of merit by penalty) PRIMARY CRITERION
Model is structurally non-identifiable.
```

# 13 Model 9 ("School")

## 13.1 Model definition: SIR-model

This model has been applied in [12].

### 13.1.1 Description

SIR model for an infulenza outbreak at an Englisch boarding School in 1978, see BMJ 1978; 1: 587.

### 13.1.2 Dynamic variables

The model contains 3 dynamic variables. The dynamics of those variables evolve according to a system of ordinary differential equations (ODE) as will be defined in the following. The following list indicates the unique variable names and their initial conditions.

- **Dynamic variable 1:** Sus
  $$[\text{Sus}](t = 0) = \text{init\_Sus}$$
- **Dynamic variable 2:** Infc
  $$[\text{Infc}](t = 0) = \text{init\_Infc}$$
- **Dynamic variable 3:** Rem
  $$[\text{Rem}](t = 0) = \text{init\_Rem}$$

### 13.1.3 Dynamic equations

The model contains 2 conversion steps. In the rate equations dynamic and input variables are indicated by square brackets. The remaining variables are model parameters that remain constant over time.

- **Process/conversion 1:**

$$\text{Sus} \xrightarrow[\text{\color{red}Rem}]{} \text{Infc}$$

$$v_1 = \frac{[\text{Infc}] \cdot [\text{Sus}] \cdot \text{k\_beta}}{[\text{Infc}] + [\text{Rem}] + [\text{Sus}]}$$

- **Process/conversion 2:**

$$\text{Infc} \longrightarrow \text{Rem}$$

$$v_2 = [\text{Infc}] \cdot \text{k\_gamma}$$

### 13.1.4 ODE system

The specified reation laws and rate equations $v$ determine an ODE system. The time evolution of the dynamical variables is calculated by solving this equation system.

$$d[\text{Sus}]/dt = -v_1$$
$$d[\text{Infc}]/dt = +v_1 - v_2$$
$$d[\text{Rem}]/dt = +v_2$$

Substituting the rates $v_i$ yields:

$d[\text{Sus}]/dt = $ 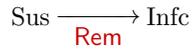

$d[\text{Infc}]/dt = $ 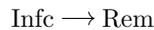 $- [\text{Infc}] \cdot \text{k\_gamma}$

$d[\text{Rem}]/dt = [\text{Infc}] \cdot \text{k\_gamma}$

The ODE system was solved by a parallelized implementation of the CVODES algorithm [4]. It also supplies the parameter sensitivities utilized for parameter estimation.

|   | name    | $\theta_{min}$ | $\hat{\theta}$ | $\theta_{max}$ | log | non-log $\hat{\theta}$ |
|---|---------|----------------|----------------|----------------|-----|------------------------|
| 1 | N_I     | -5             | +0.3763        | +4             | 1   | $+2.38 \cdot 10^{+00}$ |
| 2 | N_S     | -5             | +2.9015        | +4             | 1   | $+7.97 \cdot 10^{+02}$ |
| 3 | k_beta  | -5             | +0.2461        | +4             | 1   | $+1.76 \cdot 10^{+00}$ |
| 4 | k_gamma | -5             | -0.3140        | +4             | 1   | $+4.85 \cdot 10^{-01}$ |
| 5 | sd_Infc | -5             | +1.2966        | +4             | 1   | $+1.98 \cdot 10^{+01}$ |

**Table 12: Estimated parameter values**
$\hat{\theta}$ indicates the estimated value of the parameters. $\theta_{min}$ and $\theta_{max}$ indicate the upper and lower bounds for the parameters. The log-column indicates if the value of a parameter was log-transformed. If log $\equiv$ 1 the non-log-column indicates the non-logarithmic value of the estimate.

### 13.1.5 Derived variables

The model contains 1 derived variables. Derived variables are calculated after the ODE system was solved. Dynamic and input variables are indicated by square brackets. The remaining variables are model parameters that remain constant over time.

- **Derived variable 1:** N

$$[N](t) = [Infc] + [Rem] + [Sus]$$

### 13.1.6 Observables

The model contains only a single observable. Observables are calculated after the ODE system was solved and derived variables are calculated. Dynamic, input and derived variables are indicated by square brackets. The remaining variables are model parameters that remain constant over time. In additon to the equation for the observable, also their corresponding error model $\sigma$ is indicated.

- **Observable 1:** Infected

$$\text{Infected}(t) = [Infc]$$
$$\sigma\{\text{Infected}\}(t) = \text{sd\_Infc}$$

### 13.1.7 Conditions

Conditions modify the model according to replacement rules. New model parameters can be introduced or relations between existing model parameters can be implemented. The following list are default conditions that can be replace my experiment specific conditions defined seperately for each data set.

$$\text{init\_Infc} \rightarrow \text{N\_I}$$
$$\text{init\_Rem} \rightarrow 0$$
$$\text{init\_Sus} \rightarrow \text{N\_S}$$

## 13.2 Calibration using the "English Boarding School" data from 1978

The agreement of the model observables and the experimental data, yields a value of the objective function $\chi^2 = 97.5027$ for 14 data points in this data set. The trajectories of the dynamic variables are shown in Figure 13.

### 13.2.1 Estimated model parameters

In total 5 parameters are estimated from the experimental data. The best fit yields a value of the objective function $-2\log(L) = 125.136$ for a total of 15 data points. The model parameters were estimated by maximum likelihood estimation. In Table 12 the estimated parameter values are given. Parameters highlighted in red color indicate parameter values close to their bounds. The parameter name prefix init_ indicates the initial value of a dynamic variable.

## 13.3 Profile likelihood of model parameters

As a classical approach, identifiability of the model parameters was assessed using the profile likelihood [7]. An overview is displayed in Figure 14.

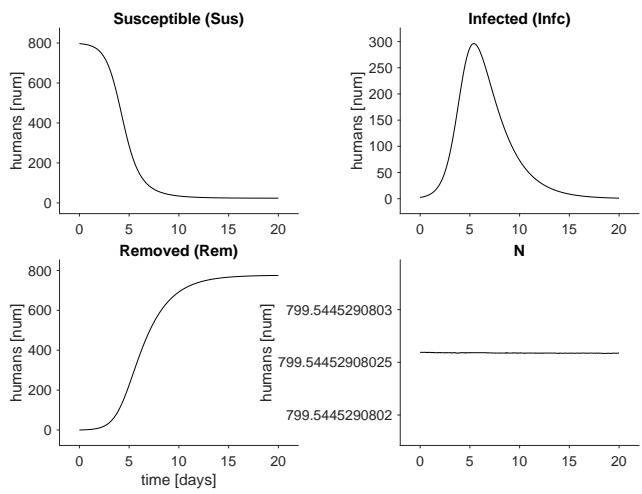

**Figure 13:** Trajectories of the dynamic variables. The dynamical behaviour is determined by numerically integrating the ODE system defined in Section 13.1.3.

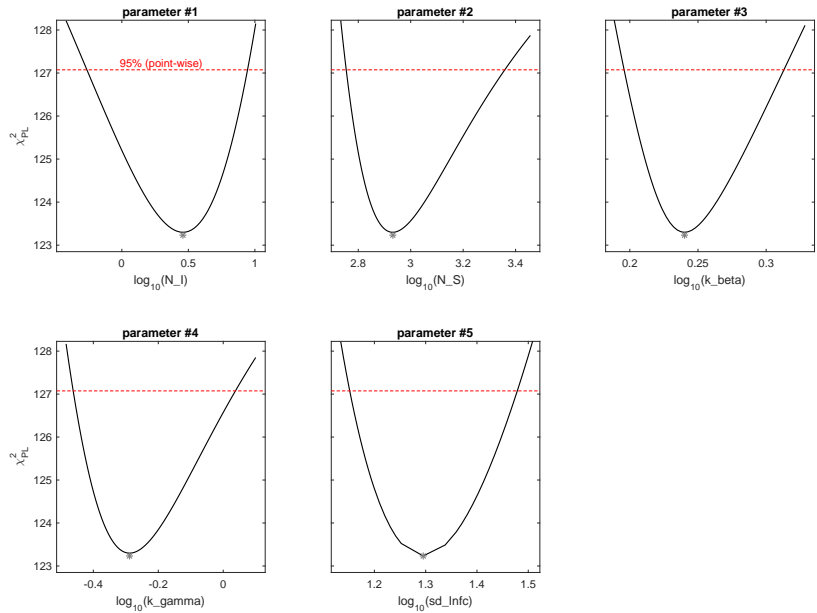

**Figure 14: Overview of the profile likelihood of the model parameters**
The solid lines indicate the profile likelihood. The broken lines indicate the threshold to assess confidence intervals. The asterisks indicate the optimal parameter values.

## 13.4 Identifiability-test

Applying the new approach for investigating identifiability yielded the following outcome.

```
>> arIdentifiabilityTest
Identifiability-test started ...
... Identifiability-test finished.


Identifiability-test was performed with radius = 1 and penalty-SD = 1.


All 5 optimization runs are in the chi2-range 6.92013e-09.

Calculations took 0.12 seconds.
[Compared to 21.04 seconds required for calcuating the likelihood profiles.]


1.0000 (increase of merit by penalty, before fitting)
0.3336 (decrease of merit by fitting)
0.1711 (movement of parameters by penalized fitting)
0.6664 (total increase of merit by penalty) PRIMARY CRITERION
Model is identifiable.
```

# 14 Model 10 ("Zika")

## 14.1 Model definition

This model has been applied in [12].

### 14.1.1 Description

"Full" Zika model (vector-borne infection SEIR-SEI model) with asymptomatcally infected humans, human-to-human infection and covalescent humans from Toensing et al, *Profile likelihood-based analyses of infectious disease models*.

### 14.1.2 Dynamic variables

The model contains 10 dynamic variables. The dynamics of those variables evolve according to a system of ordinary differential equations (ODE) as will be defined in the following. The following list indicates the unique variable names and their initial conditions.

- **Dynamic variable 1:** S_v
$$[S\_v](t = 0) = \text{init\_S\_v}$$

- **Dynamic variable 2:** E_v
$$[E\_v](t = 0) = \text{init\_E\_v}$$

- **Dynamic variable 3:** I_v
$$[I\_v](t = 0) = \text{init\_I\_v}$$

- **Dynamic variable 4:** S_h
$$[S\_h](t = 0) = \text{init\_S\_h}$$

- **Dynamic variable 5:** E_h
$$[E\_h](t = 0) = \text{init\_E\_h}$$

- **Dynamic variable 6:** I_hs
$$[I\_hs](t = 0) = \text{init\_I\_hs}$$

- **Dynamic variable 7:** I_ha
$$[I\_ha](t = 0) = \text{init\_I\_ha}$$

- **Dynamic variable 8:** I_hc
$$[I\_hc](t = 0) = \text{init\_I\_hc}$$

- **Dynamic variable 9:** R_h
$$[R\_h](t = 0) = \text{init\_R\_h}$$

- **Dynamic variable 10:** I_h_obs
$$[I\_h\_obs](t = 0) = \text{init\_I\_h\_obs}$$

### 14.1.3 Reactions

The model contains 14 "reactions". Reactions define interactions between dynamics variables and build up the ODE systems. The following list indicates the reaction laws and their corresponding reaction rate equations. Promoting rate modifiers are indicated in black above the rate law arrow. Inhibitory rate modifiers are indicated in red below the rate law arrow. In the reaction rate equations dynamic and input variables are indicated by square brackets. The remaining variables are model parameters that remain constant over time.

- **Reaction 1:**

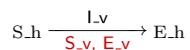

$$v_1 = \frac{[I\_v] \cdot [S\_h] \cdot \text{beta\_vh}}{[E\_v] + [I\_v] + [S\_v]}$$

- **Reaction 2:**

$$S\_h \xrightarrow[\textcolor{red}{R\_h}]{I\_hs,\ I\_ha,\ I\_hc} E\_h$$

$$v_2 = \frac{[S\_h] \cdot \text{beta\_hh} \cdot ([I\_ha] + [I\_hc] + [I\_hs])}{[E\_h] + [I\_ha] + [I\_hc] + [I\_hs] + [R\_h] + [S\_h]}$$

- **Reaction 3:**

$$E\_h \longrightarrow I\_ha$$

$$v_3 = [E\_h] \cdot \text{kappa\_as} \cdot \text{nu\_h}$$

- **Reaction 4:**

$$E\_h \longrightarrow I\_hs$$

$$v_4 = -[E\_h] \cdot \text{nu\_h} \cdot (\text{kappa\_as} - 1)$$

- **Reaction 5:**

$$\varnothing \xrightarrow{E\_h} I\_h\_obs$$

$$v_5 = -[E\_h] \cdot \text{nu\_h} \cdot (\text{kappa\_as} - 1)$$

- **Reaction 6:**

$$I\_hs \longrightarrow I\_hc$$

$$v_6 = [I\_hs] \cdot \text{gamma\_h1}$$

- **Reaction 7:**

$$I\_ha \longrightarrow I\_hc$$

$$v_7 = [I\_ha] \cdot \text{gamma\_h1}$$

- **Reaction 8:**

$$I\_hc \longrightarrow R\_h$$

$$v_8 = [I\_hc] \cdot \text{gamma\_h2}$$

- **Reaction 9:**

$$S\_v \xrightarrow[\textcolor{red}{S\_h,\ E\_h,\ I\_hc,\ R\_h}]{I\_hs,\ I\_ha} E\_v$$

$$v_9 = \frac{[S\_v] \cdot \text{beta\_hv} \cdot ([I\_ha] + [I\_hs])}{[E\_h] + [I\_ha] + [I\_hc] + [I\_hs] + [R\_h] + [S\_h]}$$

- **Reaction 10:**

$$E\_v \longrightarrow I\_v$$

$$v_{10} = [E\_v] \cdot nu\_v$$

- **Reaction 11:**

$$S\_v \longrightarrow \varnothing$$

$$v_{11} = [S\_v] \cdot mu\_v$$

- **Reaction 12:**

$$E\_v \longrightarrow \varnothing$$

$$v_{12} = [E\_v] \cdot mu\_v$$

- **Reaction 13:**

$$I\_v \longrightarrow \varnothing$$

$$v_{13} = [I\_v] \cdot mu\_v$$

- **Reaction 14:**

$$\varnothing \xrightarrow{E\_v,\ I\_v} S\_v$$

$$v_{14} = mu\_v \cdot ([E\_v] + [I\_v] + [S\_v])$$

### 14.1.4 ODE system

The specified reaction laws and rate equations $v$ determine an ODE system. The time evolution of the dynamical variables is calculated by solving this equation system.

$$\begin{aligned}
d[S\_v]/dt &= -v_9 - v_{11} + v_{14} \\
d[E\_v]/dt &= +v_9 - v_{10} - v_{12} \\
d[I\_v]/dt &= +v_{10} - v_{13} \\
d[S\_h]/dt &= -v_1 - v_2 \\
d[E\_h]/dt &= +v_1 + v_2 - v_3 - v_4 \\
d[I\_hs]/dt &= +v_4 - v_6 \\
d[I\_ha]/dt &= +v_3 - v_7 \\
d[I\_hc]/dt &= +v_6 + v_7 - v_8 \\
d[R\_h]/dt &= +v_8 \\
d[I\_h\_obs]/dt &= +v_5
\end{aligned}$$

Substituting the reaction rates $v_i$ yields:

$$d[S\_v]/dt = [N\_v] \cdot mu\_v - [S\_v] \cdot mu\_v - \frac{[S\_v] \cdot beta\_hv \cdot ([I\_ha]+[I\_hs])}{[N\_h]}$$

$$d[E\_v]/dt = \frac{[S\_v] \cdot beta\_hv \cdot ([I\_ha]+[I\_hs])}{[N\_h]} - [E\_v] \cdot nu\_v - [E\_v] \cdot mu\_v$$

$$d[I\_v]/dt = [E\_v] \cdot nu\_v - [I\_v] \cdot mu\_v$$

$$d[S\_h]/dt = -\frac{[I\_v] \cdot [S\_h] \cdot beta\_vh}{[N\_v]} - \frac{[S\_h] \cdot beta\_hh \cdot ([I\_ha]+[I\_hc]+[I\_hs])}{[N\_h]}$$

$$d[E\_h]/dt = [E\_h] \cdot nu\_h \cdot (kappa\_as - 1) - [E\_h] \cdot kappa\_as \cdot nu\_h + \frac{[I\_v] \cdot [S\_h] \cdot beta\_vh}{[N\_v]} + \frac{[S\_h] \cdot beta\_hh \cdot ([I\_ha]+[I\_hc]+[I\_hs])}{[N\_h]}$$

$$d[I\_hs]/dt = -[I\_hs] \cdot gamma\_h1 - [E\_h] \cdot nu\_h \cdot (kappa\_as - 1)$$

$$d[I\_ha]/dt = [E\_h] \cdot kappa\_as \cdot nu\_h - [I\_ha] \cdot gamma\_h1$$

$$d[I\_hc]/dt = [I\_ha] \cdot gamma\_h1 - [I\_hc] \cdot gamma\_h2 + [I\_hs] \cdot gamma\_h1$$

$$d[R\_h]/dt = [I\_hc] \cdot gamma\_h2$$

$$d[I\_h\_obs]/dt = -[E\_h] \cdot nu\_h \cdot (kappa\_as - 1)$$

The ODE system was solved by a parallelized implementation of the CVODES algorithm [4]. It also supplies the parameter sensitivities utilized for parameter estimation.

### 14.1.5 Derived variables

The model contains 2 derived variables. Derived variables are calculated after the ODE system was solved. Dynamic and input variables are indicated by square brackets. The remaining variables are model parameters that remain constant over time.

- **Derived variable 1:** N_v

$$[N\_v](t) = [E\_v] + [I\_v] + [S\_v]$$

- **Derived variable 2:** N_h

$$[N\_h](t) = [E\_h] + [I\_ha] + [I\_hc] + [I\_hs] + [R\_h] + [S\_h]$$

### 14.1.6 Observables

The model contains only a single observable. Observables are calculated after the ODE system was solved and derived variables are calculated. Dynamic, input and derived variables are indicated by square brackets. The remaining variables are model parameters that remain constant over time. In additon to the equation for the observable, also their corresponding error model $\sigma$ is indicated.

- **Observable 1:** Infected_cum

$$\text{Infected\_cum}(t) = [I\_h\_obs]$$

$$\sigma\{\text{Infected\_cum}\}(t) = \sqrt{\text{Infected\_cum}^2 \cdot sd\_rel^2 + sd\_abs^2}$$

### 14.1.7 Conditions

Conditions modify the model according to replacement rules. New model parameters can be introduced or relations between existing model parameters can be implemented. The following list are default conditions that can be replace my experiment specific conditions defined seperately for each data set.

$$\text{init\_I\_h\_obs} \rightarrow -\text{init\_I\_h} \cdot (kappa\_as - 1)$$
$$\text{init\_I\_ha} \rightarrow \text{init\_I\_h} \cdot kappa\_as$$
$$\text{init\_I\_hc} \rightarrow 0$$
$$\text{init\_I\_hs} \rightarrow -\text{init\_I\_h} \cdot (kappa\_as - 1)$$
$$\text{init\_R\_h} \rightarrow 0$$

## 14.2 Experiment: Zika_Colombia

### 14.2.1 Data and model calibration

The data set from Instituto Nacional de Salud Colombia describe the Zika infections 2015/2016 in Colombia [5] was extraced in [12] via WebPlot-Digitizer.

The agreement of the model observables and the experimental data yields a value of the objective function $\chi^2 = 840.295$ for 57 data points in this data set. The trajectories of the input and the dynamic variables of the fitted model are shown in Figure 15.

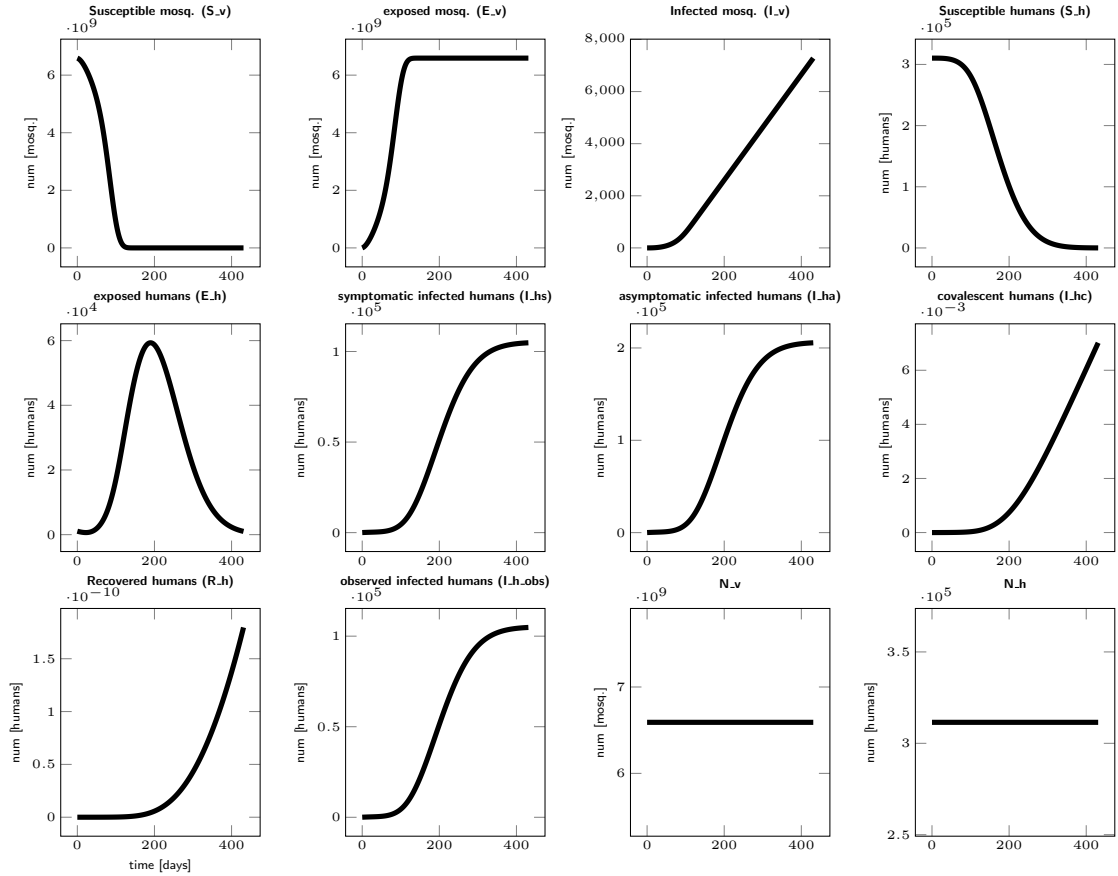

**Figure 15: Zika_Colombia trajectories of the input and dynamic variables.** The dynamical behaviour is determined by numerically integrating the ODE system defined in Section 14.1.4.

| | name | $\theta_{min}$ | $\hat{\theta}$ | $\theta_{max}$ | log | non-log $\hat{\theta}$ |
|---|---|---|---|---|---|---|
| 1 | beta_hh | -1e+01 | -9.5631 | +5 | 1 | $+2.73 \cdot 10^{-10}$ |
| 2 | beta_hv | -1e+01 | +0.3010 | +5 | 1 | $+2.00 \cdot 10^{+00}$ |
| 3 | beta_vh | -1e+01 | +4.3945 | +5 | 1 | $+2.48 \cdot 10^{+04}$ |
| 4 | gamma_h1 | -1e+01 | -9.8316 | +5 | 1 | $+1.47 \cdot 10^{-10}$ |
| 5 | gamma_h2 | -1e+01 | -9.3486 | +5 | 1 | $+4.48 \cdot 10^{-10}$ |
| 6 | init_E_h | -5 | +3.3707 | +1e+01 | 1 | $+2.35 \cdot 10^{+03}$ |
| 7 | init_E_v | -5 | -2.1202 | +1e+01 | 1 | $+7.58 \cdot 10^{-03}$ |
| 8 | init_I_h | -5 | +2.3918 | +1e+01 | 1 | $+2.46 \cdot 10^{+02}$ |
| 9 | init_I_v | -5 | -5.0000 | +1e+01 | 1 | $+1.00 \cdot 10^{-05}$ |
| 10 | init_S_h | -5 | +5.8345 | +1e+01 | 1 | $+6.83 \cdot 10^{+05}$ |
| 11 | init_S_v | -5 | +10.0000 | +1e+01 | 1 | $+1.00 \cdot 10^{+10}$ |
| 12 | kappa_as | +0 | +0.8466 | +1 | 0 | $+8.47 \cdot 10^{-01}$ |
| 13 | mu_v | -1e+01 | -9.7294 | +5 | 1 | $+1.86 \cdot 10^{-10}$ |
| 14 | nu_h | -1e+01 | -1.5094 | +5 | 1 | $+3.09 \cdot 10^{-02}$ |
| 15 | nu_v | -1e+01 | -8.2788 | +5 | 1 | $+5.26 \cdot 10^{-09}$ |
| 16 | sd_abs | -5 | -4.3789 | +3 | 1 | $+4.18 \cdot 10^{-05}$ |
| 17 | sd_rel | -5 | -1.2138 | -0.3 | 1 | $+6.11 \cdot 10^{-02}$ |

**Table 13: Estimated parameter values**
$\hat{\theta}$ indicates the estimated value of the parameters. $\theta_{min}$ and $\theta_{max}$ indicate the upper and lower bounds for the parameters. The log-column indicates if the value of a parameter was log-transformed. If log ≡ 1 the non-log-column indicates the non-logarithmic value of the estimate.

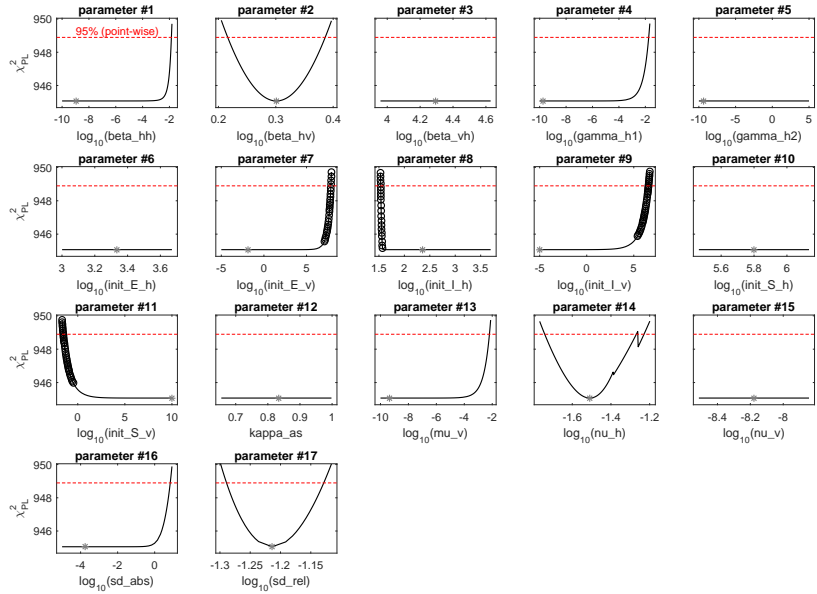

**Figure 16: Overview of the profile likelihood of the model parameters**
The solid lines indicate the profile likelihood. The broken lines indicate the threshold to assess confidence intervals. The asterisks indicate the optimal parameter values.

## 14.3 Estimated model parameters

In total 17 parameters are estimated from the experimental data. The best fit yields a value of the objective function $-2\log(L) = 946.892$ for a total of 58 data points. The model parameters were estimated by maximum likelihood estimation. In Table 13 the estimated parameter values are given. Parameters highlighted in red color indicate parameter values close to their bounds. The parameter name prefix init_ indicates the initial value of a dynamic variable.

## 14.4 Profile likelihood of model parameters

As a classical approach, identifiability of the model parameters was assessed using the profile likelihood [7]. An overview is displayed in Figure 16.

## 14.5 Identifiability-test

Applying the new approach for investigating identifiability yielded the following outcome.

```
>> arIdentifiabilityTest
Identifiability-test started ...
... Identifiability-test finished.


Identifiability-test was performed with radius = 1 and penalty-SD = 1.


All 5 optimization runs are in the chi2-range 8.7179e-06.


Calculations took 0.80 seconds.
[Compared to 846.11 seconds required for calcuating the likelihood profiles.]


1.0000 (increase of merit by penalty, before fitting)
1.0000 (decrease of merit by fitting)
1.0000 (movement of parameters by penalized fitting)
-0.0000 (total increase of merit by penalty) PRIMARY CRITERION
Model is structurally non-identifiable.
```

# 15 Model 11 ("Schwen")

## 15.1 Model: Kreutz_IR_binding

This model has been published in [10].

### 15.1.1 Description

This model describes insulin binding to the receptor in two distinct cell entities of murine hepatocytes.

### 15.1.2 Dynamic variables

The model contains 11 dynamic variables. The dynamics of those variables evolve according to a system of ordinary differential equations (ODE) as will be defined in the following. The following list indicates the unique variable names and their initial conditions.

- **Dynamic variable 1:** Ins

    $[\text{Ins}](t = 0) = \text{init\_Ins}$

- **Dynamic variable 2:** Rec1

    $[\text{Rec1}](t = 0) = \text{init\_Rec1}$

- **Dynamic variable 3:** Rec2

    $[\text{Rec2}](t = 0) = \text{init\_Rec2}$

- **Dynamic variable 4:** IR1

    $[\text{IR1}](t = 0) = \text{init\_IR1}$

- **Dynamic variable 5:** IR2

    $[\text{IR2}](t = 0) = \text{init\_IR2}$

- **Dynamic variable 6:** IR1in

    $[\text{IR1in}](t = 0) = \text{init\_IR1in}$

- **Dynamic variable 7:** IR2in

    $[\text{IR2in}](t = 0) = \text{init\_IR2in}$

- **Dynamic variable 8:** Uptake1

    $[\text{Uptake1}](t = 0) = \text{init\_Uptake1}$

- **Dynamic variable 9:** Uptake2

    $[\text{Uptake2}](t = 0) = \text{init\_Uptake2}$

- **Dynamic variable 10:** InsulinFragments

    $[\text{InsulinFragments}](t = 0) = \text{init\_InsulinFragments}$

- **Dynamic variable 11:** BoundUnspec

    $[\text{BoundUnspec}](t = 0) = \text{init\_BoundUnspec}$

### 15.1.3 Reactions

The model contains 14 reactions. Reactions define interactions between dynamics variables and build up the ODE systems. The following list indicates the reaction laws and their corresponding reaction rate equations. Promoting rate modifiers are indicated in black above the rate law arrow. Inhibitory rate modifiers are indicated in red below the rate law arrow. In the reaction rate equations dynamic and input variables are indicated by square brackets. The remaining variables are model parameters that remain constant over time.

- **Reaction 1:**

  $$\text{Ins} + \text{Rec1} \longrightarrow \text{IR1}$$

  $$v_1 = [\text{Ins}] \cdot [\text{Rec1}] \cdot \text{ka1}$$

- **Reaction 2:**

  $$\text{Ins} + \text{Rec2} \longrightarrow \text{IR2}$$

  $$v_2 = [\text{Ins}] \cdot [\text{Rec2}] \cdot \text{ka1} \cdot \text{ka2fold}$$

- **Reaction 3:**

  $$\text{Ins} \longrightarrow \text{BoundUnspec}$$

  $$v_3 = [\text{Ins}] \cdot \text{kon\_unspec}$$

- **Reaction 4:**

  $$\text{BoundUnspec} \longrightarrow \text{Ins}$$

  $$v_4 = [\text{BoundUnspec}] \cdot \text{koff\_unspec}$$

- **Reaction 5:**

  $$\text{IR1} \longrightarrow \text{Ins} + \text{Rec1}$$

  $$v_5 = [\text{IR1}] \cdot \text{kd1}$$

- **Reaction 6:**

  $$\text{IR2} \longrightarrow \text{Ins} + \text{Rec2}$$

  $$v_6 = [\text{IR2}] \cdot \text{kd1} \cdot \text{kd2fold}$$

- **Reaction 7:**

  $$\text{IR1} \longrightarrow \text{IR1in}$$

  $$v_7 = [\text{IR1}] \cdot \text{kin}$$

- **Reaction 8:**

  $$\text{IR2} \longrightarrow \text{IR2in}$$

  $$v_8 = [\text{IR2}] \cdot \text{kin2}$$

- **Reaction 9:**

  $$\text{IR1in} \longrightarrow \text{IR1}$$

  $$v_9 = [\text{IR1in}] \cdot \text{kout}$$

- **Reaction 10:**

  $$\text{IR2in} \longrightarrow \text{IR2}$$

  $$v_{10} = [\text{IR2in}] \cdot \text{kout2}$$

- **Reaction 11:**

  $$\text{IR1in} \longrightarrow \text{Rec1} + \text{InsulinFragments}$$

  $$v_{11} = [\text{IR1in}] \cdot \text{kout\_frag}$$

- **Reaction 12:**

  $$\text{IR2in} \longrightarrow \text{Rec2} + \text{InsulinFragments}$$

  $$v_{12} = [\text{IR2in}] \cdot \text{kout\_frag}$$

- **Reaction 13:**

  $$\varnothing \xrightarrow[\text{IR1}]{\text{Ins, Rec1}} \text{Uptake1}$$

  $$v_{13} = [\text{Ins}] \cdot [\text{Rec1}] \cdot \text{ka1} - [\text{IR1}] \cdot \text{kd1}$$

- **Reaction 14:**

  $$\varnothing \xrightarrow[\text{IR2}]{\text{Ins, Rec2}} \text{Uptake2}$$

  $$v_{14} = [\text{Ins}] \cdot [\text{Rec2}] \cdot \text{ka1} \cdot \text{ka2fold} - [\text{IR2}] \cdot \text{kd1} \cdot \text{kd2fold}$$

### 15.1.4 ODE system

The specified reation laws and rate equations $v$ determine an ODE system. The time evolution of the dynamical variables is calculated by solving this equation system.

$$\begin{aligned}
d[\text{Ins}]/dt &= -v_1 - v_2 - v_3 + v_4 + v_5 + v_6 \\
d[\text{Rec1}]/dt &= -v_1 + v_5 + v_{11} \\
d[\text{Rec2}]/dt &= -v_2 + v_6 + v_{12} \\
d[\text{IR1}]/dt &= +v_1 - v_5 - v_7 + v_9 \\
d[\text{IR2}]/dt &= +v_2 - v_6 - v_8 + v_{10} \\
d[\text{IR1in}]/dt &= +v_7 - v_9 - v_{11} \\
d[\text{IR2in}]/dt &= +v_8 - v_{10} - v_{12} \\
d[\text{Uptake1}]/dt &= +v_{13} \\
d[\text{Uptake2}]/dt &= +v_{14} \\
d[\text{InsulinFragments}]/dt &= +v_{11} + v_{12} \\
d[\text{BoundUnspec}]/dt &= +v_3 - v_4
\end{aligned}$$

Substituting the reaction rates $v_i$ yields:

d[Ins]/dt = [IR1] · kd1 + [BoundUnspec] · koff_unspec − [Ins] · kon_unspec − [Ins] · [Rec1] · ka1 + [IR2] · kd1 · kd2fold − [Ins] · [Rec2] · ka1 · ka2fold

d[Rec1]/dt = [IR1] · kd1 + [IR1in] · kout_frag − [Ins] · [Rec1] · ka1

d[Rec2]/dt = [IR2in] · kout_frag + [IR2] · kd1 · kd2fold − [Ins] · [Rec2] · ka1 · ka2fold

d[IR1]/dt = [IR1in] · kout − [IR1] · kin − [IR1] · kd1 + [Ins] · [Rec1] · ka1

d[IR2]/dt = [IR2in] · kout2 − [IR2] · kin2 − [IR2] · kd1 · kd2fold + [Ins] · [Rec2] · ka1 · ka2fold

d[IR1in]/dt = [IR1] · kin − [IR1in] · kout − [IR1in] · kout_frag

d[IR2in]/dt = [IR2] · kin2 − [IR2in] · kout2 − [IR2in] · kout_frag

d[Uptake1]/dt = [Ins] · [Rec1] · ka1 − [IR1] · kd1

d[Uptake2]/dt = [Ins] · [Rec2] · ka1 · ka2fold − [IR2] · kd1 · kd2fold

d[InsulinFragments]/dt = [IR1in] · kout_frag + [IR2in] · kout_frag

d[BoundUnspec]/dt = [Ins] · kon_unspec − [BoundUnspec] · koff_unspec

The ODE system was solved by a parallelized implementation of the CVODES algorithm [4]. It also supplies the parameter sensitivities utilized for parameter estimation.

### 15.1.5 Conditions

Conditions modify the model according to replacement rules. New model parameters can be introduced or relations between existing model parameters can be implemented. The following list are default conditions that can be replace my experiment specific conditions defined seperately for each data set.

$$\begin{aligned}
\text{init\_BoundUnspec} &\to 0 \\
\text{init\_IR1} &\to 0 \\
\text{init\_IR1in} &\to 0 \\
\text{init\_IR2} &\to 0 \\
\text{init\_IR2in} &\to 0 \\
\text{init\_InsulinFragments} &\to 0 \\
\text{init\_Rec1} &\to \text{ini\_R1} \\
\text{init\_Rec2} &\to \text{ini\_R1} \cdot \text{ini\_R2fold} \\
\text{init\_Uptake1} &\to 0 \\
\text{init\_Uptake2} &\to 0
\end{aligned}$$

## 15.2 Experiment: FacsData_unlog10

### 15.2.1 Experiment specific conditions

To evaluate the model for this experiment the following conditions are applied.

- **Local condition #1 (global condition #1):**

    init_Ins → 0

- **Local condition #2 (global condition #2):**

    init_Ins → 0.1

- **Local condition #3 (global condition #3):**

    init_Ins → 1

- **Local condition #4 (global condition #4):**

    init_Ins → 10

- **Local condition #5 (global condition #5):**

    init_Ins → 100

- **Local condition #6 (global condition #6):**

    init_Ins → 1000

- **Local condition #7 (global condition #7):**

    init_Ins → 10000

### 15.2.2 Experimental data and model fit

The agreement of the model observables and the experimental data, given in Table 14, yields a value of the objective function $\log Lik = -170.172$ for 34 data points in this data set.

## 15.3 Experiment: Elisa_relative_nExpID1

### 15.3.1 Experiment specific conditions

To evaluate the model for this experiment the following conditions are applied.

- **Local condition #8 (global condition #1):**

    init_Ins → 0
    km_nExpID → km_nExpID1
    offset_nExpID → offset_nExpID1
    scaleElisa_nExpID → scaleElisa_nExpID1

| time [n/a] | init_Ins | IR1_obs conc. [n/a] | IR2_obs conc. [n/a] | IRsum_obs conc. [n/a] |
|---|---|---|---|---|
| 0 | 0 | NaN | NaN | 2.35234 |
| 15 | 0.1 | NaN | NaN | 1.73224 |
| 15 | 1 | NaN | NaN | 1.76693 |
| 15 | 10 | NaN | NaN | 1.78202 |
| 1 | 100 | 2.00666 | 4.24486 | NaN |
| 2 | 100 | 1.86764 | 4.07748 | NaN |
| 5 | 100 | 2.34948 | 6.31757 | NaN |
| 15 | 100 | 2.13306 | 6.64872 | NaN |
| 30 | 100 | 2.25294 | 9.25131 | NaN |
| 1 | 1000 | 3.18628 | 13.5705 | NaN |
| 2 | 1000 | 3.26316 | 19.5628 | NaN |
| 5 | 1000 | 3.83517 | 26.415 | NaN |
| 15 | 1000 | 4.31454 | 41.5385 | NaN |
| 30 | 1000 | 4.79015 | 47.8376 | NaN |
| 1 | 10000 | 6.19759 | 61.2759 | NaN |
| 2 | 10000 | 7.05776 | 75.8475 | NaN |
| 5 | 10000 | 8.5539 | 107.576 | NaN |
| 15 | 10000 | 8.82925 | 118.061 | NaN |
| 30 | 10000 | 9.54412 | 136.186 | NaN |

Table 14: Experimental data for the experiment FacsData_unlog10

- **Local condition #9 (global condition #4):**

$$\text{init\_Ins} \to 10$$
$$\text{km\_nExpID} \to \text{km\_nExpID1}$$
$$\text{offset\_nExpID} \to \text{offset\_nExpID1}$$
$$\text{scaleElisa\_nExpID} \to \text{scaleElisa\_nExpID1}$$

- **Local condition #10 (global condition #7):**

$$\text{init\_Ins} \to 10000$$
$$\text{km\_nExpID} \to \text{km\_nExpID1}$$
$$\text{offset\_nExpID} \to \text{offset\_nExpID1}$$
$$\text{scaleElisa\_nExpID} \to \text{scaleElisa\_nExpID1}$$

### 15.3.2 Experimental data and model fit

The agreement of the model observables and the experimental data, given in Table 15, yields a value of the objective function $\log Lik = -86.037$ for 66 data points in this data set.

## 15.4 Experiment: Elisa_relative_nExpID2

### 15.4.1 Experiment specific conditions

To evaluate the model for this experiment the following conditions are applied.

- **Local condition #11 (global condition #1):**

$$\text{init\_Ins} \to 0$$
$$\text{km\_nExpID} \to \text{km\_nExpID2}$$
$$\text{offset\_nExpID} \to \text{offset\_nExpID2}$$
$$\text{scaleElisa\_nExpID} \to \text{scaleElisa\_nExpID2}$$

| time [min] | init_Ins | Insulin_obs conc. [nM] |
|---|---|---|
| 15 | 0 | 0.026231 |
| 15 | 0 | 0.032385 |
| 0.25 | 10 | 0.490769 |
| 0.25 | 10 | 0.479231 |
| 0.5 | 10 | 0.569615 |
| 0.5 | 10 | 0.535 |
| 1 | 10 | 0.731154 |
| 1 | 10 | 0.729231 |
| 2 | 10 | 0.623462 |
| 2 | 10 | 0.590769 |
| 3 | 10 | 0.963846 |
| 3 | 10 | 0.965769 |
| 4 | 10 | 1.09654 |
| 4 | 10 | 0.994615 |
| 5 | 10 | 1.16577 |
| 5 | 10 | 1.10423 |
| 7 | 10 | 1.06769 |
| 7 | 10 | 1.04462 |
| 10 | 10 | 0.615769 |
| 10 | 10 | 0.588846 |
| 15 | 10 | 0.944615 |
| 15 | 10 | 0.919615 |
| 20 | 10 | 1.17538 |
| 20 | 10 | 1.10423 |
| 25 | 10 | 1.21385 |
| 25 | 10 | 1.13885 |
| 30 | 10 | 1.29654 |
| 30 | 10 | 1.26577 |
| 40 | 10 | 1.36192 |
| 40 | 10 | 1.27346 |
| 50 | 10 | 0.486923 |
| 50 | 10 | 0.467692 |
| 60 | 10 | 0.390769 |
| 60 | 10 | 0.371538 |
| 0.25 | 10000 | 2850.77 |
| 0.25 | 10000 | 2762.31 |
| 0.5 | 10000 | 2250.77 |
| 0.5 | 10000 | 2031.54 |
| 1 | 10000 | 2796.92 |
| 1 | 10000 | 2720 |
| 2 | 10000 | 3439.23 |
| 2 | 10000 | 3223.85 |
| 3 | 10000 | 3343.08 |
| 3 | 10000 | 3189.23 |
| 4 | 10000 | 3416.15 |
| 4 | 10000 | 3281.54 |
| 5 | 10000 | 2927.69 |
| 5 | 10000 | 2920 |
| 7 | 10000 | 3154.62 |
| 7 | 10000 | 2935.38 |
| 10 | 10000 | 2462.31 |
| 10 | 10000 | 2350.77 |
| 15 | 10000 | 3550.77 |
| 15 | 10000 | 3081.54 |
| 20 | 10000 | 3212.31 |
| 20 | 10000 | 3162.31 |
| 25 | 10000 | 3196.92 |
| 25 | 10000 | 3246.92 |
| 30 | 10000 | 3146.92 |
| 30 | 10000 | 3146.92 |
| 40 | 10000 | 3996.92 |
| 40 | 10000 | 3646.92 |
| 50 | 10000 | 3116.15 |
| 50 | 10000 | 3043.08 |
| 60 | 10000 | 2631.54 |
| 60 | 10000 | 2543.08 |

Table 15: Experimental data for the experiment Elisa_relative_nExpID1

- **Local condition #12 (global condition #4):**

$$\text{init\_Ins} \to 10$$
$$\text{km\_nExpID} \to \text{km\_nExpID2}$$
$$\text{offset\_nExpID} \to \text{offset\_nExpID2}$$
$$\text{scaleElisa\_nExpID} \to \text{scaleElisa\_nExpID2}$$

- **Local condition #13 (global condition #5):**

$$\text{init\_Ins} \to 100$$
$$\text{km\_nExpID} \to \text{km\_nExpID2}$$
$$\text{offset\_nExpID} \to \text{offset\_nExpID2}$$
$$\text{scaleElisa\_nExpID} \to \text{scaleElisa\_nExpID2}$$

### 15.4.2 Experimental data and model fit

The agreement of the model observables and the experimental data, given in Table 16, yields a value of the objective function $\log Lik = -160.284$ for 66 data points in this data set.

## 15.5 Experiment: Elisa_relative_nExpID3

### 15.5.1 Experiment specific conditions

To evaluate the model for this experiment the following conditions are applied.

- **Local condition #14 (global condition #1):**

$$\text{init\_Ins} \to 0$$
$$\text{km\_nExpID} \to \text{km\_nExpID3}$$
$$\text{offset\_nExpID} \to \text{offset\_nExpID3}$$
$$\text{scaleElisa\_nExpID} \to \text{scaleElisa\_nExpID3}$$

- **Local condition #15 (global condition #4):**

$$\text{init\_Ins} \to 10$$
$$\text{km\_nExpID} \to \text{km\_nExpID3}$$
$$\text{offset\_nExpID} \to \text{offset\_nExpID3}$$
$$\text{scaleElisa\_nExpID} \to \text{scaleElisa\_nExpID3}$$

- **Local condition #16 (global condition #5):**

$$\text{init\_Ins} \to 100$$
$$\text{km\_nExpID} \to \text{km\_nExpID3}$$
$$\text{offset\_nExpID} \to \text{offset\_nExpID3}$$
$$\text{scaleElisa\_nExpID} \to \text{scaleElisa\_nExpID3}$$

### 15.5.2 Experimental data and model fit

The agreement of the model observables and the experimental data, given in Table 17, yields a value of the objective function $\log Lik = -18.1078$ for 54 data points in this data set.

| time [min] | init_Ins | Insulin_obs conc. [nM] |
|---|---|---|
| 15 | 0 | 0.021957 |
| 15 | 0 | 0.022826 |
| 0.25 | 10 | 1.10543 |
| 0.25 | 10 | 1.04239 |
| 0.5 | 10 | 1.17717 |
| 0.5 | 10 | 1.1337 |
| 1 | 10 | 1.34239 |
| 1 | 10 | 1.27283 |
| 2 | 10 | 1.34022 |
| 2 | 10 | 1.29457 |
| 3 | 10 | 1.00544 |
| 3 | 10 | 1.01196 |
| 4 | 10 | 1.05326 |
| 4 | 10 | 1.04457 |
| 5 | 10 | 1.20978 |
| 5 | 10 | 1.11848 |
| 7 | 10 | 1.44457 |
| 7 | 10 | 1.37935 |
| 10 | 10 | 1.33587 |
| 10 | 10 | 1.29022 |
| 15 | 10 | 1.07283 |
| 15 | 10 | 1.07935 |
| 20 | 10 | 1.24022 |
| 20 | 10 | 1.21413 |
| 25 | 10 | 1.68152 |
| 25 | 10 | 1.71848 |
| 30 | 10 | 1.64022 |
| 30 | 10 | 1.68152 |
| 40 | 10 | 1.625 |
| 40 | 10 | 1.52717 |
| 50 | 10 | 1.83804 |
| 50 | 10 | 1.49456 |
| 60 | 10 | 1.42065 |
| 60 | 10 | 1.39674 |
| 0.25 | 100 | 18.8424 |
| 0.25 | 100 | 18.1359 |
| 0.5 | 100 | 21.5163 |
| 0.5 | 100 | 22.125 |
| 1 | 100 | 17.3424 |
| 1 | 100 | 16.6793 |
| 2 | 100 | 14.6576 |
| 2 | 100 | 13.962 |
| 3 | 100 | 20.0489 |
| 3 | 100 | 19.9402 |
| 4 | 100 | 9.20109 |
| 4 | 100 | 8.2337 |
| 5 | 100 | 22.6359 |
| 5 | 100 | 22.4946 |
| 7 | 100 | 13.4076 |
| 7 | 100 | 12.7228 |
| 10 | 100 | 14.9946 |
| 10 | 100 | 14.7446 |
| 15 | 100 | 15.1685 |
| 15 | 100 | 14.8315 |
| 20 | 100 | 20.7337 |
| 20 | 100 | 20.038 |
| 25 | 100 | 21.2337 |
| 25 | 100 | 20.0707 |
| 30 | 100 | 21.3315 |
| 30 | 100 | 20.8098 |
| 40 | 100 | 25.1359 |
| 40 | 100 | 25.9511 |
| 50 | 100 | 23.288 |
| 50 | 100 | 23.9293 |
| 60 | 100 | 15.4402 |
| 60 | 100 | 15.0489 |

Table 16: Experimental data for the experiment Elisa_relative_nExpID2

| time [min] | init_Ins | Insulin_obs conc. [nM] |
|---|---|---|
| 15 | 0 | 0.008336 |
| 15 | 0 | 0.011193 |
| 0.25 | 10 | 0.070243 |
| 0.25 | 10 | 0.0731 |
| 0.5 | 10 | 0.062386 |
| 0.5 | 10 | 0.063814 |
| 2 | 10 | 0.054529 |
| 2 | 10 | 0.057386 |
| 3 | 10 | 0.102386 |
| 3 | 10 | 0.1181 |
| 4 | 10 | 0.052386 |
| 4 | 10 | 0.050243 |
| 5 | 10 | 0.067386 |
| 5 | 10 | 0.075957 |
| 7 | 10 | 0.1081 |
| 7 | 10 | 0.1181 |
| 10 | 10 | 0.130957 |
| 10 | 10 | 0.136671 |
| 20 | 10 | 0.265243 |
| 20 | 10 | 0.222386 |
| 25 | 10 | 0.243814 |
| 25 | 10 | 0.220957 |
| 40 | 10 | 0.1131 |
| 40 | 10 | 0.1281 |
| 50 | 10 | 0.2981 |
| 50 | 10 | 0.265243 |
| 60 | 10 | 0.318814 |
| 60 | 10 | 0.332386 |
| 0.25 | 100 | 1.06185 |
| 0.25 | 100 | 1.09148 |
| 0.5 | 100 | 1.28037 |
| 0.5 | 100 | 1.16556 |
| 2 | 100 | 2.56185 |
| 2 | 100 | 2.51 |
| 3 | 100 | 4.32111 |
| 3 | 100 | 4.15074 |
| 4 | 100 | 1.36556 |
| 4 | 100 | 1.17296 |
| 5 | 100 | 4.99889 |
| 5 | 100 | 4.32482 |
| 7 | 100 | 3.60259 |
| 7 | 100 | 3.43222 |
| 10 | 100 | 5.02111 |
| 10 | 100 | 5.14333 |
| 20 | 100 | 4.35074 |
| 20 | 100 | 4.26926 |
| 25 | 100 | 7.93593 |
| 25 | 100 | 7.72852 |
| 40 | 100 | 5.89148 |
| 40 | 100 | 5.31741 |
| 50 | 100 | 4.73222 |
| 50 | 100 | 4.6137 |
| 60 | 100 | 6.13963 |
| 60 | 100 | 5.8063 |

Table 17: Experimental data for the experiment Elisa_relative_nExpID3

## 15.6 Experiment: Elisa_relative_nExpID4

### 15.6.1 Experiment specific conditions

To evaluate the model for this experiment the following conditions are applied.

- **Local condition #17 (global condition #1):**

$$\text{init\_Ins} \rightarrow 0$$
$$\text{km\_nExpID} \rightarrow \text{km\_nExpID4}$$
$$\text{offset\_nExpID} \rightarrow \text{offset\_nExpID4}$$
$$\text{scaleElisa\_nExpID} \rightarrow \text{scaleElisa\_nExpID4}$$

- **Local condition #18 (global condition #5):**

$$\text{init\_Ins} \rightarrow 100$$
$$\text{km\_nExpID} \rightarrow \text{km\_nExpID4}$$
$$\text{offset\_nExpID} \rightarrow \text{offset\_nExpID4}$$
$$\text{scaleElisa\_nExpID} \rightarrow \text{scaleElisa\_nExpID4}$$

- **Local condition #19 (global condition #7):**

$$\text{init\_Ins} \rightarrow 10000$$
$$\text{km\_nExpID} \rightarrow \text{km\_nExpID4}$$
$$\text{offset\_nExpID} \rightarrow \text{offset\_nExpID4}$$
$$\text{scaleElisa\_nExpID} \rightarrow \text{scaleElisa\_nExpID4}$$

### 15.6.2 Experimental data and model fit

The agreement of the model observables and the experimental data, given in Table 18, yields a value of the objective function $\log Lik = -159.667$ for 66 data points in this data set.

## 15.7 Estimated model parameters

In total 30 parameters are estimated from the experimental data. The best fit yields a value of the objective function $-2\log(L) = -54.9408$ for a total of 293 data points. The model parameters were estimated by maximum likelihood estimation. In Table 19 the estimated parameter values are given. Parameters highlighted in red color indicate parameter values close to their bounds. The parameter name prefix init_ indicates the initial value of a dynamic variable.

## 15.8 Profile likelihood of model parameters

As a classical approach, identifiability of the model parameters was assessed using the profile likelihood [7]. An overview is displayed in Figure 17.

## 15.9 Identifiability-test

Applying the new approach for investigating identifiability yielded the following outcome.

| time [min] | init_Ins | Insulin_obs conc. [nM] |
|---|---|---|
| 15 | 0 | 1.45517 |
| 15 | 0 | 1.14483 |
| 0.25 | 100 | 39.0069 |
| 0.25 | 100 | 37.2483 |
| 0.5 | 100 | 35.6276 |
| 0.5 | 100 | 34.3517 |
| 1 | 100 | 37.7655 |
| 1 | 100 | 36.3517 |
| 2 | 100 | 40.2828 |
| 2 | 100 | 39.1448 |
| 3 | 100 | 29.5586 |
| 3 | 100 | 27.9034 |
| 4 | 100 | 32.0069 |
| 4 | 100 | 31.1793 |
| 5 | 100 | 30.2828 |
| 5 | 100 | 29.6621 |
| 7 | 100 | 34.4207 |
| 7 | 100 | 33.731 |
| 10 | 100 | 39.1103 |
| 10 | 100 | 36.731 |
| 15 | 100 | 38.3172 |
| 15 | 100 | 36.4552 |
| 20 | 100 | 32.9724 |
| 20 | 100 | 31.8 |
| 25 | 100 | 33.7655 |
| 25 | 100 | 32.9724 |
| 30 | 100 | 34.1103 |
| 30 | 100 | 32.6276 |
| 40 | 100 | 34.1793 |
| 40 | 100 | 32.5586 |
| 50 | 100 | 39.1793 |
| 50 | 100 | 35.6966 |
| 60 | 100 | 30.7655 |
| 60 | 100 | 29.2483 |
| 0.25 | 10000 | 4486.9 |
| 0.25 | 10000 | 4366.21 |
| 0.5 | 10000 | 3297.24 |
| 0.5 | 10000 | 3128.28 |
| 1 | 10000 | 4076.55 |
| 1 | 10000 | 3524.83 |
| 2 | 10000 | 5400.69 |
| 2 | 10000 | 5362.76 |
| 3 | 10000 | 4135.17 |
| 3 | 10000 | 3942.07 |
| 4 | 10000 | 4438.62 |
| 4 | 10000 | 4317.93 |
| 5 | 10000 | 3673.1 |
| 5 | 10000 | 3566.21 |
| 7 | 10000 | 4817.93 |
| 7 | 10000 | 4707.59 |
| 10 | 10000 | 4169.66 |
| 10 | 10000 | 3893.79 |
| 15 | 10000 | 5124.83 |
| 15 | 10000 | 3793.79 |
| 20 | 10000 | 4238.62 |
| 20 | 10000 | 3986.9 |
| 25 | 10000 | 4197.24 |
| 25 | 10000 | 3945.52 |
| 30 | 10000 | 3880 |
| 30 | 10000 | 3686.9 |
| 40 | 10000 | 4531.72 |
| 40 | 10000 | 4555.86 |
| 50 | 10000 | 5224.83 |
| 50 | 10000 | 4786.9 |
| 60 | 10000 | 5342.07 |
| 60 | 10000 | 5176.55 |

Table 18: Experimental data for the experiment Elisa_relative_nExpID4

|   | name | $\theta_{min}$ | $\hat{\theta}$ | $\theta_{max}$ | log | non-log $\hat{\theta}$ |
|---|------|------|------|------|-----|----------|
| 1 | IR_obs_std | -5 | -1.3041 | -1 | 1 | $+4.96 \cdot 10^{-02}$ |
| 2 | fragments | +0 | +1.0000 | +1 | 0 | $+1.00 \cdot 10^{+00}$ |
| 3 | ini_R1 | -5 | +1.7693 | +3 | 1 | $+5.88 \cdot 10^{+01}$ |
| 4 | ini_R2fold | -5 | +1.1927 | +3 | 1 | $+1.56 \cdot 10^{+01}$ |
| 5 | ka1 | -5 | -2.1182 | +3 | 1 | $+7.62 \cdot 10^{-03}$ |
| 6 | ka2fold | -5 | +0.2061 | +3 | 1 | $+1.61 \cdot 10^{+00}$ |
| 7 | kd1 | -5 | +1.2804 | +3 | 1 | $+1.91 \cdot 10^{+01}$ |
| 8 | kd2fold | -5 | +0.3368 | +3 | 1 | $+2.17 \cdot 10^{+00}$ |
| 9 | kin | -5 | -0.4139 | +3 | 1 | $+3.86 \cdot 10^{-01}$ |
| 10 | kin2 | -5 | -0.2710 | +3 | 1 | $+5.36 \cdot 10^{-01}$ |
| 11 | km_nExpID1 | +3 | +8.0000 | +8 | 1 | $+1.00 \cdot 10^{+08}$ |
| 12 | km_nExpID2 | +3 | +8.0000 | +8 | 1 | $+1.00 \cdot 10^{+08}$ |
| 13 | km_nExpID3 | +3 | +8.0000 | +8 | 1 | $+1.00 \cdot 10^{+08}$ |
| 14 | km_nExpID4 | +3 | +8.0000 | +8 | 1 | $+1.00 \cdot 10^{+08}$ |
| 15 | koff_unspec | -5 | +1.0153 | +3 | 1 | $+1.04 \cdot 10^{+01}$ |
| 16 | kon_unspec | -5 | +1.3195 | +3 | 1 | $+2.09 \cdot 10^{+01}$ |
| 17 | kout | -5 | -1.3409 | +3 | 1 | $+4.56 \cdot 10^{-02}$ |
| 18 | kout2 | -5 | -1.4491 | +3 | 1 | $+3.56 \cdot 10^{-02}$ |
| 19 | kout_frag | -5 | -1.9578 | +3 | 1 | $+1.10 \cdot 10^{-02}$ |
| 20 | offset | -5 | +1.1636 | +3 | 1 | $+1.46 \cdot 10^{+01}$ |
| 21 | offset_nExpID1 | -5 | -1.5999 | +5 | 1 | $+2.51 \cdot 10^{-02}$ |
| 22 | offset_nExpID2 | -5 | -1.6627 | +5 | 1 | $+2.17 \cdot 10^{-02}$ |
| 23 | offset_nExpID3 | -5 | -2.1357 | +5 | 1 | $+7.32 \cdot 10^{-03}$ |
| 24 | offset_nExpID4 | -5 | +0.1948 | +5 | 1 | $+1.57 \cdot 10^{+00}$ |
| 25 | scale | -5 | -0.8789 | +3 | 1 | $+1.32 \cdot 10^{-01}$ |
| 26 | scaleElisa_nExpID1 | +0.1 | +0.5380 | +1 | 0 | $+5.38 \cdot 10^{-01}$ |
| 27 | scaleElisa_nExpID2 | +0.1 | +0.5697 | +1 | 0 | $+5.70 \cdot 10^{-01}$ |
| 28 | scaleElisa_nExpID3 | +0.1 | +0.1000 | +1 | 0 | $+1.00 \cdot 10^{-01}$ |
| 29 | scaleElisa_nExpID4 | +0.1 | +1.0000 | +1 | 0 | $+1.00 \cdot 10^{+00}$ |
| 30 | std | -5 | -0.5826 | +3 | 1 | $+2.61 \cdot 10^{-01}$ |

**Table 19: Estimated parameter values**
$\hat{\theta}$ indicates the estimated value of the parameters. $\theta_{min}$ and $\theta_{max}$ indicate the upper and lower bounds for the parameters. The log-column indicates if the value of a parameter was log-transformed. If log $\equiv 1$ the non-log-column indicates the non-logarithmic value of the estimate.

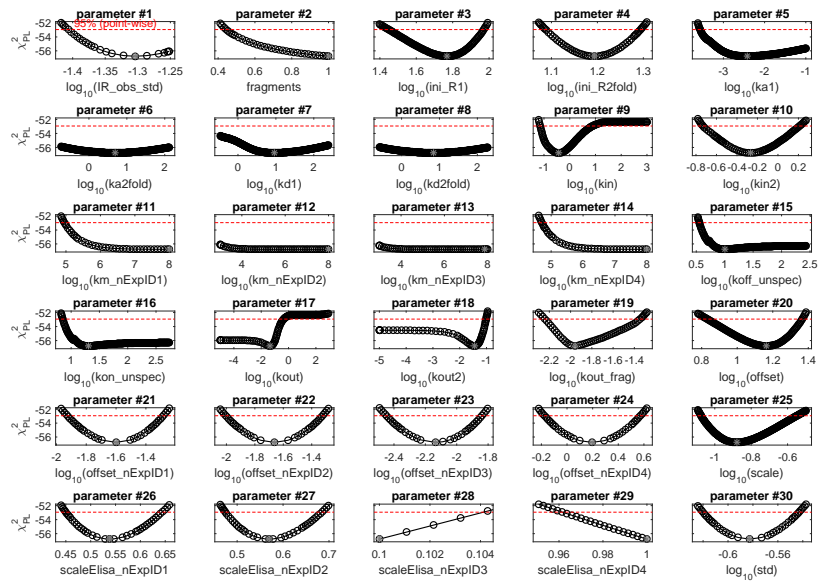

**Figure 17: Overview of the profile likelihood of the model parameters**
The solid lines indicate the profile likelihood. The broken lines indicate the threshold to assess confidence intervals. The asterisks indicate the optimal parameter values.

```
>> arIdentifiabilityTest
Identifiability-test started ...
... Identifiability-test finished.


Identifiability-test was performed with radius = 1 and penalty-SD = 1.


All 4 optimization runs with random intial guesses are in the chi2-range 1.3209e-06.


Calculations took 8.40 seconds.
[Compared to 9002.76 seconds required for calcuating the likelihood profiles.]


1.0000 (increase of merit by penalty, before fitting)
1.1396 (decrease of merit by fitting)
1.0006 (movement of parameters by penalized fitting)
-0.1396 (total increase of merit by penalty) PRIMARY CRITERION
Model is structurally non-identifiable.
```



\end{document}